\documentclass[usenatbib,useAMS]{mnras}

\usepackage{array}

\usepackage{enumerate}

\usepackage{aas_macros}

\usepackage{amsmath}
\usepackage{mathtools}
\usepackage{IEEEtrantools}
\usepackage{float}

\usepackage{txfonts}

\usepackage[applemac]{inputenc}
\usepackage[T1]{fontenc} 

\usepackage[english]{babel}
\usepackage{booktabs}
\usepackage{multirow}

\usepackage{savesym}
\savesymbol{iint}
\savesymbol{iiint}
\savesymbol{iiiint}
\savesymbol{idotsint}

\usepackage{amssymb,amsmath}

\usepackage{amstext}
\usepackage{graphicx}
\usepackage{caption}
\usepackage{subcaption}
\usepackage[np,noautolanguage]{numprint}

\usepackage{bigstrut}

\usepackage{scalefnt}
\usepackage{letltxmacro}
\usepackage{color}

\usepackage{mathrsfs}

\hypersetup{breaklinks, colorlinks, linktocpage=true, linkcolor=blue, citecolor=blue, filecolor=black, urlcolor=blue}

\usepackage{relsize}

\usepackage{etoolbox}
\makeatletter

\patchcmd{\NAT@citex}
  {\@citea\NAT@hyper@{%
     \NAT@nmfmt{\NAT@nm}%
     \hyper@natlinkbreak{\NAT@aysep\NAT@spacechar}{\@citeb\@extra@b@citeb}%
     \NAT@date}}
  {\@citea\NAT@nmfmt{\NAT@nm}%
   \NAT@aysep\NAT@spacechar\NAT@hyper@{\NAT@date}}{}{}

\patchcmd{\NAT@citex}
  {\@citea\NAT@hyper@{%
     \NAT@nmfmt{\NAT@nm}%
     \hyper@natlinkbreak{\NAT@spacechar\NAT@@open\if*#1*\else#1\NAT@spacechar\fi}%
       {\@citeb\@extra@b@citeb}%
     \NAT@date}}
  {\@citea\NAT@nmfmt{\NAT@nm}%
   \NAT@spacechar\NAT@@open\if*#1*\else#1\NAT@spacechar\fi\NAT@hyper@{\NAT@date}}
  {}{}

\makeatother

\bibpunct{(}{)}{;}{a}{}{,} 

\usepackage{threeparttable}

\numberwithin{equation}{section}

\newcolumntype{L}[1]{>{\raggedright\let\newline\\\arraybackslash\hspace{0pt}}m{#1}}
\newcolumntype{C}[1]{>{\centering\let\newline\\\arraybackslash\hspace{0pt}}m{#1}}
\newcolumntype{R}[1]{>{\raggedleft\let\newline\\\arraybackslash\hspace{0pt}}m{#1}}

\newcommand{\unsim}{\mathord{\sim}}

\newcommand{\vect}[1]{\boldsymbol{#1}}

\renewcommand{\bmath}[1]{\mbox{ \boldmath $\!#1\!$ \unboldmath}}
\renewcommand{\mathbfss}[1]{\mathsf{\mathbf{#1}}}

\newcommand{\txn}[1]{\textnormal{#1}}

\newcommand{\Planck}{\textit{Planck}}

\LetLtxMacro{\oldtextsc}{\textsc}
\renewcommand{\textsc}[1]{\oldtextsc{\scalefont{1.2}#1}}

\newcommand{\cloudy}{\textsc{cloudy}}
\newcommand{\pybeagle}{\textsc{pyp-beagle}}
\newcommand{\beagle}{\textsc{beagle}}
\newcommand{\multinest}{\textsc{multinest}}
\newcommand{\eazy}{\textsc{eazy}}
\newcommand{\bpz}{\textsc{bpz}}
\newcommand{\getdist}{\textsc{getdist}}

\newcommand{\lambdao}{\hbox{$\lambda_\txn{obs}$}}

\newcommand{\zbpz}{\hbox{$z_\textsc{bpz}$}}
\newcommand{\zeazy}{\hbox{$z_\textsc{eazy}$}}
\newcommand{\zbeagle}{\hbox{$z_\textsc{beagle}$}}

\newcommand{\SN}{\hbox{$\txn{S}/\txn{N}$}}
\newcommand{\SNtot}{\hbox{${(\SN)}_\txn{tot}$}}
\newcommand{\SNave}{\hbox{$\langle \SN \rangle$}}
\newcommand{\Gyr}{\hbox{$\txn{Gyr}$}}

\newcommand{\ergs}{\hbox{$\txn{erg}\,\txn{s}^{-1}\,\txn{\AA}^{-1}$}}

\newcommand{\thetab}{\hbox{$\bmath{\Theta}$}}
\newcommand{\Db}{\hbox{$\bmath{D}$}}

\newcommand{\M}{\hbox{$\txn{M}$}}

\newcommand{\zref}{\hbox{$z_\txn{ref}$}}
\newcommand{\zspec}{\hbox{$z_\txn{spec}$}}

\newcommand{\sigNMAD}{\hbox{$\sigma_\txn{NMAD}$}}

\newcommand{\K}{\hbox{$K$}}

\newcommand{\Lya}{\hbox{Ly$\alpha$}}

\newcommand{\Mstar}{\hbox{$\txn{M}_{\ast}$}}
\newcommand{\Msun}{\hbox{$\txn{M}_{\sun}$}}

\newcommand{\ism}{\textsc{ism}}
\newcommand{\bc}{\textsc{bc}}

\newcommand{\tprime}{\hbox{$t^{\prime}$}}
\newcommand{\tform}{\hbox{$t_\txn{form}$}}
\newcommand{\zform}{\hbox{$z_\txn{form}$}}
\newcommand{\aFebra}{\hbox{$[\alpha/\txn{Fe}]$}}
\newcommand{\aFe}{\hbox{$\alpha/\txn{Fe}$}}
\newcommand{\FeH}{\hbox{$[\txn{Fe}/\txn{H}]$}}

\newcommand{\Zism}{\hbox{$Z_\ism$}}
\newcommand{\Zsun}{\hbox{$Z_\odot$}}
\newcommand{\Zyoung}{\hbox{$Z_\txn{young}$}}
\newcommand{\Us}{\hbox{$U_\txn{S}$}}
\newcommand{\xid}{\hbox{$\xi_\txn{d}$}}
\newcommand{\vsys}{\hbox{$v_\txn{sys}$}}

\newcommand{\muLSF}{\hbox{$\mu_\txn{LSF}$}}
\newcommand{\sigLSF}{\hbox{$\sigma_\txn{LSF}$}}

\newcommand{\Omegam}{\hbox{$\Omega_{\mathrm{m}}$}}

\newcommand{\Omegal}{\hbox{$\Omega_{\Lambda}$}}

\newcommand{\pvalue}{\hbox{$p\txn{-value}$}}
\newcommand{\Nrep}{\hbox{$\txn{N}_\txn{rep}$}}
\newcommand{\N}{\hbox{$\txn{N}$}}
\newcommand{\Nout}{\hbox{$\txn{N}_\txn{out}$}}
\newcommand{\yrep}{\hbox{$\bmath{y}^\txn{rep}$}}

\newcommand{\OLF}{\hbox{$\textnormal{OLF}$}}

\newcommand{\tauLthT}{\hbox{$\hat{\tau}_\lambda(t^\prime,\theta)$}}

\newcommand{\nVism}{\hbox{$n^\ism_V$}}
\newcommand{\nLism}{\hbox{$n^\ism_\lambda$}}
\newcommand{\tauLism}{\hbox{$\hat{\tau}^\ism_\lambda$}}
\newcommand{\tauLismTh}{\hbox{$\hat{\tau}^\ism_\lambda(\theta)$}}
\newcommand{\tauLbc}{\hbox{$\hat{\tau}^\bc_\lambda$}}
\newcommand{\tauV}{\hbox{$\hat{\tau}_V$}}
\newcommand{\tauVbc}{\hbox{$\hat{\tau}^\bc_V$}}
\newcommand{\tauVism}{\hbox{$\hat{\tau}^\ism_V$}}

\newcommand{\tauLismTt}{\hbox{$\hat{\tau}^\ism_\lambda(\theta,\tprime)$}}
\newcommand{\Tthin}{\hbox{$\txn{t}_\txn{thin}$}}
\newcommand{\Tthick}{\hbox{$\txn{t}_\txn{thick}$}}
\newcommand{\Tbulge}{\hbox{$\txn{t}_\txn{bulge}$}}
\newcommand{\tauLthinth}{\hbox{$\hat{\tau}^\txn{thin}_\lambda(\theta)$}}
\newcommand{\tauLthickth}{\hbox{$\hat{\tau}^\txn{thick}_\lambda(\theta)$}}
\newcommand{\tauLbulgeth}{\hbox{$\hat{\tau}^\txn{bulge}_\lambda(\theta)$}}

\newcommand{\Hi}{\mbox{H\,{\sc i}}}
\newcommand{\Hii}{\mbox{H\,{\sc ii}}}

\newcommand{\Tssp}{\hbox{$\txn{t}_\txn{SSP}$}}
\newcommand{\Tstart}{\hbox{$\txn{t}_\txn{start}$}}
\newcommand{\Tend}{\hbox{$\txn{t}_\txn{end}$}}
\newcommand{\tausfr}{\hbox{$\tau_\txn{SFR}$}}

\newcommand{\Tuniverse}{\hbox{$\txn{t}_{\textsc{u},z}$}}

\title[Modelling and interpreting galaxy SEDs with BEAGLE]{Modelling and interpreting spectral energy distributions of galaxies with BEAGLE}
\author[J. Chevallard]
{Jacopo~Chevallard$^{1}$\thanks{E-mail: jchevall@cosmos.esa.int}\thanks{ESA Research Fellow} and
St\'ephane Charlot$^{2}$
\\
\\
$^{1}$Scientific Support Office, Directorate of Science and Robotic Exploration, ESA/ESTEC, Keplerlaan 1, 2201 AZ Noordwijk, The Netherlands\\
$^{2}$Sorbonne Universit\'es, UPMC-CNRS, UMR7095, Institut d'Astrophysique de Paris, F-75014, Paris, France
}

\begin{document}

\date{Submitted to MNRAS on }

\maketitle

\label{firstpage}

\begin{abstract}

We present a new-generation tool to model and interpret spectral energy distributions (SEDs) of galaxies, which incorporates in a consistent way the production of radiation and its transfer through the interstellar and intergalactic media. This flexible tool, named \beagle\ (for BayEsian Analysis of GaLaxy sEds), allows one to build mock galaxy catalogues as well as to interpret any combination of photometric and spectroscopic galaxy observations in terms of physical parameters. The current version of the tool includes versatile modeling of the emission from stars and photoionized gas, attenuation by dust and accounting for different instrumental effects, such as spectroscopic flux calibration and line spread function. We show a first application of the \beagle\ tool to the interpretation of broadband SEDs of a published sample of $\unsim10^4$ galaxies at redshifts $0.1 \lesssim z\lesssim8$. We find that the constraints derived on photometric redshifts using this multi-purpose tool are comparable to those obtained using public, dedicated photometric-redshift codes and quantify this result in a rigorous statistical way. We also show how the post-processing of \beagle\ output data with the Python extension \pybeagle\ allows the characterization of systematic deviations between models and observations, in particular through posterior predictive checks. The modular design of the \beagle\ tool allows easy extensions to incorporate, for example, the absorption by neutral galactic and circumgalactic gas, and the emission from an active galactic nucleus, dust and shock-ionized gas. Information about public releases of the \beagle\ tool will be maintained on \url{http://www.jacopochevallard.org/beagle}.\end{abstract}


\section{Introduction}\label{sec:intro}

Over the last 15 years, our understanding of how galaxies form and evolve has improved substantially. The combination of technological and theoretical progress has brought this field into a new era: advances in observational techniques (e.g. multi-object spectroscopy, efficient near-infrared CCDs) have enabled multi-wavelength observations of large samples of galaxies out to the highest redshifts, while the steady rise of computational power and refinement of numerical techniques have fostered new approaches (e.g. semi-analytic models, hydro-dynamic simulations) to model the formation and evolution of galaxies. This progress has led to a general consensus about the main physical ingredients required to describe the evolution of the galaxy population \citep[e.g.][]{Lu2014, Vogelsberger2014, Gonzalez2014, Henriques2015, Schaye2015}:  collapse and hierarchical growth of dark matter haloes; accretion of baryons onto these haloes; conversion of baryons into stars; feedback of massive stars and active galactic nuclei (AGN) on star formation; supernova- and AGN-driven outflows of metal-enriched gas; infall of both pristine and metal-enriched gas onto galaxies. The large-scale environment can also affect galaxy properties, in particular, by providing quenching mechanisms (e.g., tidal or ram-pressure stripping, strangulation; e.g. \citealt{Lagos2014, Rafieferantsoa2015}), and through its influence on the merger rate \citep[e.g.][]{Lackner2012, Rafieferantsoa2015} and galactic spins \citep[e.g.][]{Hahn2010, Codis2012}. Although these different ingredients are present in many galaxy formation models, we still lack a detailed quantification of their respective roles in shaping the properties of galaxies. This is because of the complexity inherent in galaxy physics, which combines gravity, radiation hydro-dynamics, magnetic fields and high-energy physics, acting on scales from less than a pc (e.g., for the formation of proto-stellar cores) to over a Mpc (e.g., for environmental effects). For this reason, `first-principles' simulations of galaxy formation remain far beyond the reach of current computational capabilities. Instead, small-scale baryonic physics is generally subsumed into sub-grid prescriptions, which vary from model to model \citep[e.g.][]{Scannapieco2012, Vogelsberger2013, Haas2013a, Haas2013b, Torrey2014, Crain2015}. The appropriateness of such prescriptions, and hence, our ability to understand galaxy formation, must be assessed by comparing simulated and observed galaxy properties.

Comparing the predictions of galaxy formation models with observations requires one to relate properties pertaining to the evolution of baryons in dark-matter haloes, such as gas cooling and star formation, with observables, such as  ultraviolet, optical and infrared spectral energy distributions (SEDs). This can be achieved using models of stellar population synthesis and of the transfer of starlight through the interstellar and intergalactic media \citep[e.g.][]{Tinsley1978, Bruzual1983, Arimoto1987, Guiderdoni1987, Buzzoni1989, Bressan1994, Worthey1994, Leitherer1995, Fioc1997, BC03, Maraston2005, Vazdekis2010, Conroy2010b, Maraston2011}. This spectral modelling involves several additional components, such as the stellar initial mass function, prescriptions for the evolution, spectral properties and release of heavy elements by individual stars of different initial masses and chemical compositions, and prescriptions for the influence of the interstellar medium (ISM) and the intergalactic medium (IGM) on stellar radiation. Galaxy SEDs can be computed in this way from the histories of star formation and chemical enrichment predicted by galaxy formation models. The interpretation of photometric and spectroscopic galaxy observations with such model SEDs to constrain stellar masses, metallicities, star formation histories and ionized-gas properties is at the base of most galaxy evolution studies. 

Two major limitations, often neglected, affect this type of analysis: the adoption of oversimplified models to describe the wide variety of observed galaxy SEDs and the presence of `systematic' model uncertainties. This second limitation has been addressed in several studies already \citep[e.g.][]{Charlot1996, Cervino2000, Percival2009, Conroy2009, Conroy2010a, Conroy2010b}. The difficulty of precisely quantifying systematic model uncertainties has led to mainly qualitative conclusions, leaving the problem unsolved. The first limitation is easier to tackle, for example, by using more physically realistic models of galaxy SEDs and combining these with advanced statistical techniques to extract physical constraints from data. This appears as the most promising route to fully exploit the information gathered by modern photometric and spectroscopic galaxy surveys. Yet, the several tools proposed so far to interpret galaxy SEDs in terms of physical parameters do not allow one to fully exploit the high quality of modern data. For example, most existing approaches rely on the adoption of a rigid physical model (e.g., analytic, two-parameter star formation histories combined with a standard dust attenuation curve and the assumption that all stars in a galaxy have the same metallicity) to describe galaxy SEDs \citep[e.g.][]{Bolzonella2010, Wuyts2011, Ilbert2013, Muzzin2013, Hernan2013, Bauer2013, Lundgren2014, Mortlock2015, Kawinwanichakij2015, Kochiashvili2015}. Even with the inclusion of superimposed bursts of star formation \citep[e.g.][]{Kauffmann2003, Gallazzi2005, Pozzetti2007, Gallazzi2009, daCunha2010}, this does not allow a physically consistent description of the contributions by stars, gas and dust to the integrated emission from a galaxy, nor the inclusion of a potential AGN component (a notable exception is the approach of \citealt{Pacifici2012}, who incorporate star formation and chemical enrichment histories from numerical simulations of galaxy formation and emission from photo-ionized gas). Also, current spectral analysis tools are generally optimized to interpret either photometric or spectroscopic observations of galaxies, but not arbitrary combinations thereof. Finally, most existing tools suffer from additional limitations: many focus on the selection of `best-fitting' parameters rather than on the uncertainties associated with these parameters (e.g., chi-square minimisation techniques; \citealt{Arnouts1999, Bolzonella2000, Kriek2009}); when this is not the case, the number of free parameters that can be explored is generally limited (e.g., with grid-based Bayesian techniques; \citealt{daCunha2008, Noll2009, Pacifici2012}); and when more sophisticated (e.g., Markov Chain Monte Carlo, hereafter MCMC) techniques allow the exploration of more parameters, instrumental effects are generally not incorporated in the analysis \citep[e.g.][]{Serra2011, Acquaviva2011, Han2014}. 


In  this paper, we introduce a new-generation tool to interpret galaxy SEDs, \beagle\ (for BayEsian Analysis of GaLaxy sEds), which incorporates several main novelties. The modular design of this tool, written in Fortran 2003/08, allows one to easily combine, in a physically consistent way, different prescriptions for the production of starlight in galaxies and its transfer through the ISM (absorption and emission by gas, attenuation by dust) and the IGM (absorption by gas). Other modules to be implemented in the future include the infrared emission from dust and the emission from an AGN. The \beagle\ tool includes several possible prescriptions to describe the star formation and chemical enrichment histories of galaxies, ranging from simple analytic functions to the predictions of sophisticated galaxy formation models. The flexible parametrization of these and other model parameters allows one to build mock galaxy catalogues as well as to interpret any combination of photometric and spectroscopic observation of galaxies by adapting model complexity (i.e., number of free parameters) to the data, without sacrificing coherence. Moreover, the adopted Bayesian framework allows the characterisation of complex, non-linear correlations among model parameters in high-dimensional parameters spaces, comparisons between competing models, the description within a hierarchical framework of single objects as well as populations of objects, and a reduction of the parameter space using informative priors based on high-quality observations. Finally, posterior predictive checks comparing model predictions with observations enable one to identify model failures, which can drive future model developments.

In Section~\ref{sec:model} below, we outline the modular, highly versatile approach used to describe the physical properties of stars, gas and dust in the \beagle\ tool. In Section~\ref{sec:beagle}, we describe in detail the statistical approach and output products arising from the analysis of  galaxy SEDs with this tool. In Section~\ref{sec:SED_fitting} we present an example application of \beagle\ to the interpretation of broadband SEDs of galaxies in a wide range of redshifts and discuss the results of such modelling in Section~\ref{sec:discussion}. We also introduce also a Python-based extension to post-process \beagle\ results, named \pybeagle\ (for PYthon Postprocessing of \beagle). We compare our approach with existing SED fitting codes in Section~\ref{sec:comparison} and summarize our conclusions in Section~\ref{sec:conclusions}. Throughout the paper, we adopt a present-day solar metallicity $\Zsun=0.01524$ (corresponding to a zero-age metallicity of 0.017, see Table~3 of \citealt{Bressan2012}) and the latest constraints on cosmological parameters from  \Planck, i.e. $\Omegal = 0.6911$,  $\Omegam = 0.3089$, and $H_0 = 67.74$ \citep[see last column `TT,TE,EE+lowP+lensing+ext' of Table 4 of][]{Planck2015}.

\section{Astrophysical ingredients of BEAGLE}\label{sec:model}

Our main aim in this paper is to design a general, astrophysically sound framework to fit any combination of photometric and spectroscopic galaxy observations, as well as to produce and analyse mock catalogues of galaxy SEDs. The need for a single framework to study both true and mock galaxy data arises from the growing role played by simulations in the optimal preparation and exploitation of  modern surveys. Simulations are important to optimise the observational strategy (e.g. signal-to-noise thresholds, spectral resolution) needed to answer a given scientific question, but also to characterise the performances and systematics of the increasingly complex instruments mounted on new ground-based and space-based telescopes. 

The ultraviolet, optical and infrared SEDs of galaxies include contributions by stars, gas, dust and potentially an AGN. Since these components are physically linked to one another, they must be described in a consistent way in spectral models of galaxies. In this section, we present our formalism to model the production of starlight in galaxies and its transfer through the ISM and the IGM. We appeal to the isochrone synthesis technique introduced by \citet{Charlot1991}.\footnote{The term `isochrone' refers to the location of coeval stars with homogeneous chemical composition and different masses in the Hertzsprung-Russell diagram.} In this approach, the luminosity (in units of \ergs) emitted at wavelength $\lambda$ by a galaxy at time $t$ from the onset of star formation can be expressed as
\begin{equation}\label{eq:sed_model}
L_\lambda(t) = \int_0^t d\tprime \, \psi(t-\tprime) \, S_\lambda[\tprime,\vect{Z}(t-\tprime)] \, T^\ism_\lambda(t,t^\prime) \, ,
\end{equation} 
where $\psi(t-\tprime)$ is the star formation rate at time $t-\tprime$ (which traces the star formation history), $S_\lambda[\tprime,\vect{Z}(t-t^\prime)]$ the luminosity emitted per unit wavelength per unit mass by a simple stellar population (SSP) of age $t^\prime$ and chemical composition $\vect{Z}(t-t^\prime)$ (each element of this vector corresponding to a different chemical element), and $T^\ism_\lambda(t,t^\prime)$ the transmission function of the ISM. The function $S_\lambda(\tprime,\vect{Z})$ can be expressed as \citep[e.g.][]{Conroy2009}
\begin{equation}\label{eq:lumin}
S_\lambda(\tprime,\vect{Z}) = \int_{m_\txn{low}}^{m_\txn{up}} dm \, \phi(m) \, \Lambda_\lambda[L_\txn{bol}(m,\vect{Z},\tprime),T_\txn{eff}(m,\vect{Z},\tprime),\vect{Z}] \, ,
\end{equation}
where $m$ is the stellar mass, $\phi(m)$ the stellar initial mass function [IMF; defined such that $\phi(m)dm$ is the number of stars born with masses between $m$ and $m+dm$] with lower and upper mass cutoffs $m_\txn{low}$ and $m_\txn{up}$, and  $\Lambda_\lambda[L_\txn{bol}(m,\vect{Z},\tprime),T_\txn{eff}(m,\vect{Z},\tprime),\vect{Z}]$  the spectral energy distribution of a star with bolometric luminosity $L_\txn{bol}(m,\vect{Z},\tprime)$, effective temperature $T_\txn{eff}(m,\vect{Z},\tprime)$, and chemical composition $\vect{Z}$.

Following \citet{Charlot2001}, we express the transmission function in the ISM as
\begin{equation}\label{eq:transmiss}
T^\ism_\lambda(t,t^\prime) =  T^+_\lambda(t,t^\prime) \, T^0_\lambda(t,t^\prime) \, ,
\end{equation}
where $T^+_\lambda(t,t^\prime)$ and $T^0_\lambda(t,t^\prime)$ are the transmission functions of the ionized gas and the neutral ISM, respectively. Here $T^+_\lambda(t,t^\prime)$ accounts for both the absorption and the emission of photons in the ionized gas, i.e., $T^+_\lambda(t,t^\prime)$ will be close to zero at wavelengths shorter than the H-Lyman limit and greater than unity at those corresponding to emission lines. The transmission functions of the neutral interstellar gas and the IGM will be discussed separately in Sections~\ref{sec:dust} and \ref{sec:IGM} below.

For a galaxy at redshift $z$, the spectral flux density (in units of $\txn{erg}\,\txn{s}^{-1}\,\txn{\AA}^{-1}\,\txn{cm}^{-2}$) reaching the observer at wavelength $\lambda_\txn{obs} = \lambda\,(1+z)$ is related to the luminosity $L_\lambda$ emitted at rest wavelength $\lambda$ (equation~\ref{eq:sed_model}) by the relation
\begin{equation}\label{eq:flux}
F_{\lambda_\txn{obs}} = \frac{L_\lambda}{4\pi d_L(z)^2}\frac{1}{1+z} \, T^\txn{IGM}_{\lambda, z}  \, ,
\end{equation}
where $d_L$ is the luminosity distance at redshift $z$, $T^\txn{IGM}_{\lambda, z}$ the (redshift-dependent) transmission function of the IGM, and the factor $1/(1+z)$ ensures energy conservation, by accounting for the wavelength stretching $d\lambdao = (1+z)\,d\lambda$. 

In the next sections, we describe our prescriptions for the emission from stars and the ionized gas, the star formation and chemical enrichment histories, changes in the $\alpha$-element to iron abundance ratio (\aFe), attenuation by dust and IGM absorption. We stress that a main feature allowed by the modular design of the \beagle\ tool is that the prescriptions adopted here for these different physical ingredients can be easily replaced by alternative ones. We do not discuss here our prescriptions for an AGN component and infrared dust emission, which are currently being incorporated and will be the subject of a future release (Section~\ref{sec:conclusions}). 

\subsection{Stellar population synthesis code}\label{sec:stelpops}

We adopt the latest version of the \citet{BC03} code to describe the emission of stellar populations at wavelengths between 91 \AA\ and 190 \micron\ (equation~\ref{eq:lumin}). This version of the code incorporates updated stellar evolutionary tracks \citep{Bressan2012, Marigo2013} and the MILES library of observed optical stellar spectra \citep{Sanchez2006} to describe the properties of stars in the Hertzsprung-Russell diagram [$ \Lambda_\lambda(L_\txn{bol},T_\txn{eff},\vect{Z})$ in equation~\eqref{eq:lumin}]. We compute in this way SSPs in wide ranges of ages, $t^\prime \in [10^5, 2\times10^{10}]$ yr, and metallicities, $Z \in [0.0001,0.04]$ (i.e. $\FeH \in [-2.18, 0.42] $).\footnote{Here $Z$ stands for the sum of the elements of vector $\vect{Z}$, which corresponds to the mass fraction in all elements heavier than helium.}

\subsection{Stellar initial mass function}\label{sec:imf}

The stellar IMF [term $\phi(m)$ in equation~\ref{eq:lumin}] controls the relative contributions by stars of different masses to the integrated SED of a galaxy. Recently, several studies have claimed the presence of systematic variations of the IMF, such as a potential steepening in early-type galaxies \citep[e.g.,][]{vanDokkum2010, Conroy2012b, Cappellari2012, LaBarbera2013, Sonnenfeld2015}, for which theoretical motivations exist \citep[e.g.,][]{Chabrier2014}. The implication of these claims for stellar population synthesis modelling are being complicated by the fact that different studies produce inconsistent results \citep{Smith2014, Zieleniewski2015} and may depend on underlying assumptions \citep{Tang2015, McConnell2016}. In this paper, we adopt a fixed IMF corresponding to the Galactic-disc IMF of \citet{Chabrier2003}, with lower and upper mass cutoffs $m_\txn{low} = 0.1$ and $m_\txn{up} = 100$ \Msun. We note that the \citet{Chabrier2003} IMF is very similar to the `universal' IMF of \citet{Kroupa2001} and differs from the classical \citet{Salpeter1955} IMF [$\phi(m)\propto m^{-2.35}$] through the turn-over of the distribution $\phi(m)$ at stellar masses $m<1 \; \Msun$.

\subsection{$\alpha$ element-to-iron abundance ratio}

Many spectral evolution models available in the literature (e.g. P\'egase, \citealt{Fioc1997, LeBorgne2004}; \textsc{galaxev}, \citealt{BC03}; \textsc{fsps}, \citealt{Conroy2009, Conroy2012a}; \citealt{Vazdekis2010}; MaStro, \citealt{Maraston2011}) rely on libraries of observed spectra of solar-neighbourhood stars (e.g. ELODIE, \citealt{Prugniel2001, Prugniel2007}; STELIB, \citealt{LeBorgne2003}; Indo-US, \citealt{Valdes2004}; MILES, \citealt{Sanchez2006}). This is because current theoretical spectral libraries suffer from persisting inaccuracies linked to, for example, the inclusion of line blanketing, incomplete line lists, the accuracy of quantum-mechanical calculations, the treatment of convection in stellar interiors (see e.g. \citealt{Martins2007}). The problem of appealing to observed spectral libraries is that the abundances of solar-neighbourhood stars are tied to the particular star formation and chemical enrichment history at our location in the Milky Way. This potentially biases the interpretation with such libraries of galaxies with different star formation and chemical enrichment histories, such as massive, early-type galaxies, which have typically larger \aFe\ ratio than star-forming, late-type galaxies like the Milky Way \citep[e.g.][]{Milone2000, Worthey2014}.

A way to overcome the limitations of observed spectral libraries is to adopt theoretical libraries to compute differential properties of stellar populations with non-scaled solar element abundance ratios relative to the well-calibrated scaled-solar case. Following \citet[see also \citealt{Vazdekis2015}]{Walcher2009}, we account for the dependence of the function $S_\lambda$ on \aFe\ ratio in equation~\eqref{eq:sed_model} by writing
\begin{equation}\label{eq:alpha}
S_\lambda(\tprime,Z,\aFebra) = \frac{S^\txn{theor}_\lambda(\tprime,Z,\aFebra)}{S^\txn{theor}_\lambda(\tprime,Z)} S_\lambda(\tprime,Z) \, ,
\end{equation}
where $S_\lambda(\tprime,Z)$ is the luminosity per unit wavelength of an SSP of age $\tprime$ and metallicity $Z$ computed using an empirical library of stellar spectra (Section~\ref{sec:stelpops}), for which $\aFe\approx\aFe_\odot$ (i.e. $\aFebra\approx0$), and $S^\txn{theor}_\lambda(\tprime,Z)$ and $S^\txn{theor}_\lambda(\tprime,Z,\aFebra)$ are the theoretical predictions for SSps of same age $\tprime$ and metallicity $Z$ for scaled-solar ($\aFebra=0$) non-scaled solar ($\aFebra\neq0$) $\alpha$ element-to-iron abundance ratio, respectively. We adopt for $S^\txn{theor}_\lambda(\tprime,Z)$ and $S^\txn{theor}_\lambda(\tprime,Z,\aFebra)$ the predictions of \citet[][as updated by \citealt{Coelho2014}]{Coelho2007}, which  
cover the wavelength range $\lambda \in [0.13, 100]$~\micron\ at low resolution, and $\lambda \in [0.25, 0.9]$~\micron\ at high resolution, ages between 30 Myr and 14 Gyr, iron abundances $\FeH \in [-1, 0.2]$ and $\alpha$ elements-to-iron abundance ratios $\aFebra \in [0, 0.4]$.

\subsection{Nebular emission}\label{sec:nebular}

We adopt the recent photoionization models of star-forming galaxies of \citet[][who follow the prescription of \citealt{Charlot2001}]{Gutkin2016} to describe the transmission function of the ionized gas $T^+_\lambda(t,t^\prime)$ in equation~\eqref{eq:transmiss}. These models combine the stellar population synthesis code described in Section~\ref{sec:stelpops} above with the latest version of the photoionization code \cloudy\ (version 13.3, last described by \citealt{Ferland2013}). In this approach, the ensemble of \Hii\ regions and the diffuse gas ionized by young stars throughout a galaxy are described by means of effective (i.e. galaxy-wide) parameters. The main adjustable parameters of the photoionized gas are the interstellar metallicity, \Zism, the typical ionization parameter of a newly ionized \Hii\ region, \Us\ (which characterizes the ratio of ionizing-photon to gas densities at the edge of the Stroemgren sphere), and the dust-to-metal (mass) ratio, \xid\ (which characterizes the depletion of metals on to dust grains).  We adopt the large grid of models computed by \citet{Gutkin2016} for $\Zism \in [0.0001,0.04]$ (i.e., $\FeH \in [-2.18, 0.42] $), $\log \Us \in [-4, -1]$ and  $\xid \in [0.1, 0.5]$. These provide the nebular (lines+continuum) emission at all SSP ages less than $t'=10\,$Myr (equation~\ref{eq:transmiss}), over which 99.9 per cent of the ionizing photons are released by a single stellar generation. Hence, in this prescription, $T^+_\lambda(t,t^\prime)=1$ for $t'>10\,$Myr. We do not include in the present paper the nebular emission produced by faint, post-asymptotic-giant-branch stars, shocks and a potential AGN component (these will be implemented in future releases of the \beagle\ tool).

\subsection{Star formation and chemical enrichment histories}\label{sec:sfh_ceh}

The star formation history $\psi(t-\tprime)$ and chemical enrichment history $Z(t-\tprime)$ entering the expression of the spectral energy distribution $L_\lambda(t)$ of a galaxy (equation~\ref{eq:sed_model}) depend on the complex physical processes affecting baryons trapped in dark-matter potential wells (see Section~\ref{sec:intro}). In principle, one should be able to constrain $\psi$ and $Z$ by comparing the observed SED of the galaxy to the predictions of spectral evolution models corresponding to different star formation and chemical enrichment histories \citep[e.g.][]{Heavens2004, Panter2007, Cid2007, Pacifici2013, Tojeiro2013}. This requires an approach including the broadest possible ranges in $\psi$ and $Z$, enabling one to identify the parametrization best describing the observations as well as potential biases linked to the adoption of less appropriate assumptions \citep[e.g.][]{Pacifici2015}. 

A common way to parametrize the star formation and chemical enrichment histories of galaxies is to adopt a smooth, analytic (e.g. constant, exponentially declining or rising) function to describe $\psi(t-\tprime)$ and a fixed metallicity $Z$ to describe the chemical composition of stars of all ages in a galaxy. Although such parametrization may be adequate to describe the average properties of different galaxy classes (e.g. massive galaxies of different morphological types in the nearby Universe), they cannot account for the complex star formation and chemical enrichment histories of individual galaxies in a hierarchical universe. As an example, the Milky Way is known to have experienced different bursts of star formation in the past \citep[e.g.][]{Rocha2000b} and to currently harbour stars spanning a broad range of chemical compositions \citep[e.g.][]{Rocha2000a}.\footnote{The superposition of `stochastic' bursts on top of smooth star formation histories, has been shown to better reproduce the observed spectroscopic properties of individual galaxies \citep[e.g.][]{Kauffmann2003, Brinchmann2004}, but still assuming that all stars in a given galaxy have a fixed metallicity.} 

This complexity may be more appropriately modelled by appealing to non-parametric descriptions of $\psi(t-\tprime)$ and $Z(t-\tprime)$. This requires discretising the evolution in a finite number of `star formation periods' (or `bins', each spanning a range of ages), adjusting the mass fraction and metallicity in each period to best reproduce a given observation (see, e.g.; \textsc{starlight}, \citealt{Cid2004, Cid2005}; \textsc{moped}, \citealt{Heavens2000}; \textsc{steckmap}, \citealt{Ocvirk2006} and \textsc{vespa}, \citealt{Tojeiro2007}). A common limitation of this approach is the large number of free parameters involved and their correlations, which requires high-\SN\ spectroscopic observations across a broad wavelength range to produce useful constraints \citep[e.g.][]{Tojeiro2013, McDermid2015, Lopez2016, Citro2016}. An alternative approach was recently proposed by \citet{Pacifici2012}, who adopt physically motivated prescriptions for $\psi(t-\tprime)$ and $Z(t-\tprime)$ extracted from the semi-analytic post-treatment of a cosmological dark-matter simulation. They show that this provides a powerful new means of interpreting observed SEDs of galaxies in terms of physical parameters \citep[e.g.][]{Pacifici2013, Pacifici2015}. They also show how the dependence of $\psi(t-\tprime)$ and $Z(t-\tprime)$ on the adopted galaxy formation model can be alleviated by resampling the model star formation and chemical enrichment histories to widen the range of evolutionary stages probed at any redshift, at the expense of weakening the link between the resulting library of star formation and chemical enrichment histories and the input cosmological and galaxy formation model \citep{Pacifici2012}.  

We incorporate in \beagle\ all of the above approaches to describe the functions $\psi(t-\tprime)$ and $Z(t-\tprime)$ in equation~\eqref{eq:sed_model}. We achieve this by means of a general parametrization of the star formation and chemical enrichment histories of a galaxy, described as a succession of a flexible number of star formation periods, and by interpolating on the fly the grid of input SSPs (Section~\ref{sec:stelpops}) to compute $S_\lambda[\tprime,\vect{Z}(t-\tprime)]$ in equation~\eqref{eq:lumin} for any combination of age and metallicity. The shapes of $\psi(t-\tprime)$ and $Z(t-\tprime)$ in these components can be drawn from analytic functions or from different flavours of galaxy formation models, such as phenomenological and semi-analytic models and hydro-dynamic simulations. We note that such a parametrization allows one to easily separate the past history of star formation, imprinted in the spectral signatures of low-mass, long-lived stars, from the `current' (i.e., averaged over the last 10--100 Myr) star formation, traced by young massive stars, which controls the nebular and far-ultraviolet emission (ignoring the contribution from shocks and a potential AGN). This parametrization is also appropriate to describe the stochastic nature of the star formation process in galaxies. A main advantage of the flexible approach presented here is the possibility to adapt the complexity of the description of the star formation and chemical enrichment histories of a galaxy to the type and quality of the available data (e.g. photometry versus spectroscopy, low- versus high-\SN). 

\subsection{Dust attenuation}\label{sec:dust}

\renewcommand{\arraystretch}{1.5}
\begin{table*}
\begin{threeparttable}

	\centering
	\begin{tabular}{l c c c c}
\toprule

\multirow{2}{*}{\textbf{Prescription for dust attenuation}}	    &  \multicolumn{3}{c}{\textbf{Effects accounted for}}	& \multirow{2}{*}{\textbf{Adjustable parameters}} \\     

 \cmidrule{2-4} 

						    & Geometry\tnote{a} & Age dependence\tnote{b} & Inclination\tnote{c} & \\

\midrule

Mean extinction curve (MW, LMC, SMC)	 &    no			&  no 		   & no  			& \tauV \\

\citet{Calzetti1994}						&      yes 			& yes (implicitly)  &  no 			& \tauV 	\\

\citet{Charlot2000}						&      yes 			& yes  		  &  no 			& \tauV, $\mu$ 	\\

\citet{Chevallard2013} `quasi-universal' relation	&      yes 			& yes   		&  yes (implicitly) 	& \tauV, $\mu$  	\\

\citet{Chevallard2013} `full model'			&      yes 			& yes  	 	&  yes 			& \tauV, $\mu$, $\theta$, \Tthin, \Tthick, \Tbulge  	\\

\bottomrule
	\end{tabular}
\begin{tablenotes}
\item [a] Influence of the relative distributions of stars and dust on attenuation (a `no' refers to a uniform foreground screen model)
\item [b] Influence of stellar age on attenuation (enhanced attenuation of young stars in their birth clouds relative to older stars)
\item [c] Influence of disc inclination on attenuation (when some stars and dust components are distributed in discs)
\end{tablenotes}
\caption{Different prescriptions for dust attenuation implemented in the \beagle\ tool (see Section~\ref{sec:dust} for details).}
\label{tab:dust_models}	
\end{threeparttable}
\end{table*}

In the previous sections, we have presented our prescriptions for the emission from stars and ionized gas, changes in heavy-element abundance ratios and the star formation and chemical enrichment histories of a galaxy in the \beagle\ tool. We now turn to the effect of dust attenuation on the stellar and nebular emission, expressed by the function $T^0_\lambda(t,t^\prime)$ in equation~\eqref{eq:transmiss}. The signatures of this effect compete with  those of age and metallicity in galaxy SEDs \citep[e.g.][]{Wise1996, Papovich2001, Guo2011}. The importance of a careful description of dust properties is also that this may be used to constrain the mechanisms of dust production (e.g., winds from asymptotic-giant-branch stars, supernova explosions; \citealt{Hofner2009, Cherchneff2010}) and destruction (e.g. shocks; \citealt{Jones2004, Jones2011}) from the analysis of different types of galaxies at various redshifts.

We follow the standard terminology and refer by attenuation (or `effective absorption') to the combined effects of absorption and scattering in and out of the line of sight to a galaxy caused by both local and global geometric effects, the term extinction being reserved for photon absorption along and scattering out of a single line of sight \citep[e.g.][]{Charlot2000}. The dependence of extinction on wavelength (i.e. the extinction curve) has been measured in the Milky Way, the Small and the Large Magellanic Clouds \citep[MW, SMC and LMC;][]{Prevot1984,Bouchet1985,Clayton1985,Cardelli1988, Cardelli1989}.  These studies show that, in the Milky Way, the ultraviolet-to-near infrared extinction curve varies along different lines of sight, while the strength of the characteristic absorption bump near 2175~\AA\ drops from the average MW, to the LMC, to the SMC extinction curves. These extinction curves are commonly used to model the effect of attenuation by dust on galaxy SEDs, although this amounts to assuming that dust in unrealistically distributed in a uniform screen between the source and the observer.\footnote{A special case in which the use of mean extinction curves is physically motivated is the analysis of occulting galaxy pairs. In this case, dust in the foreground galaxy attenuates the light coming from the background one, hence motivating the adoption of a screen geometry to describe the effect of dust on the light emitted by the background galaxy \citep[e.g.][]{Holwerda2007, Holwerda2013}.} Such an assumption may introduce unwanted biases in the interpretation of galaxy SEDs. It also neglects the fact that young stars in their birth clouds are typically more attenuated than older stars in galaxies \citep[e.g.][]{Silva1998, Charlot2000}.
  
We adopt a more physically motivated prescription for attenuation by dust and express the transmission function of the neutral ISM (equations~\ref{eq:sed_model} and \ref{eq:transmiss}) as a function of stellar age $t^\prime$ and galaxy inclination $\theta$,
\begin{equation}
T^0_\lambda(t,t^\prime) \equiv T^0_\lambda(t^\prime,\theta) \,.
\end{equation}
Here we have dropped the dependence of $T^0$ on galaxy age $t$, i.e., we assume that attenuation by dust depends only on the current ISM properties and on the age distribution of stellar populations in a galaxy. We can rewrite the transmission function as
\begin{equation}
\label{eq:t0dust}
T^0_\lambda(t^\prime,\theta) = \exp[-\tauLthT] \,,
\end{equation}
where \tauLthT\ is the attenuation (or effective absorption) optical depth of the dust affecting photons emitted at wavelength $\lambda$ in all directions by all stars and gas in the galaxy, which emerge in the direction $\theta$ from the normal to the equatorial plane of the galaxy (assuming azimuthal symmetry). 

We use the general parametrization in equation~\eqref{eq:t0dust} to implement in a flexible way in the \beagle\ tool different prescriptions for $\tauLthT$, summarized in Table~\ref{tab:dust_models}. These include simple mean extinction curves (MW, LMC, SMC), the starburst attenuation curve of \citet{Calzetti1994}, the two-component dust model of \citet[][hereafter CF00]{Charlot2000} and new prescriptions by \citet[][hereafter C13]{Chevallard2013}. We indicate in Table~\ref{tab:dust_models} how these different prescriptions account for thee major features affecting dust attenuation in galaxies: the distribution of dust relative to stars (geometry); the enhanced attenuation of young stars in their birth clouds relative to older stars; and disc inclination.\footnote{We do not consider explicitly here changes in the size distribution and optical properties of dust grains, as these are implicitly included in Table~\ref{tab:dust_models}, either through the difference between the MW, LMC and SMC extinction curves or the possibility to change the slope of the attenuation curve in the CF00 and C13 prescriptions. Such changes are expected to have less effect on integrated galaxy SEDs than geometry, age dependence and inclination \citep[e.g.,][]{Granato2000, Fontanot2009, Chevallard2013}, except perhaps around the 2175\,\AA\ absorption feature \citep[e.g.][]{Conroy2010dust}.} The mean MW, LMC and SMC extinction curves often adopted in galaxy spectral analyses do not account for any of these features. The (angle-averaged) attenuation curve of \citet{Calzetti1994}, empirically determined from the spectra of 39 nearby starburst and blue-compact galaxies, implicitly incorporates the effects of geometry (as indicated by the shallower slope of this curve relative to standard extinction curves) and of the dependence of attenuation on stellar age (as indicated by the factor of $\unsim2$ difference in the attenuation affecting line and continuum photons in this sample). The main limitation of this prescription is that it was derived by neglecting the strong dependence of the slope of the attenuation curve on dust optical depth (see section~3.1.2 of C13). In Table~\ref{tab:dust_models}, only the C13 prescriptions include the dependence of attenuation on galaxy inclination.

We now describe in slightly more detail the implementation in the \beagle\ tool of the CF00 and C13 dust prescriptions. The CF00 prescription is based on an angle-averaged, two-component model, which accounts for the fact that stars are born in dense molecular clouds, which dissipate on a timescale of typically 10\,Myr. The dust attenuation optical depth in this model can be expressed as	
\begin{equation}\label{eq:tau_cf00}
\hat{\tau}_\lambda(t^\prime)=\left\{ \begin{array}{l l}
\tauLbc + \tauLism & \txn{for} \hspace{3mm} t^\prime \leqslant 10\,\txn{Myr}\,,\\
\tauLism  &  \txn{for} \hspace{3mm} t^\prime > 10\,\txn{Myr} \,, \end{array}\right.
\end{equation}
where $\tauLbc$ and $\tauLism$ are the dust optical depths in stellar birth clouds and the ambient (diffuse) ISM, respectively. The attenuation curves in these two components are parametrized as
\begin{align}
\tauLbc &= \tauVbc\ \left ( \frac{\lambda}{0.55\,\micron }\right ) ^ {-n^\bc_V} \, ,\\
\tauLism &= \tauVism\ \left ( \frac{\lambda}{0.55\,\micron }\right ) ^ {-n^\ism_V} \, , \label{eq:tauLism}
\end{align}
where $\tauVbc$ and $\tauVism$ are the $V-$band attenuation optical depths in stellar birth clouds and in the ambient ISM and, following \citet[see also \citealt{daCunha2008}]{Wild2007}, we adopt $n^\bc_V = 1.3$ and $n^\ism_V = 0.7$. We parametrize this model in terms of the total $V$-band attenuation optical depth, $ \tauV=\tauVbc+\tauVism$, and the fraction of this arising from dust in the ambient ISM, $\mu=\tauVism/ \tauV$, which is typically constrained in the range $[0.3, 0.5]$ \citep[e.g.][]{Wild2007}.

Recently, C13 proposed a new approach to account for the effects of inclination and dust/star geometry on the attenuation of galaxy SEDs. C13 incorporate the generic predictions of different types of sophisticated models of radiative transfer in dusty media into the two-component dust model of CF00. C13 show that these predictions can be subsumed in a quasi-universal relation between $V$-band attenuation optical depth in the diffuse ISM and shape of the attenuation curve. This relation, which accounts for the effects of dust/star geometry (including ISM clumpiness) and galaxy inclination, exhibits a steepening of the attenuation curve (from more starburst-like to more MW-like) at increasing dust optical depth (as a consequence of either a rise in the amount of dust or a higher inclination). A fit to a wide range of models yields (see fig. 4 of C13)
\begin{equation}\label{eq:n_tau_fit}
\nVism =  \frac{ 2.8 }{ 1+3 \sqrt{\tauVism} }\hspace{10mm} (\pm25\,\txn{percent}) \,,
\end{equation}
where the typical scatter is indicated in parentheses. The above expression was derived at optical wavelengths. C13 show that, by adopting a wavelength-dependent exponent of the power law in equation~\eqref{eq:tauLism}, one can reproduce the generic predictions of radiative transfer models over the entire wavelength range from the near ultraviolet to the near infrared, neglecting the 2175 \AA\ bump. This can be achieved by rewriting equation~\eqref{eq:tauLism} as
\begin{equation}\label{eq:pow_law}
\tauLismTh = \tauVism \left ( \frac{\lambda}{0.55\,\micron}\right )^{-n^\ism_\lambda} \, ,
\end{equation} 
where the exponent of the power law is a linear function of wavelength,
\begin{equation}\label{eq:n_lambda}
\nLism(\tauVism) = \nVism+ b \,(\lambda/\micron -0.55) \,,
\end{equation} 
valid over the range $0.1 \le \lambda \le 2.5$ \micron, and the coefficient $b$ is given by
\begin{equation}\label{eq:b_tau_fit}
b =  0.3-0.05 \, \tauVism\hspace{12mm} (\pm10\,\txn{percent}) \, .
\end{equation}
We note that the above implementation of the C13 prescription does not require more parameters than the original CF00 model, since equations~\eqref{eq:n_tau_fit}--\eqref{eq:b_tau_fit} depend only on the $V$-band attenuation optical depth in the diffuse ISM, \tauVism\ (see column `Adjustable parameters' of Table~\ref{tab:dust_models}).	

Finally, we also implement in the \beagle\ tool the more sophisticated dust prescription proposed by C13 (referred to as `full model' in Table~\ref{tab:dust_models}), which allows one to explicitly express the dependence of attenuation in the ambient ISM on the viewing angle $\theta$. This is achieved by associating stars in different age ranges with the thin-disc, thick-disc and bulge components of the versatile model of radiative transfer of \citet{Tuffs2004}. In this case, we rewrite equation~\eqref{eq:tauLism} as
\begin{equation}
\tauLismTt=\left\{ \begin{array}{l l}
\tauLthinth & \hspace{3mm} \txn{for} \hspace{3mm} t^\prime \leqslant \Tthin \,,\\
\tauLthickth & \hspace{3mm} \txn{for} \hspace{3mm} \Tthin < t^\prime \leqslant \Tthick \,,\\
\tauLbulgeth  & \hspace{3mm} \txn{for} \hspace{3mm} \Tthick < t^\prime \leqslant \Tbulge \,, \\
\,\,0 & \hspace{3mm} \txn{for} \hspace{3mm}  t^\prime>\Tbulge\,,\end{array}\right.
\label{eq:tau_T04ism}
\end{equation}
where \tauLthinth, \tauLthickth\ and \tauLbulgeth\ are the attenuation curves at inclination $\theta$ for a thin stellar disc, thick stellar disc and bulge in the \citet{Tuffs2004} model. We stress that equations~\eqref{eq:n_tau_fit}--\eqref{eq:tau_T04ism} affect only the dust attenuation arising from the diffuse ISM (the term \tauLism\ in equation~\ref{eq:tau_cf00}), while the birth-cloud component (\tauLbc\ in equation~\ref{eq:tau_cf00}) is treated in C13 as in the original CF00 model. We also emphasise that the flexible modular structure of the \beagle\ tool enables one to easily substitute the predictions of any other model of radiative transfer for those of the \citet{Tuffs2004} one.

\subsection{IGM absorption}\label{sec:IGM}

\begin{figure}
	\centering
	\resizebox{\hsize}{!}{\includegraphics{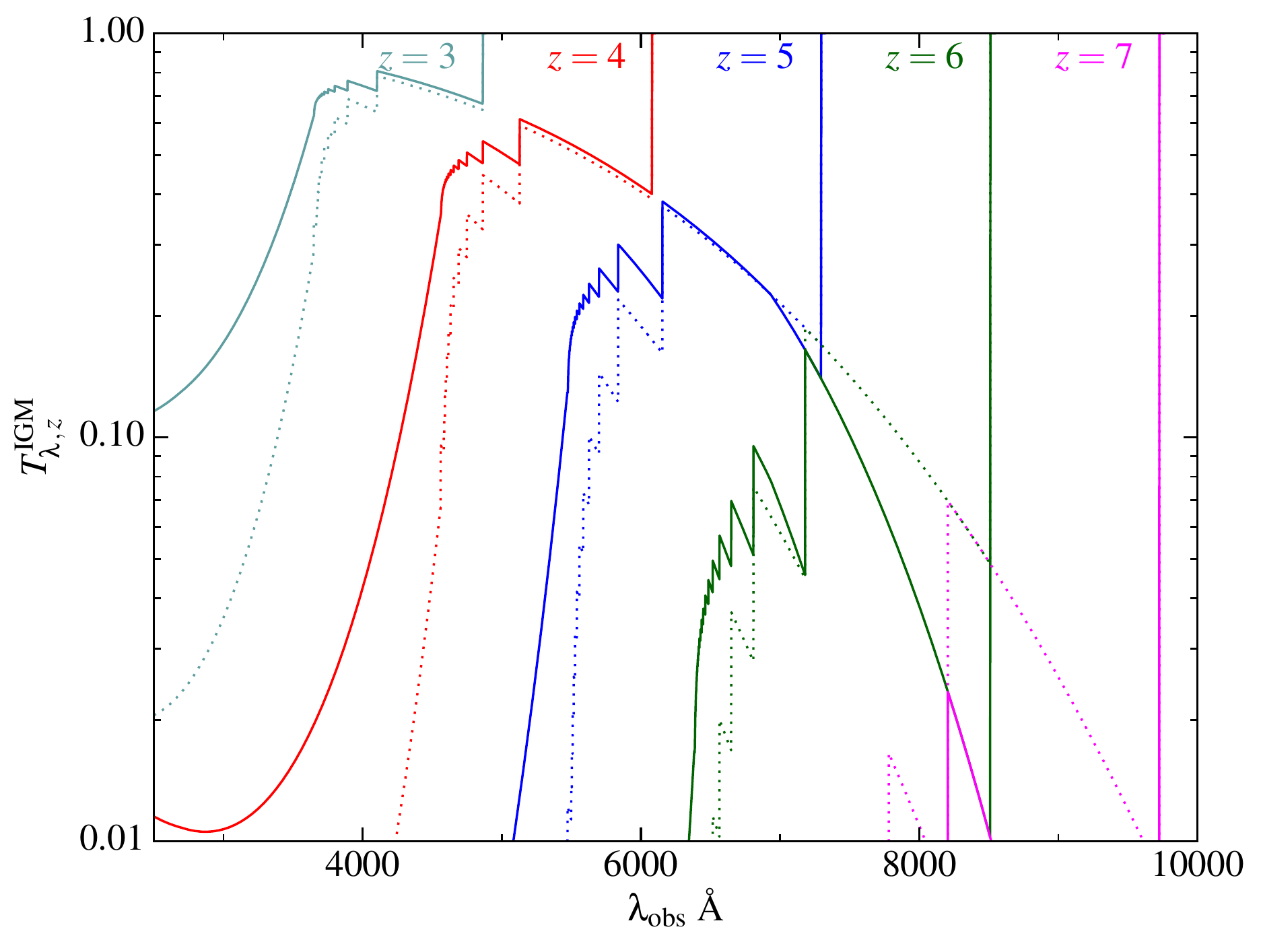}}
	\caption{Predictions of the IGM absorption models of \citet[][dotted lines]{Madau1995} and \citet[][solid lines]{Inoue2014} for sources at different redshifts.}
	\label{fig:IGM}
\end{figure} 

We now consider the absorption of photons emerging from galaxies by gas in the IGM along the line of sight. This is modelled through the transmission function $T^\txn{IGM}_{\lambda, z}$ of the IGM in equation~\eqref{eq:flux}. The IGM is composed primarily of hydrogen and helium and contains three main cloud components: the \Lya\ forest, Lyman-limit systems (LLS) and damped \Lya\ systems (DLA), in order of increasing \Hi\ column density and decreasing number density \citep[e.g.][]{Rauch1998, Peroux2003, Wolfe2005}. The \Lya\ forest consists mainly of primordial gas, while LLS and DLA, which are thought to be associated with haloes and galaxies, are enriched in metals. Neutral hydrogen is the primary contributor to $T^\txn{IGM}_{\lambda, z}$, which can thus be estimated from the \Hi\ column density distributions and number density evolutions of the different cloud components.

\citet{Madau1995} proposed a first, simple analytic model to compute the contributions to $T^\txn{IGM}_{\lambda, z}$ by \Hi\ in the \Lya\ forest and LSS, neglecting the much rarer DLA. This model includes photoelectric absorption of Lyman-continuum photons and blanketing by Lyman-series lines (from the combined absorption in many intervening clouds) of the background galaxy SED, which produces a characteristic `staircase' profile of $T^\txn{IGM}_{\lambda, z}$ as a function of observed wavelength (Fig.~\ref{fig:IGM}). Recently, \citet{Inoue2014} updated this widely used model by revising the \Hi\ column density distributions and number density evolutions of the \Lya\ forest and LSS and by adding the contribution to $T^\txn{IGM}_{\lambda, z}$ by \Hi\ in DLA. In Fig.~\ref{fig:IGM}, we compare the IGM transmission functions predicted by this model (solid lines) and the \citet{Madau1995} model (dotted lines), for background sources at different redshifts. For sources at $z\lesssim 5$, the model of \citet{Madau1995} predicts a lower transmission than that of \citet{Inoue2014} at all rest wavelengths $\lambdao/(1+z) < \lambda_{\txn{Ly}\beta}$, while at $z\gtrsim6$, the trend is partially reversed. For sources at $z=7$, the \citet{Inoue2014} model predicts that nearly all photons emitted at rest wavelengths $\lambdao/(1+z) < \lambda_{\txn{Ly}\alpha}$ are absorbed by the IGM, while 1--7 per cent of the emission is transmitted at $\lambdao/(1+z) \in [\lambda_{\txn{Ly}\beta}, \lambda_{\txn{Ly}\alpha}]$ in the \citet{Madau1995} model. It is important to note that, because of the steepness of the IGM transmission curves in Fig.~\ref{fig:IGM}, these differences between the two models can translate into differences of up to $\unsim1$\,mag in the observed colours of high-redshift galaxies (see fig.~8 of \citealt{Inoue2014}). 

The analytic prescriptions of \citet{Madau1995} and \citet{Inoue2014} for $T^\txn{IGM}_{\lambda, z}$ are both limited by the fact that they pertain to averages over infinite numbers of sight lines, while individual galaxies probe single lines of sight through the IGM. In the future, we plan to account for variations in $T^\txn{IGM}_{\lambda, z}$ along different lines of sight in the \beagle\ tool by appealing to the prescription of \citet{Harrison2011}.

\subsection{Line-of-sight velocity distribution}\label{sec:losvd}

The precise fitting of spectroscopic galaxy observations requires one to also account for the effects of stars and gas  kinematics on the emergent SED. We implement this feature in the \beagle\ tool by introducing a flexible description of the  line-of-sight velocity distribution (LOSVD) taken from \citet[][see also \citealt{Gerhard1993}]{vanDerMarel1993}. This consists in decomposing the LOSVD into orthogonal functions via a Gauss-Hermite series, which enables the clean modelling of deviations from pure Gaussian line profiles (the orthogonality of the Hermite polynomials minimising correlations among the adjustable coefficients). In terms of the standardised variable $x=(v-\vsys)/\sigma$, we therefore express the LOSVD as
\begin{equation}\label{eq:losvd}
\mathcal{L}_x(\vsys,\sigma,h_3,h_4) = \frac{1}{\sigma\sqrt{2\pi}}\exp\left(-\frac{x^2}{2}\right) \, \Big[ 1+ h_3 H_3(x) + h_4 H_4(x) \Big] \, ,
\end{equation}
where \vsys\ is the galaxy systemic velocity, $\sigma$ the velocity dispersion, and $H_3(x)$ and $H_4(x)$ the Hermite polynomials of order 3 and 4,
\begin{align}
H_3(x)   = & \frac{1}{\sqrt{3}} \left(2\sqrt{2}x^3-3\sqrt{2}x \right) \,,\\
H_4(x)   = &  \frac{1}{\sqrt{24}} \left(4x^4-12x^2+3 \right) \, .
\end{align} 
The adjustable coefficients $h_3$ and $h_4$ measure, respectively, asymmetric and symmetric deviations from pure Gaussian LOSVD. They can be determined separately for stars and gas. 

\subsection{Instrumental effects}\label{sec:instrum}

So far, we have discussed the production of starlight in galaxies and its transfer through the interstellar and the intergalactic media. In this section, we address important instrumental effects altering the SED of a galaxy observed through a telescope: the spectral response and the spectroscopic flux calibration. In a Bayesian framework, accounting for instrumental effects  is straightforward so long as these can be parametrized, as this amounts to adding `nuisance' parameters.\footnote{A nuisance parameter is a parameter of no direct interest, but which must be included in the analysis to obtain reliable inference about the parameters of interest.} In the next subsections, we describe two instrumental effects incorporated in the \beagle\ tool : the instrumental spectral response (i.e. the line spread function) and the spectroscopic flux calibration. Other effects potentially biasing the interpretation of galaxy SEDs, such as the calibration of photometric zero points, require a different approach. For example, a rigorous treatment of zero-point offsets can be performed by adopting a hierarchical Bayesian modelling approach. This and other instrumental effects will be investigated in future work.

\subsubsection{Line spread function}

The line spread function, noted $\mathcal{L}^\txn{LSF}_{\lambda_\txn{obs}}$, describes the instrumental spectral response, which relates the spectral flux density $F_{\lambda_\txn{obs}}$ reaching the telescope at wavelength $\lambda_\txn{obs}$ (equation~\ref{eq:flux}) to that effectively measured by the observer, $F^\prime_{\lambda_\txn{obs}}$, in such a way that
\begin{equation}\label{eq:lsf_convol}
F^\prime_{\lambda_\txn{obs}} = F_{\lambda_\txn{obs}} \otimes \mathcal{L}^\txn{LSF}_{\lambda_\txn{obs}} \, ,
\end{equation}
where the symbol $\otimes$ indicates convolution. A common first-order approximation is to model the line spread function as a Gaussian function, 
\begin{equation}\label{eq:lsf}
\mathcal{L}^\txn{LSF}_{\lambda_\txn{obs}}(\muLSF, \sigLSF) = \frac{1}{\sigLSF\sqrt{2\pi}}\exp\left[-\frac{1}{2}\left(\frac{\lambdao-\muLSF}{\sigLSF}\right)^2\right] \,,
\end{equation}
with mean $\muLSF=0$ and dispersion $\sigLSF(\lambdao)=\lambdao/R$, where $R=\lambdao/\Delta\lambdao$ is the instrumental spectral resolution. In practice, this may provide only a poor approximation of the true line spread function, which depends on, for example, the intrinsic light profile of the source, the point spread function and the width of the spectroscopic aperture (slit or fibre). Ideally, therefore, the line spread function should be determined on an object-by-object basis. Another complication is that, for spectral analyses involving models based on observed stellar spectral libraries, the line spread function affecting the original stellar spectra should also be accounted for to perform meaningful comparisons with galaxy observations \citep[see e.g.][]{Koleva2009}.

In the \beagle\ framework, we adopt a flexible parametrization of the line spread function, in which the parameters $\muLSF$ and $\sigLSF$ in equation~\eqref{eq:lsf} are taken to be polynomials of adjustable degree in $\lambda_\txn{obs}$, i.e.
\begin{align}
 \muLSF(\lambdao) & = \sum_{i=0}^n c_i \, \mathcal{P}_i(\lambdao-\lambda_\txn{obs}^{c}) \, , \\
 \sigLSF(\lambdao) & = \sum_{i=0}^n d_i \, \mathcal{P}_i(\lambdao-\lambda_\txn{obs}^{c}) \, ,
\end{align}
where $n$ indicates the degree of the polynomial, $c_i$ and $d_i$ the coefficients of the polynomial, and $\mathcal{P}_i$ the $i$-th term of the polynomial expanded around the central wavelength $\lambda_\txn{obs}^{c}$. The coefficients of the polynomials $c_i$ and $d_i$ are treated as nuisance parameters and marginalised out when computing the posterior probability distributions of the model parameters of interest. This ensures that uncertainties arising from an inaccurate knowledge of the line spread function are correctly propagated to the statistical constraints on interesting model parameters. Moreover, the possibility for $\muLSF(\lambdao)$ to differ from zero enables one to straightforwardly account for any inaccurate wavelength calibration.
 
\subsubsection{Spectroscopic flux calibration}

A major challenge in the reduction of spectroscopic data is to achieve a reliable flux calibration, both absolute and relative. We refer here by `absolute' to the calibration of an observed SED on an absolute flux scale using observations of standard stars (and accounting for any required aperture correction). By `relative', we refer to the calibration of the flux at any wavelength with respect  another. The quality of the absolute calibration determines how well galaxy properties depending on total flux, such as stellar mass and star formation rate, can be evaluated. The quality of the relative flux calibration can potentially affect all galaxy properties, as it alters the shape of the SED. Obtaining an accurate relative flux calibration is challenging because of the contamination by several factors, such as the wavelength dependences of the point spread function and the galaxy light profile.

Two main approaches are generally considered to deal with inaccurate relative flux calibrations: one is to continuum-normalise the observed and model spectra before comparing them \citep[e.g.][]{Wolf2007, Spiniello2014}; the other is to introduce a smooth correction function by which to multiply the model continuum, in order to bring it in agreement with the observed one \citep[e.g.][]{Kelson2000, Koleva2009}. We implement both strategies in the \beagle\ tool to account for inaccurate relative flux calibrations. We allow either the observed and model spectra to be continuum-normalised, or a series of Legendre polynomials of adjustable order to be used to correct the model continuum shape, the  coefficients of this series being treated as nuisance parameters (as in the case of the LOSVD in section~\ref{sec:losvd}, we choose orthogonal polynomials to minimise correlation between adjustable parameters). This second, more flexible approach presents several advantages over a continuum normalisation, which requires a fixed determination of the continuum, is highly sensitive to noise and erases valuable information contained in the spectral continuum shape. The subtlety is to select the smallest possible order of the series of Legendre polynomials able to account for the continuum-shape mismatch between model and observed spectra, while preserving informative spectral features such as the 4000\,\AA\ break and broad molecular absorption lines. This order is typically around 2 to 3.

\section{Main characteristics of BEAGLE}\label{sec:beagle}

In Section~\ref{sec:model}, we presented the astrophysical ingredients used to model galaxy SEDs in the framework of the \beagle\ tool. These include, at the present time: the consistent modelling of the emission from stars and ionized gas by means of combined stellar population synthesis and photo-ionization codes; a prescription to account for the effects of changes in the $\alpha$ element-to-iron abundance ratio on stellar population spectra; different prescriptions for the star formation and chemical histories of galaxies, dust attenuation and IGM absorption; and simple analytic models of the LOSVD and instrumental effects. In this section, we describe the statistical approach at the basis of the \beagle\ tool, the combined implementation of the different model ingredients and the main output products. 

\subsection{Statistical approach}\label{sec:statapp}

Our main goal in this paper is to develop a new-generation tool for the analysis of galaxy SEDs. In addition to enabling the production of mock catalogues of any spectroscopic and photometric galaxy properties, this tool must allow one to derive statistical constraints on a wide range of galaxy physical parameters from observed SEDs. To achieve this, we appeal to the modelling approach outlined in Section~\ref{sec:model} to perform statistical inference on physical parameters from observed galaxy samples. A plethora of approaches have been proposed in the literature to perform such inference on the basis of various statistical methods (e.g., minimum $\chi^2$, maximum likelihood, Bayesian; see Section~\ref{sec:intro}). To select the approach optimal for our purpose, we require the constraints on galaxy physical parameters derived through an inference process to allow:

\begin{itemize} 
	\item a rigorous propagation of observational errors into statistical constraints on model parameters, to compute realistic uncertainties in these parameters; 
	\item a full characterisation of correlations among model parameters, to deal with parameter degeneracies and multi-modal solutions; 
	\item a proper accounting of instrumental effects, to minimise the impact of instrumental systematics on inference products.
\end{itemize}
In addition to these requirements, the full exploitation of high-quality data will require complex models with multiple parameters. The choice of the optimal statistical approach therefore also depends on the computational and memory requirements of such multi-dimensional models.

In the framework of the \beagle\ tool, we adopt a Bayesian approach to perform inference on galaxy physical parameters from observed SEDs. This approach satisfies the above requirements through the characterisation of prior and posterior probability distributions of model parameters, the consideration of nuisance parameters, and a precise framework to perform model comparisons and hierarchical analyses of multi-level observational constraints using multi-parameter models. This kind of approach is now routinely employed to interpret large astrophysical datasets in the context of parameter spaces of very high dimensions ($>10^7$ parameters, e.g. \citealt{Jasche2015}) by appealing to efficient computational techniques (e.g., Markov chain Monte Carlo, Hamiltonian Monte Carlo, Nested Sampling).

In practice, Bayes' theorem allows one to perform inference on a set of model parameters by combining information obtained from an experiment (through the `likelihood function') with any prior knowledge about the parameters (through the `prior probability distribution'). Past and current knowledge is therefore combined in the posterior probability distribution of the parameters, which can be expressed as \citep[e.g.][]{Jeffreys1961}
\begin{align}
\txn{posterior} & = \frac{\txn{prior} \hspace{0.2cm} \cdot \hspace{0.2cm} \txn{likelihood}}{\txn{evidence}} \, , \\
 \intertext{and, mathematically, as}
P(\thetab \mid \Db, H) &= \frac{P(\thetab \mid H) \; P(\Db \mid \thetab, H)}{\int P(\thetab \mid H) \; P(\Db \mid \thetab, H) \, d\thetab} \,\label{eq:bayes} ,
\end{align}
where \thetab\ refers to a set of parameters of a model (hypothesis) $H$, \Db\ to a dataset, and the denominator (i.e. the evidence, or marginal likelihood) is often written simply as $P(\Db \mid H)$. For simplicity, in the remainder of this paper, we refer to the prior distribution of a model $H$ with parameters \thetab\ as $\pi(\thetab)\equiv P(\thetab \mid H)$, and to the likelihood function of a dataset \Db\ given a model $H$ with parameters \thetab\ as $\mathcal{L}(\thetab)\equiv P(\Db \mid \thetab, H)$. This function depends on both the physical model and the statistical description of the noise in the data (e.g., Gaussian, Poisson). Given a prior probability distribution, which reflects our belief about the model parameters before considering the dataset \Db, equation~\eqref{eq:bayes} enables rigorous statistical constraints on these parameters from observations at any \SN\ ratio. These constraints do not require any assumption about the shape of the posterior probability distribution (unlike, e.g., the Gaussian shape implicitly assumed when estimating confidence intervals with the criterion $\Delta\chi^2<1$). We note that, as described in Section~\ref{sec:instrum}, instrumental systematics can also be incorporated in a Bayesian approach, by means of nuisance parameters, which can be marginalised out when computing the posterior probability distributions of the model parameters of interest \citep[see, e.g., the treatment of nuisance parameters in][]{Planck2015}.

Another main interest of equation~\eqref{eq:bayes} is the possibility to fully characterise complex, non-linear correlations among model parameters, an achievement often ignored in statistical approaches focusing on simple point-wise (e.g. minimum $\chi^2$) estimates of best-fitting model parameters. Such correlations can reveal parameter degeneracies and lead to multi-modal solutions, i.e., to different parameter combinations providing similarly good fits to the data (which one may then try to break by appealing to new observables; see, e.g., fig.~19 of \citealt{Planck2015}). The problem is particularly acute, for example, in the context of galaxy redshift determinations from deep photometric observations (see Section~\ref{sec:photoz}). A Bayesian analysis provides a rigorous solution to this problem, through the methodical comparison of models populating the different modes of a posterior probability distribution (Section~\ref{sec:multiple_solutions}). Such statistically-driven comparison allows the selection of the best subset of models able to account for a dataset, beyond the selection of simply the best set of model parameters.

It is worth discussing briefly the role of the prior probability distribution entering equation~\eqref{eq:bayes}. In most situations, the data entering equation~\eqref{eq:bayes} through the likelihood function will be informative enough for the prior probability distribution to have a negligible impact on inference results. In the case of poorly informative data, the influence of the prior can be studied by testing different choices and by comparing the prior and posterior probability distributions of model parameters.\footnote{We refer the interested user to \citet{Loredo2012} for an interesting discussion about common misconceptions regarding Bayesian methods, including the role of prior distributions.} As an example, the analysis of low-\SN\ photometric observations does not allow one to put strong constraints on all model parameters, but the presence of a prior term in equation~\eqref{eq:bayes} allows one to properly incorporate our ignorance about the value of these parameters into their posterior probability distribution (see Section~\ref{sec:PDF} below). The current version of the \beagle\ tool allows the user to choose, for any adjustable parameter of the model, between uniform (linear or logarithmic), Gaussian (linear or logarithmic) or Cauchy (to allow for broader tails) prior distributions. Additional prior distributions will be made available in the future. Finally, we note that prior probability distributions can also be used in a hierarchical way in Bayesian analyses, by adopting `hyper-priors' with their own `hyper-parameters' to describe the parameter priors. This enables a multi-level analysis, which reduces the final uncertainties in model parameters by allowing data to share information with one another \citep[see, e.g.,][]{Sonnenfeld2015}. 

\subsection{Model implementation}\label{sec:parametrization}

\begin{figure*}
	\centering
	\resizebox{\hsize}{!}{\includegraphics{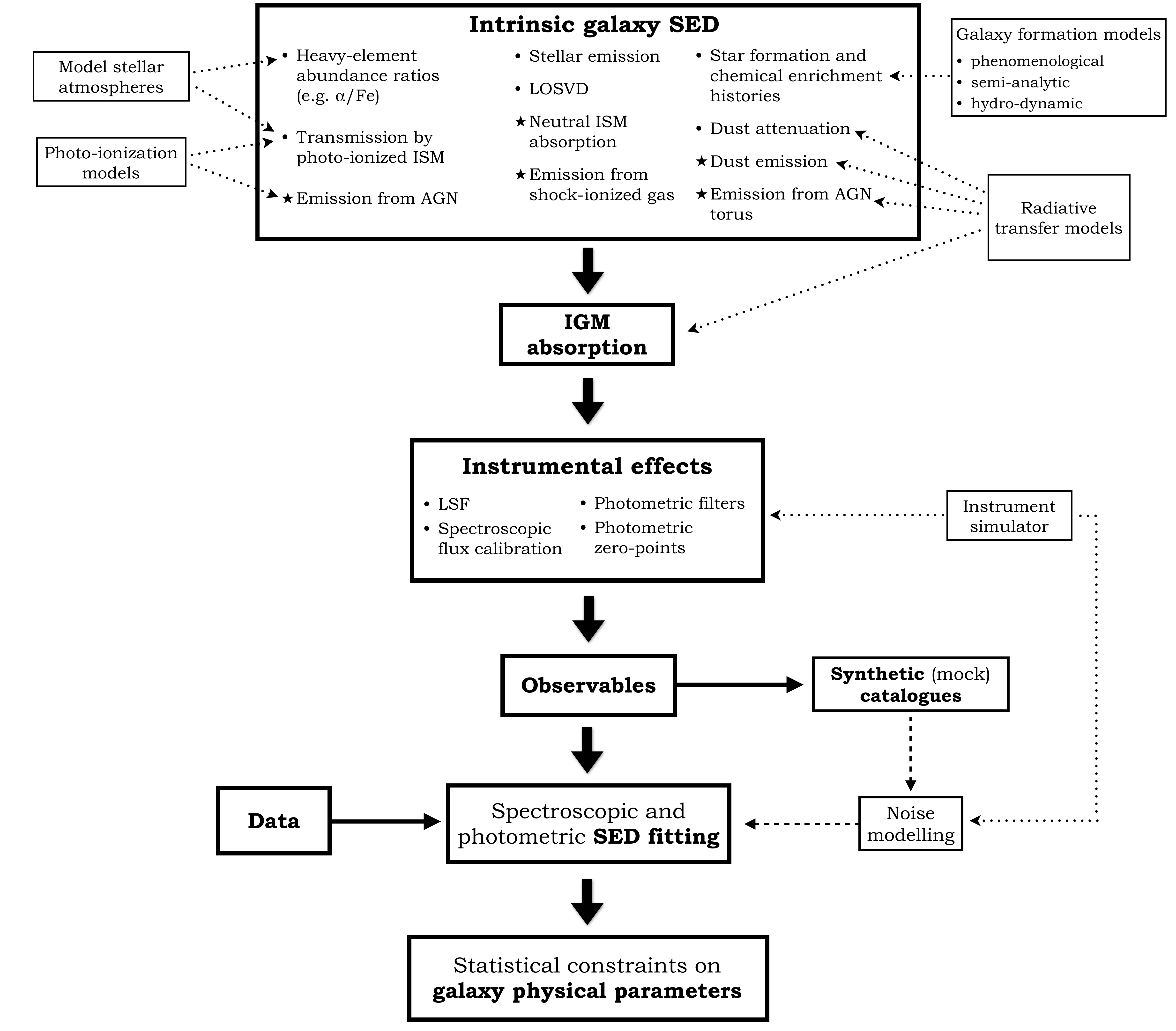}}
	\caption{Workflow diagram showing the different building blocks of the \beagle\ tool. The astrophysical ingredients indicated with stars are currently being implemented and will soon be available. Dotted arrows indicate how external models can be incorporated into \beagle\ to inform the various ingredients. See Section~\ref{sec:model} for details.}
	\label{fig:workflow}
\end{figure*} 

We now describe our strategy to implement the various astrophysical ingredients of \beagle\ (Section~\ref{sec:model}) into the Bayesian statistical framework outlined in Section~\ref{sec:statapp}, with the aim of creating a flexible, physically motivated tool for the analysis of galaxy SEDs. The flexibility of this tool is crucial to overcome a major challenge faced by any SED modelling approach: to adapt to a wide variety of data (from photometric to spectroscopic; from high- to low-\SN; at diverse spectral resolution over different wavelength ranges) without neither sacrificing model completeness nor over-fitting data with too many uncontrolled adjustable parameters. 

Fig.~\ref{fig:workflow} shows our implementation in a fully integrated tool of the different astrophysical ingredients described in Section~\ref{sec:model}. This workflow diagram highlights the different steps leading to either the production of synthetic catalogues of galaxy SEDs or the quantitative interpretation of observed SEDs with model ones. The ingredients used to define an intrinsic galaxy SED are shown at the top, and the external inputs used to inform these ingredients on the sides (e.g., galaxy formation and spectral evolution models). The light emerging from the model galaxy is then processed through IGM absorption and instrumental effects to produce observables directly comparable with data. 

As a complement to Fig.~\ref{fig:workflow}, Table~\ref{tab:adjust_par} summarises the adjustable parameters used to compute galaxy observables in the \beagle\ tool. To achieve the flexibility required to handle different types of (both true and pseudo) observations, we allow each adjustable parameter to be either `free', `fixed' or `dependent'. Free parameters are drawn from prior probability distributions (e.g. uniform, normal, log-normal), both when building synthetic catalogues of galaxy SEDs and when fitting an observation, in which case they enter equation~\eqref{eq:bayes} as elements of array \thetab. Alternatively, an adjustable parameter can be fixed to a default (or standard) value, both when producing synthetic observations and when fitting an observed SED. This is useful to preserve the full coherence of a multi-parameter model when dealing with observations with low constraining power, such as limited-band photometry at moderate \SN. Finally, the value of an adjustable parameter can also be set to depend on other adjustable parameters, for example, through an analytic relation. 

We note that, from a Bayesian point view, the consideration of an adjustable parameters as free, fixed or dependent corresponds simply to different choices of prior probability distributions. The prior probability distribution of a fixed parameter is a Dirac $\delta$ function, as is that of a dependent parameter, only conditional in this case on the values of other parameters. In the current version of the \beagle\ tool, the prior probability distribution of each model parameter is independent of the others, which makes the generalised  Bayesian formulation of the above three parameter classes not easy to implement. In the future, we will expand the flexibility of prior probability distributions to allow the choice of any type of joint and conditional probability distributions. This will enable the incorporation in the inference process of a priori relations between different galaxy physical parameters (e.g., the mass-metallicity relation), accounting at the same time for the scatter about such relations.

\renewcommand{\arraystretch}{1.35}
\begin{table*}
\begin{threeparttable}

	\centering
	\begin{tabular}{L{0.15\linewidth-2\tabcolsep} C{0.15\linewidth-2\tabcolsep} C{0.15\linewidth-2\tabcolsep} L{0.50\linewidth-2\tabcolsep}}
\toprule

\multicolumn{1}{c}{\textbf{Physical module}}	    									& \multicolumn{1}{c}{\textbf{Adjustable parameters}}	 & \multicolumn{1}{c}{\textbf{Default range}}	& \multicolumn{1}{c}{\textbf{Physical meaning}}   \\     

\midrule

\multirow{11}{*}{\begin{minipage}{2.5cm}Star formation and chemical enrichment histories\end{minipage}} 	    &  $\Tssp/\Gyr$   &  $[0.01,\Tuniverse\tnote{a}\hspace{3pt}]$ 	& Age of stars when approximating a galaxy by a simple stellar population\tnote{b} \\

													    &  $\Tstart/\Gyr$, $\Tend/\Gyr$   		& $[0,\Tuniverse\tnote{a}\hspace{3pt}]$ 		& Start/end look-back times of an arbitrary number of star formation periods\tnote{c} \\     
													    
													    &  $\txn{M}/ \Msun$		&  $[0,10^{12}]$  		& Mass of stars formed during a star formation period \\

													    &  $\tausfr/\Gyr$ 			&  $[0.01,2\times\Tuniverse\tnote{a}\hspace{3pt}]$ 			 & Star formation timescale during a star formation period (for analytic star formation histories, such as exponentially declining, delayed, rising, etc.) \\

													    &   $\log(Z/\Zsun)$ 	&  $[-2,0.25]$				& Metallicity of stars formed during a star formation period \\

													    &  \aFebra  		&  $[0,0.4]$ 					&  $\alpha$/Fe ratio of stars formed during a star formation period (relative to Solar)\tnote{d} \\

													    &  $\Delta{t}_\txn{SFR}/\txn{yr}$	&  $[10^7,10^8]$ 		& Duration of the current episode of star formation \\

													    &  $\log (\psi_\txn{S} / \txn{yr}^{-1})$  		&  $[-12,-7]$ 		& Specific star formation rate (averaged over $\Delta{t}_\txn{SFR}$) \\

													    &  $\log(\Zyoung/\Zsun)$		&   $[-2,0.25]$ 		& Metallicity of stars younger than 10\,Myr \\

													    & $z$		                                  &   $[0,15]$  		& Redshift of observation 	\\
													    
													    &  \zform \, (\tform) 		&   $[z,50]$  		& Formation redshift (look-back time) of the first stellar generation \\  \\

\multirow{6}{*}{Dust attenuation}	   					    &  \tauV  		&  $[0,5]$ 	& $V-$band attenuation optical depth\tnote{e,f} \\

													    &  $\mu$  	& $[0,1]$	& Fraction of \tauV\ arising from the dust in the diffuse ISM\tnote{e,f}  \\

													    &  $\theta /\txn{deg}$ 	&  $[0,90]$ 		& Galaxy inclination\tnote{f} \\

													    &  $\Tthin/\Gyr$  	&  $[0.1,\tform]$ 			& Stars with ages $t\le\Tthin$ pertain to thin stellar disc\tnote{f} \\

													    &  $\Tthick/ \Gyr$ 	&  $[\Tthin,\tform]$ 				& Stars with ages $\Tthin \le t \le\Tthick$ pertain to thick stellar disc\tnote{f} \\

													    &  $\Tbulge/ \Gyr$ 	&  $[\Tthick,\tform]$ 			& Stars with ages $\Tthick \le t \le\Tbulge$ pertain to stellar bulge\tnote{f} \\    \\

\multirow{3}{*}{Nebular emission}	   						    &  $\log U$  		&  $[-4,-1]$ 	& Effective galaxy-wide ionization parameter\tnote{g} \\

													    &  $\log(\Zism /\Zsun)$ 	& $[-2,0.25]$	&  Effective galaxy-wide interstellar metallicity\tnote{g} \\

													    &  \xid  		&  $[0.1,0.5]$ 		&  Effective galaxy-wide dust-to-metal mass ratio\tnote{g} \\ \\

\multirow{4}{*}{Kinematics\tnote{h}}	   					    &  $\vsys/\txn{km}\,\txn{s}^{-1}$ 		&  $[0,10^4]$	& Systemic velocity \\

													    &  $\sigma/\txn{km}\,\txn{s}^{-1}$ 	& $[0,400]$		& Velocity dispersion \\

													    &  $h_3$  	&  $[-1,1]$ 		& Coefficient of the 3rd-order Hermite polynomial\tnote{i} \\

													    &  $h_4$  	&  $[-1,1]$ 		& Coefficient of the 4th-order Hermite polynomial\tnote{i} \\
													    												    													         													   																   
\bottomrule
	\end{tabular}
\begin{tablenotes}
\item [a] This indicates the age of the Universe at redshift $z$, assuming a cosmological model and a fixed set of cosmological parameters. 
\item [b] Although we de not favour the adoption of SSPs to describe galaxy star formation and chemical enrichment histories, we have implemented this model to allow an easier comparison with previous analysis tools adopting SSPs (and for spectral analyses of individual star clusters).
\item [c] As defined in Section~\ref{sec:sfh_ceh}.
\item [d] \citet{Coelho2007,  Coelho2014, Walcher2009}.
\item [e] \citet{Charlot2000}.
\item [f] \citet{Chevallard2013}.
\item [g]  \citet{Charlot2001, Pacifici2012, Gutkin2016}.
\item [h] Different kinematic parameters can be used to describe the LOSVD of stellar and nebular emission. 
\item [i] \citet{vanDerMarel1993, Gerhard1993}.
\end{tablenotes}
\caption{Summary of the adjustable parameters available to build the intrinsic SED of a galaxy in the \beagle\ tool. This table does not include the adjustable parameters used to describe instrumental effects (see Section~\ref{sec:instrum}).}
\label{tab:adjust_par}	
\end{threeparttable}
\end{table*}

The number of adjustable parameters of the \beagle\ tool in Table~\ref{tab:adjust_par} is fairly large. Since many of these parameters cannot be easily set to standard values nor related to other parameters, in typical situations, many will have to be considered as free parameters. In practice, the actual number of free parameters will be chosen on a case by case basis as a function of the available data. It is worth briefly pausing here to comment on a common misconception regarding the influence of the number of free parameters on a statistical analysis. It is often stated that, when fitting a model to a dataset, the number of free parameters should not exceed that of (independent) data points. In reality, this statement is true only for linear models, i.e., models depending linearly on free parameters (e.g., polynomials, linear least-squares). In the more general case of non-linear models, such as in the \beagle\ tool (Fig.~\ref{fig:workflow}), the rule does not apply. While simple models should generally be preferred over more complex ones at equal predictive power, in some situations, having more free parameters than data points may be recommended to account for uncertainties about these parameters.\footnote{See the interesting discussion about the relative numbers of free parameters versus data points at \url{https://jakevdp.github.io/blog/2015/07/06/model-complexity-myth/}.}

Another potential source of ambiguity relates to the parametrization of galaxy physical properties in SED fitting tools. In standard spectral analyses, galaxies are often characterized simply in terms of stellar mass, age and metallicity. The stellar mass of a galaxy reflects the integral of the star formation history.\footnote{\label{foot:mass}A fraction of the stellar mass is actually returned to the ISM through stellar winds and SNe explosions during the evolution of the stellar population. This fraction, and hence the mass currently locked into stars, is computed using stellar population synthesis models and is recorded in \beagle\ as the quantity \Mstar, which differs from the model parameter \M\ reflecting the integral of the star formation history.} Thus, a same galaxy stellar mass could result from an infinite number of different star formation histories. This implies that, in turn, the distributions of stellar age and metallicity will depend on the specific star formation and chemical enrichment histories of a galaxy. The definition of global galaxy age and metallicity at fixed stellar mass is therefore ambiguous. In this context, it is customary to define light- and mass-weighted ages and metallicities, which are also integral quantities computed from the star formation and chemical enrichment histories of a galaxy. Finally, we note that, in many spectral analyses, the age of a galaxy is defined as the age of the oldest stars in that galaxy. This quantity is most relevant to studies of young galaxies near the reionization epoch, as it is otherwise difficult to constrain, the oldest stars tending to be out-shined by younger ones.

\subsection{Output products}\label{sec:output}

In this section, we briefly describe the output products of the \beagle\ tool (we refer the reader to the code manual for a more detailed description of these products).\footnote{The \beagle\ tool will be released in the near future as an open-source project. To be informed about the code release, please visit and register at \url{http://www.jacopochevallard.org/beagle}. In the meantime, interested users should contact the corresponding author of this paper.} These are organised in FITS files with multiple extensions, each extension containing information about a physical module (e.g., star formation and chemical enrichment histories, dust). We use a standard output format to produce synthetic SED catalogues, in the sense that the properties of different simulated galaxies are written on different rows. We describe in more detail here the output products of a model fit to an observed galaxy SED, which are more specific to the statistical approach inherent in the \beagle\ tool. 

In the output FITS extension produced for a given physical module by a spectral fit, the entries on each row are the properties predicted by a model with free parameters drawn from the posterior probability distribution of equation~\eqref{eq:bayes} using a dedicated Bayesian algorithm (e.g. \multinest; see Section~\ref{sec:UVUDF_model}). The posterior probability distributions of the parameters themselves are reported in a separate extension of the output FITS file. This generalised format allows one to interpret in a probabilistic way the constraints on not only the model free parameters, but also all other physical quantities included in the output files, which we refer to as `derived quantities'. For example, the uncertainties in the mass-weighted age and metallicity of a galaxy can be computed from the values of these quantities across all rows of the corresponding output file, weighted by the posterior probability value associated to each row. This approach allows one to easily compute the probability distribution of any theoretical or observable quantity predicted by the model (e.g. mass-weighted age, rate of ionizing photons, ultraviolet spectral slope, broadband magnitude, emission-line luminosity, absorption-line strength), as resulting from the posterior probability distribution of the adjustable parameters. 

Another advantage of the above approach is that it simplifies the adoption of posterior predictive checks to quantify how well model predictions match the data in a given observable (see Section~\ref{sec:PPC} for an example). This can be readily extended to the study of residuals, which in this context are no more a point-wise estimate of the difference between model predictions and observations, but rather a probabilistic distribution of such differences. We believe that this will be a powerful means of identifying model failures and driving the development of better models with the \beagle\ tool.

\section{Photometric SED analysis with BEAGLE}\label{sec:SED_fitting}

In this section, we present a first application of the \beagle\ tool to interpret the broadband SEDs of a published sample of about $10^4$ galaxies at redshifts $0.1 \lesssim z\lesssim8$. The observational challenge of gathering large spectroscopic samples of galaxies at high redshifts is forcing much progress in the field of galaxy formation and evolution to rely on photometric surveys. Interpreting broadband SEDs is therefore a common task in galaxy evolution studies. Historically, the derivation of galaxy physical parameters from broadband galaxy SEDs has often been decoupled from that of redshift. In fact, redshift is considered to be the most robust quantity that can be constrained from broadband photometric data, while determinations of galaxy physical parameters are expected to be more model-dependent \citep[e.g.][]{Dahlen2013, Mobasher2015}. Dedicated photometric-redshift codes generally rely on a small library of `representative' SEDs of different types of galaxies, either built using spectral evolution models (e.g. \eazy\, \citealt{Brammer2008}) or consisting of a small number of observed galaxy SEDs (e.g. \bpz\, \citealt{Benitez2000};  \textsc{lephare}, \citealt{Arnouts1999, Ilbert2006}). Such libraries enable the extraction of only limited information about physical parameters from photometric SEDs of galaxies.\footnote{Other approaches to estimate photometric redshifts also exist: `data-driven' methods are based on the application of machine-learning techniques to determine the relation between galaxy colours and redshift \citep[e.g.][]{Collister2004, Hogan2015}, while `clustering-based' methods are based on the redshift evolution of spatial correlations between galaxies \citep[e.g.][]{Schmidt2013,Rahman2016}. We do not discuss such approaches here, as they typically do not provide any information about galaxy physical parameters.} In contrast, most codes designed to constrain galaxy physical parameters from photometric data require an independent determination of redshift, often estimated using one of the above photometric-redshift codes (e.g. \textsc{fast}, \citealt{Kriek2009}; \textsc{cigale}, \citealt{Burgarella2005, Noll2009}). 

In the \beagle\ tool, we follow an alternative approach, similar that adopted by \citet[][see also \textsc{hyperz}, \citealt{Bolzonella2000, Pozzetti2007}]{Acquaviva2015}, and consider redshift just as an additional model parameter to be constrained along with the other physical parameters when fitting a galaxy SED. As we shall see below, a major advantage of this approach is to allow a rigorous propagation of the uncertainty on photometric redshift to the statistical constraints on other galaxy physical parameters, accounting at the same time for any potential correlation between redshift and other parameters. In addition, our statistical approach, based on the \multinest\ algorithm \citep[a Bayesian analysis tool based on the Nested Sampling algorithm of \citealt{Skilling2006}; see Appendix~\ref{app:multinest} and][]{Feroz2009}, can naturally deal with multi-modal solutions, i.e. with the occurrence of different combinations of parameters yielding similar posterior probability distributions (Section~\ref{sec:statapp}). This can often be the case in determinations of photometric redshifts of faint sources \citep[e.g.][]{Edmondson2006}. We now describe the photometric sample of distant galaxies we appeal to for our study (Section~\ref{sec:UVUDF_data}) and the results of SED fitting of this sample using the \beagle\ tool (Section~\ref{sec:UVUDF_model} for redshift and Section~\ref{sec:PDF} for other physical parameters).

\subsection{Galaxy sample}\label{sec:UVUDF_data}

\begin{figure}
	\centering
	\resizebox{\hsize}{!}{\includegraphics{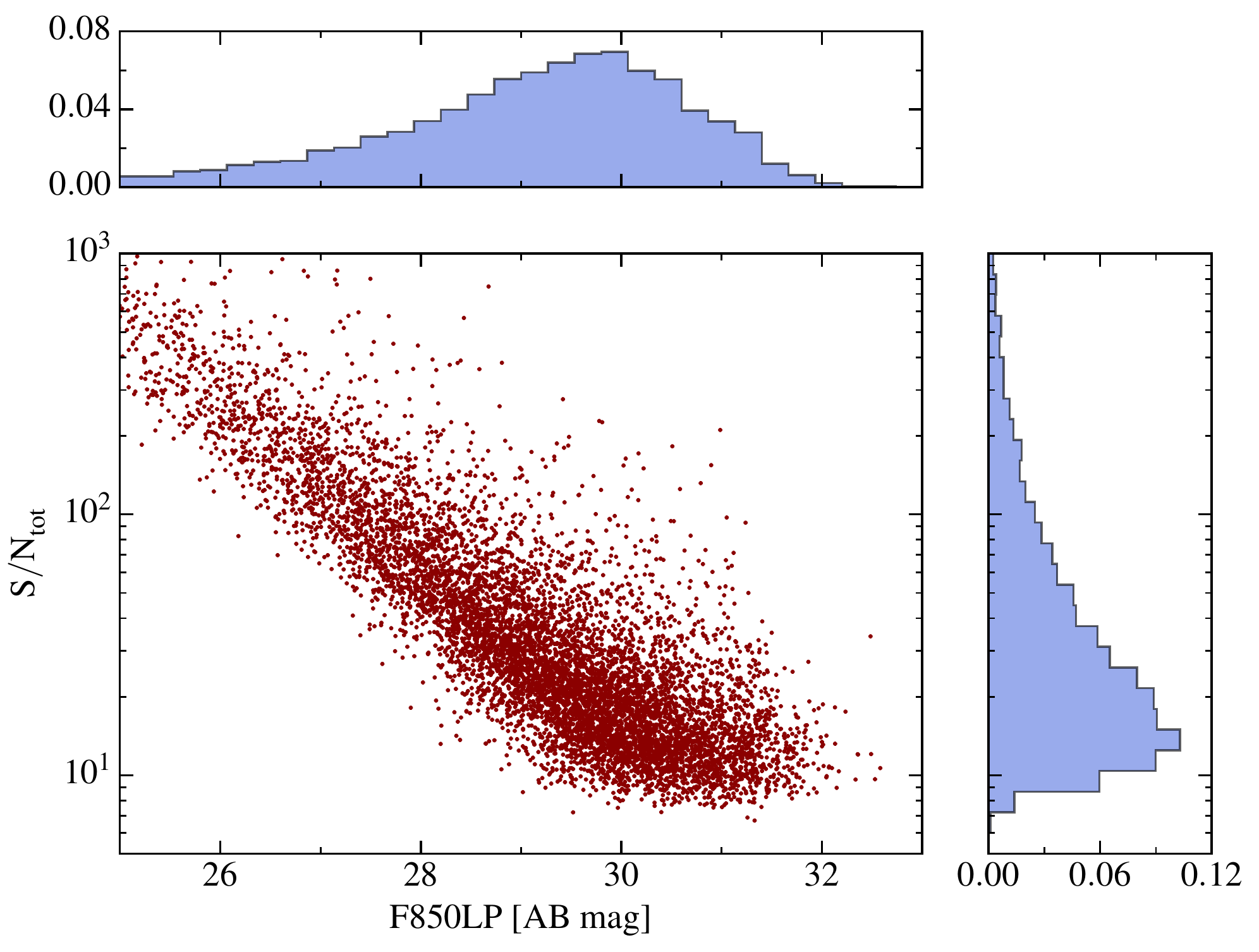}}
	\caption{Relation between total signal-to-noise ratio, computed by summing in quadrature the \SN\ of all bands with available measurements, and observed ACS/WFC F850LP magnitude, along with their marginal distributions, for all galaxies in the UVUDF catalogue.}
	\label{fig:UVUDF_S_to_N}
\end{figure}

\begin{figure}
	\centering
	\resizebox{\hsize}{!}{\includegraphics{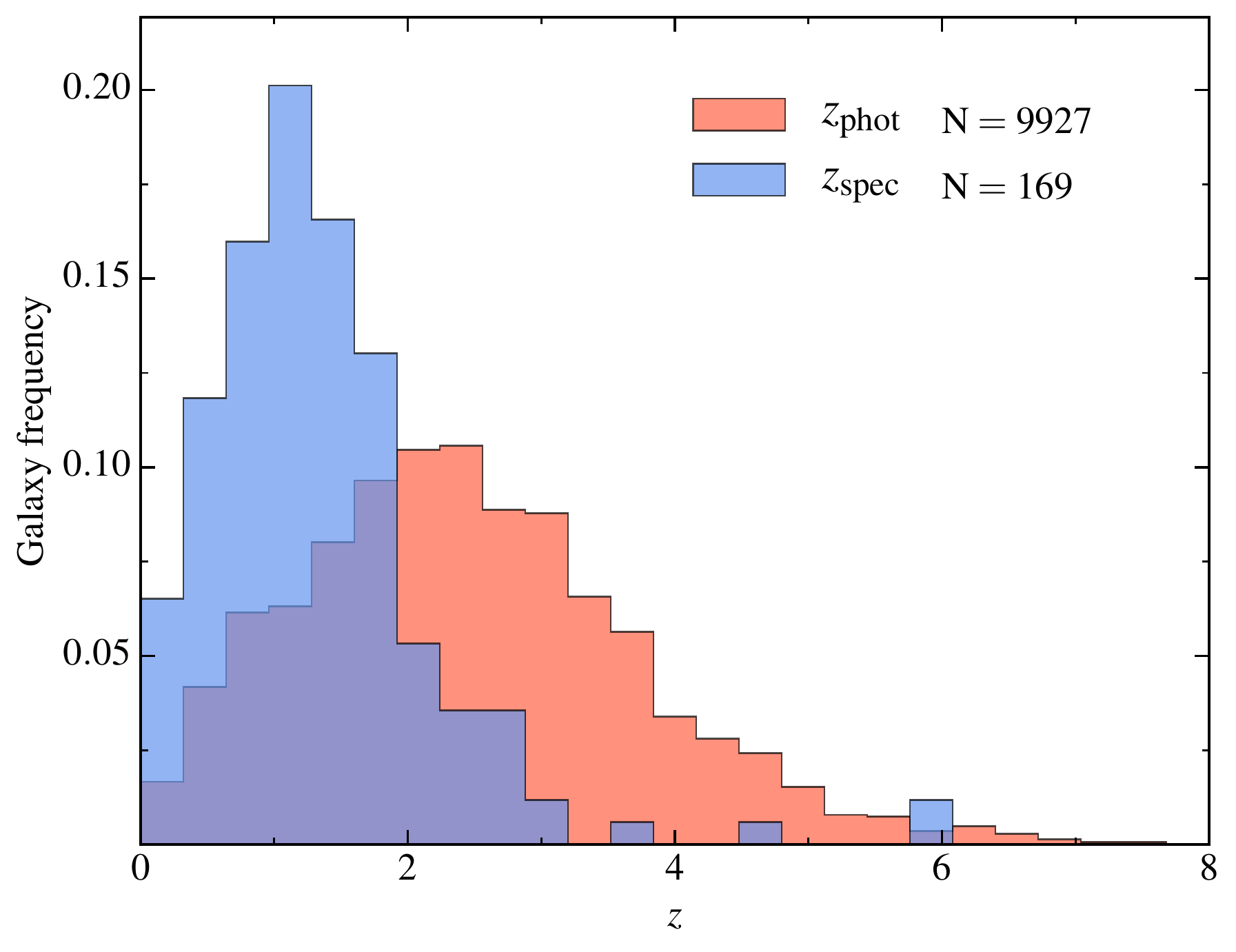}}
	\caption{Frequency distribution of redshift for the full UVUDF photometric catalogue (in red) and for galaxies with spectroscopic redshift measurements (in blue). The photometric redshifts are computed with the \beagle\ tool (see Section~\ref{sec:photoz}).}
	\label{fig:UVUDF_z}
\end{figure}

To illustrate our approach to broadband SED fitting, we appeal to the photometric galaxy catalogue assembled by \citet[][hereafter `UVUDF' catalogue]{Rafelski2015}. This consists of (up to) 11 bands at ultraviolet to near-infrared wavelengths for 9927 galaxies in the Hubble Ultra Deep Field \citep{Beckwith2006}. The ultraviolet data come from observations in three HST/WFC3 filters, F225W, F275W and F336W \citep[observations in ][]{Teplitz2013}, reanalysed by \citet{Rafelski2015} to improve the photometric and astrometric calibrations. The optical data come from observations in four ACS filters, F435W, F606W, F775W and F850LP \citep[imaging in ][]{Beckwith2006}, and the near-infrared ones in four WFC3 bands, F105W, F125W, F140W and F160W, from the UDF09, UDF12 \citep{Oesch2010a, Oesch2010b, Bouwens2011, Koekemoer2013, Ellis2013} and CANDELS GOODS-S programs \citep{Grogin2011, Koekemoer2011}. To maximise the depth of their catalogue, \citet{Rafelski2015} combine, pixel by pixel, the 4 optical and 4 near-infrared images to create a `detection image'. They identify sources in this image by running a standard extraction algorithm with different settings, varying the detection and deblending threshold parameters, then merging in a single catalogue the sources identified with the different settings. Finally, to produce a homogeneous source catalogue, they extract PSF-corrected aperture-matched photometry in all bands, and compute the total \citep{Kron1980} flux in each band by applying an aperture correction to the isophotal flux. 

We show in Fig.~\ref{fig:UVUDF_S_to_N} the relation between total signal-to-noise ratio \SNtot, computed by summing in quadrature the \SN\ of all bands with available measurements, and observed ACS/F850LP magnitude, along with the marginal distributions of these quantities, for all UVUDF galaxies. We select the F850LP band as it is the one with the largest fraction (9919/9927) of detected objects (defined as entries in the catalogue with positive measurements of both flux and flux error). The catalogue peaks around $m_\txn{AB}=29.5$, with a typical \SNtot\ in the range $\sim[10, 30]$ and a long tail of bright objects with $\SNtot \gtrsim 100$. For a small fraction (about 1.7 per cent) of the galaxies in the UVUDF catalogue, reliable spectroscopic redshifts are available from the literature. In Fig.~\ref{fig:UVUDF_z}, we compare the frequency distribution of redshift for this spectroscopic subsample (in blue) to that for the full photometric sample (in red), using photo-$z$ estimates obtained with the \beagle\ tool (see Section~\ref{sec:UVUDF_model} below). Aside from the low number of spectroscopic detections, Fig.~\ref{fig:UVUDF_z} highlights the different redshift distributions of the two samples. The spectroscopic sample peaks around $z\unsim1$ and has very few galaxies at $z\gtrsim 3$, while the photometric sample displays a broader distribution, which extends out to $z\gtrsim 6$.  

\subsection{Modelling approach}\label{sec:UVUDF_model}

\begin{table}
	\centering
	\begin{tabular}{C{0.45\columnwidth-2\tabcolsep} C{0.45\columnwidth-2\tabcolsep} }
\toprule

\multicolumn{1}{c}{Parameter}	    &  \multicolumn{1}{c}{Prior range} \\     

\midrule

$\log (\txn{M}/\Msun)$		& 	$[5,12]$ \\

$\log (\tausfr / \txn{yr})$		& 	$[7,10.5]$ \\

$\log (Z / \Zsun)$		& 	$[-2.2,0.25]$ \\

$\log (\psi_\txn{S} / \txn{yr}^{-1})$		& 	$[-14,-7]$ \\

$\tauV$		& 	$[0.001,5]$ \\

$z$		& 	$[0,15]$ \\

\zform 		&   $[z,50]$ \\ 

\bottomrule
	\end{tabular}
\caption{Prior distributions of the 7 free parameters adopted to interpret the photometric SEDs of UVUDF galaxies in Sections~\ref{sec:UVUDF_model} and \ref{sec:PDF}. A galaxy is assumed to form over a single star formation period extending from \zform\ to $z$ (see Table~\ref{tab:adjust_par} for a description of the different parameters).}
\label{tab:model1_priors}	
\end{table}

To analyse the photometric SEDs of galaxies in the UVUDF catalogue with the \beagle\ tool, we adopt the following prescriptions for the adjustable parameters listed in Table~\ref{tab:adjust_par}. For simplicity, we describe the star formation histories of model galaxies as delayed exponential functions over a single star formation period, $\psi(t^\prime)\,\propto\,t^\prime\,\exp(-t^\prime/\tau_\txn{SFR})$ for $t^\prime\leq\Tuniverse-\tform$, and assume that all stars in a given galaxy have the same metallicity $Z$. We account for the stochastic nature of star formation by adding a `current' burst of star formation to describe stars assembled in the last $\Delta{t}_\txn{SFR}=10$\,Myr of a galaxy star formation history. The strength of the burst component is parametrized in terms of the specific star formation rate $\psi_\txn{S}$. We compute the photoionization of interstellar gas by young stars in the burst as described in Section~\ref{sec:nebular}. For simplicity, in the absence of spectroscopic constraints, we adopt the same metallicity for the interstellar gas as for the stars ($\Zism=Z$), a fixed (intermediate) dust-to-metal mass ratio ($\xid = 0.3$), and the following relation between \Zism\ and $\log \Us$, derived from the analysis of SDSS galaxies by \citet[][and Carton et al. 2016, in prep]{Brinchmann2004}:
\begin{equation}\label{eq:Zgas_logU}
\log \Us=
\left\{ \begin{array}{l l}
-3.638 + 0.055\,x  +0.680\,x^2 & \txn{for } x \leqslant -0.04 \, , \\
 -3.640  &  \txn{for } x > -0.04 \,, 
\end{array} \right .
\end{equation}
where $x=\log (\Zism/\Zsun) $.
We use the prescription of \citet{Chevallard2013} to describe attenuation by dust, fixing the fraction of the dust optical depth arising from the diffuse ISM at $\mu = 0.4$ \citep{Wild2011}. Together with the mass $\txn{M}$ of stars formed, which provides the absolute scaling of the luminosity, and the redshifts of observation and formation, $z$ and \zform, these represent 7 free parameters, in the sense defined in Section~\ref{sec:parametrization}. We also include absorption by the IGM using the prescription of \citet[][see Section~\ref{sec:IGM} above]{Inoue2014}.

To compute the posterior probability distribution $P(\thetab \mid \Db, H)$ of the free model parameters favoured by the observations of a given UVUDF galaxy, we must specify the likelihood function $\mathcal{L}(\thetab)$ and prior distribution $\pi(\thetab)$ entering the right-hand side of equation~\eqref{eq:bayes}. We adopt flat linear prior distributions for $z$, \zform, $\log \txn{M}$, $\log \tausfr$, $\log Z$ and $\log \psi_\txn{S}$, and a flat logarithmic prior distribution for \tauV, within the ranges reported in Table~\ref{tab:model1_priors}. We do not introduce any luminosity function-based prior in this analysis, since we wish to investigate the presence of multi-modal solutions in the posterior distribution of model parameters in the absence of any external information. To build the likelihood function, we model the observed fluxes $\bmath{y}$ of the UVUDF galaxy as a multi-variate Gaussian random variable, with mean given by the prediction $\bmath{\hat{y}}(\bmath{\thetab}^k)$ of our model for a set of parameters $\thetab^k = [\txn{M}, \tausfr, Z, \psi_\txn{S}, \tauV, z, \zform]^k$, and noise described by a diagonal covariance matrix $\mathbfss{\Sigma}$.\footnote{We note that the use of a Gaussian likelihood function to describe data originating from the difference between two Poisson processes (i.e. the difference between source+background and background counts on a detector) is justified only in the limit of large counts, in which case the Poisson distributions can be approximated by Gaussian ones. The presence of objects with negative fluxes in the UVUDF catalogue suggests that this approximation may not be valid for the faintest sources. In those cases, a better approach would be introduce a background term in the likelihood function and use a Poisson distribution to describe the combined source+background signal \citep[e.g.][]{Thompson1999}. Unfortunately, the information in the UVUDF catalogue is not sufficient to allow us to perform such an analysis.} We therefore write the likelihood function of that galaxy as
\begin{equation}\label{eq:likelihood}
- 2\, \ln \mathcal{L}(\thetab^k) = \sum_i \left [ \frac{y_i-\hat{y}_i(\thetab^k)}{\sigma_i} \right ]^2 \, ,
\end{equation} 
where the summation index $i$ runs over all bands observed, even in the absence of detection (i.e. with negative flux after an uncertain background subtraction), and the $\sigma_i$'s are the diagonal elements of matrix $\mathbfss{\Sigma}$. Bands with no detection are of crucial importance to track the absorption of radiation blueward of hydrogen Ly$\alpha$ by the IGM (and blueward of the Lyman limit by the ISM) at high redshift.

The parameter $\sigma_i$ in equation~\eqref{eq:likelihood} is not purely the observational error. In fact, when fitting broadband photometry with spectral evolution models, it is customary to introduce an additional error term to account for uncertainties potentially unaccounted for in the observed fluxes (linked to, e.g., background subtraction, flux calibration, aperture effects) and for (unquantified) systematic uncertainties in model predictions \citep[e.g.,][]{Brammer2008,Dahlen2013,Acquaviva2015}. We account for this effect by adding a 2-per-cent relative error in quadrature to the fluxes in all photometric bands, i.e., we write $\sigma_i = \sqrt{\sigma_\txn{obs}^2+(\sigma_0\,y_i)^2}$, where $\sigma_\txn{obs}$ is the observational error and $\sigma_0 = 0.02$.\footnote{After experimenting with values in the range $0.01\leq\sigma_0\leq0.04$, we find that fixing $\sigma_0 = 0.02$ minimises the dispersion between \zbeagle\ and \zspec, as measured by the quantity \sigNMAD\ (equation~\ref{eq:sigNMAD}).} Although this additional error term does not correct for potential biases originating from systematic uncertainties, it reduces their impact on the results by widening the posterior probability distribution of model parameters. We note that adopting $\sigma_0 = 0.02$ translates into increasing flux errors in each band by factors of 1.08, 1.16 and 1.28 for $\SN=20$, 30 and 40, respectively. We can estimate the typical \SN\ of a UVUDF band starting from Fig.~\ref{fig:UVUDF_S_to_N}, by considering the quantity $\SNave \sim \SNtot /\sqrt{\txn{N}_\txn{bands}}$, where $\txn{N}_\txn{bands}$ varies from 8 to 11 depending on the galaxy. This leads to $\SNave \lesssim10$ for most galaxies, implying that only the few brightest UVUDF galaxies might be significantly affected by the addition of this error term. These also tend to be the lowest-redshift galaxies with spectroscopic redshifts (Fig.~\ref{fig:UVUDF_z}). Without the additional error term, therefore, the results for the high- and low-\SN\ subsamples could be dominated by different error sources: (uncontrolled) systematics originating from both data and models for the former, and \SN\ ratio of the data for the latter. In this case, the conclusions drawn from, for example, the comparison between spectroscopic and photometric redshifts for bright galaxies, would be irrelevant to the fainter sample. 

\begin{figure}
	\centering
	\resizebox{\hsize}{!}{\includegraphics{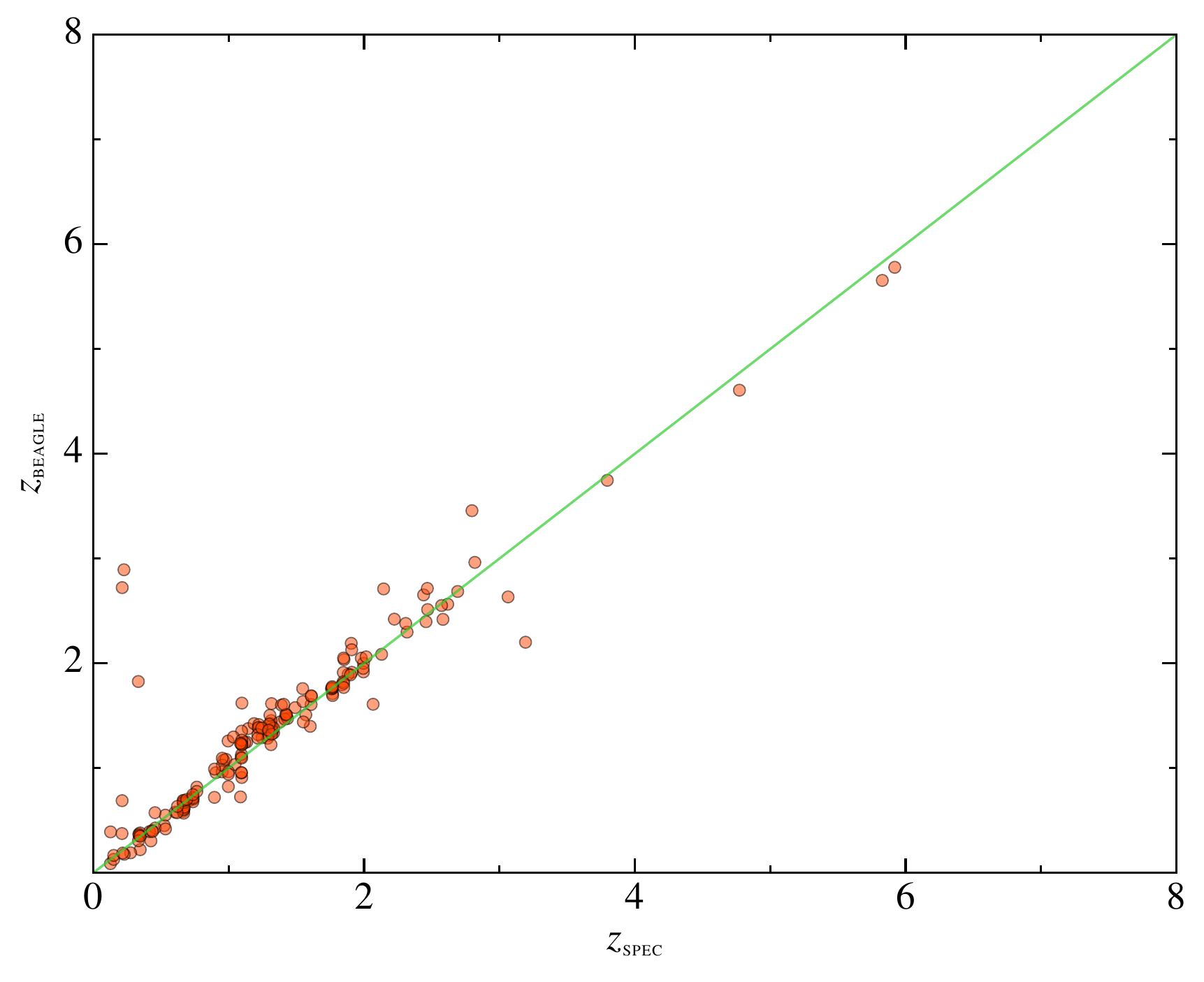}}
	\caption{Photometric redshift \zbeagle\ derived with the \beagle\ tool plotted against spectroscopic redshift \zspec, for the 169 galaxies with spectroscopic detection in the UVUDF catalogue. The green line shows the identity relation.}
	\label{fig:zbeagle_vs_zspec}
\end{figure} 

To compute the posterior probability distribution of the model parameters in Table~\ref{tab:model1_priors} (and of other derived quantities; see Section~\ref{sec:output}) with the \beagle\ tool, we adopt \multinest, an efficient Bayesian inference tool based on the Nested Sampling algorithm of \citet[see Appendix \ref{app:multinest} for detail]{Skilling2006}. We start by focusing on photometric redshift estimates, with the purpose of comparing these with estimates derived using the dedicated photometric redshift codes \bpz\ and \eazy\ in the UVUDF catalogue. To this goal, we adopt as a measurement of $z$ the mean of the marginal posterior probability distribution provided by the \beagle\ tool. In the case of multiple-redshift (i.e. multi-modal) solutions, we identify the mode with highest local evidence, as computed using \multinest, and take the posterior mean within that mode to be a measurement of $z$ (see Section~\ref{sec:multiple_solutions} for a more detailed discussion of multi-modal solutions). 

\subsection{Photometric redshifts}\label{sec:photoz}

\begin{table}
\begin{threeparttable}
	\centering
	\begin{tabular}{R{0.30\columnwidth-2\tabcolsep} R{0.20\columnwidth-2\tabcolsep} C{0.20\columnwidth-2\tabcolsep} C{0.30\columnwidth-2\tabcolsep}}
\toprule

Comparison 	    &	Bias\tnote{a} & \sigNMAD\tnote{b}	& Outlier fraction\tnote{c} \\     

\midrule

$\zbpz - \zspec$	& $0.005$	&	$0.028$ 	&  $2.4 \%$		 \\

$\zeazy - \zspec$	& $-0.013$	&	$0.030$ 	&  $5.9 \%$		 \\

$\zbeagle - \zspec$	& $0.007$	&	$0.047$ 	&  $7.1 \%$		 \\

$\zbeagle - \zbpz$	& $0.005$ &	$0.042$ 	& $8.3 \%$	 \\

$\zbeagle - \zeazy$	& $0.039$	&	$0.058$ 	& $7.3 \%$		 \\

\bottomrule
\end{tabular}
\begin{tablenotes}
	\item [a] Median of the distribution of $(z-\zref) / (1+\zref)$, where \zref\ stands for \zspec, \zbpz\ or \zeazy, depending on the comparison.
	\item [b] Computed using equation~\eqref{eq:sigNMAD} by substituting \zspec, \zbpz\ or \zeazy\ for \zref, depending on the comparison.
	\item [c] Fraction of galaxies with $\lvert \Delta z \rvert / (1+\zref) > 0.15$ for the comparisons $\zbpz-\zspec$, $\zeazy-\zspec$ and $\zbeagle-\zspec$, and with $\lvert \Delta z \rvert / (1+\zref)> 0.25$ in the other two cases.
\end{tablenotes}
\caption{Bias, normalised median absolute deviation and outlier fraction for the comparisons between \beagle-derived photometric redshifts and spectroscopic (169 galaxies) and \bpz- and on \eazy-derived photometric redshifts (9927 galaxies) from the UVUDF catalogue.}
\label{tab:model1_photo_z}	
\end{threeparttable}
\end{table}

In their original study, \citet{Rafelski2015} estimate the photometric redshifts of all UVUDF sources by appealing to two standard codes, \bpz\ \citep{Benitez2000} and \eazy\ \citep{Brammer2008}. Both codes rely on a small set of template galaxy SEDs computed using the \textsc{pegase} population synthesis code \citep{Fioc1997}, although with different prescriptions for the contamination of broadband fluxes by nebular emission lines. The \bpz\ code relies on the original prescription of \textsc{pegase} for nebular emission, while the \eazy\ code incorporates a simplified model relating the $\txn{Ly}\alpha$, $\txn{H}\alpha$, $\txn{H}\beta$, $\txn{H}\gamma$, $[\txn{O}\textsc{ii}]\,\lambda 3727$ and $[\txn{O}\textsc{iii}]\,\lambda\lambda 4959,\, 5007$ line luminosities to the star formation rate \citep[see][]{Brammer2011}. The \bpz\ code also includes a Bayesian prior, based on previously measured galaxy luminosity functions, to help constrain redshift estimates in cases of multiple solutions.

\citet{Rafelski2015} use the subsample of 169 galaxies with spectroscopic redshifts in the UVUDF catalogue to assess the quality of photometric redshifts derived using the \bpz\ and \eazy\ codes, although this quality check is limited by construction to redshifts $z\lesssim3$ (Fig.~\ref{fig:UVUDF_z}). They compute for both \bpz\ and \eazy\ the scatter of the difference $\Delta z$ between photometric redshift and spectroscopic redshift, which they quantify through the normalized median absolute deviation
\begin{equation}\label{eq:sigNMAD}
\sigNMAD = 1.48 \times \txn{median} \left \lvert \frac{\Delta z -\txn{median}(\Delta z)}{1+\zref} \right \rvert \, ,
\end{equation} 
where the reference redshift is the spectroscopic one ($\zref=\zspec$), and the factor $1.48$ ensures that \sigNMAD\ be equal to the standard deviation for a Gaussian distribution. \citet{Rafelski2015} find $\sigNMAD = 0.028$ with the \bpz\ code and  $\sigNMAD = 0.030$ with the \eazy\ code. Following \citet{Brammer2008}, they define the fraction of outliers (OLF) as the fraction of galaxies with $\lvert \Delta z \rvert / (1+\zspec) >5\sigNMAD$, i.e. $\lvert \Delta z \rvert / (1+\zspec)>0.15$. This yields $\OLF\in[1.2\%,4.2\%]$ (4 outliers) with the \bpz\ code, and $\OLF\in[4.1\%,8.4\%]$ (10 outliers) with the \eazy\ code, where the interval in brackets indicates the 68 per cent confidence range computed assuming a Poisson distribution.\footnote{Given $k$ observations originating from a Poisson process with mean $m$, the confidence level $1-\alpha$ on the mean can be computed as $\tfrac{1}{2}\chi^2(\alpha/2;2k)\le m \le \tfrac{1}{2}\chi^2(1-\alpha/2;2k+2)$, where $\chi^2(p;n)$ indicates the quantile function of the chi-squared distribution with $n$ degrees of freedom.} We follow \citet{Rafelski2015} and define the OLF in the comparison between \beagle-derived and spectroscopic redshifts as the fraction of galaxies with $\lvert \Delta z \rvert / (1+\zspec) > 0.15$. We increase the threshold from $0.15$ to $0.25$ when comparing \beagle-derived photometric redshifts with \bpz- and \eazy-derived ones, as $\sigNMAD$ in those cases is typically around 0.05 (see Table~\ref{tab:model1_photo_z}).

Fig.~\ref{fig:zbeagle_vs_zspec} shows that, for the 169 galaxies with spectroscopic detections in the UVUDF catalogue, the photometric redshifts computed as described in Section~\ref{sec:UVUDF_model} above using the \beagle\ tool agree well with the spectroscopic ones. The corresponding normalized median absolute deviation is $\sigNMAD = 0.047$, and the fraction of outliers $\txn{OLF}\in[5.1 \%, 9.8\%]$ (12 outliers; see Section~\ref{sec:UVUDF_data}). The value of $\sigNMAD$ is larger than that obtained using both the \bpz\ and \eazy\ codes. This is likely because the two standard codes \bpz\ and \eazy\ rely on restricted sets of spectral templates optimised for the determination of photometric redshifts, while we consider a full model spanning a broad parameter space to describe the emission from a galaxy. Thus, in our approach, a large number of templates corresponding to different sets of parameters can potentially be consistent with the observed fluxes within the errors, which tends to increase the dispersion in the photometric redshifts derived for a galaxy at a given spectroscopic redshift. In return, the \beagle\ tool has the advantage of providing valuable constraints on galaxy physical properties other than redshift (Figs~\ref{fig:posterior_A} and \ref{fig:posterior_B} below), as well as unique insight into the fundaments of photometric redshift determinations (Section~\ref{sec:discussion}). Regarding this last point, for example, we can identify those galaxies with multi-modal posterior probability distributions and quantify the integrated probability in each mode (Section~\ref{sec:multiple_solutions}). We note that, when considering only those galaxies with a single significant mode (i.e., 107 out of 169 galaxies exhibiting $2\, \ln \K > 10$, where \K\ is the `Bayes factor' defined by equation~\ref{eq:bayes_factor}), we obtain $\sigNMAD = 0.037$ and a single outlier, object \#4721.

It is instructive to investigate the origin of the 12 outliers in the comparison between \zbeagle\ and \zspec. For 5 galaxies (\#1990, \#8292, \#21130, \#22245,  \#50714), the \beagle\ tool identifies two redshift solutions of comparable probability, for which the second solution matches the spectroscopic redshift (see Section~\ref{sec:multiple_solutions} for an in-depth discussion of multimodal solutions). To gain insight into the origin of the remaining 7 outliers, we run a query at the location of each galaxy to obtain its classification in the Simbad\footnote{\url{http://simbad.u-strasbg.fr/simbad/}} database. We also visually examine the SED and image of each outlier from the UVUDF website.\footnote{\url{http://asd.gsfc.nasa.gov/UVUDF/catalogs.html}} Two outliers (\#7024 and \#10157) appear to be contaminated by nearby objects. One (\#4721), classified in Simbad as an AGN, exhibits a strong ultraviolet upturn, which cannot be reproduced by the AGN-free model adopted here (but see Section~\ref{sec:conclusions}). The SEDs of 2 galaxies (\#4658 and \#31320) with $\zspec\sim0.2$ lack ultraviolet observations, and therefore do not sample any strong spectral continuum break useful to constrain photometric redshifts (Lyman, Balmer; these galaxies are outliers in the \bpz-based analysis too). For one outlier (\#4562, $\zspec=2.15$), classified as a `near-IR' source in Simbad and with no ultraviolet observation, we obtain the same photometric redshift ($\approx 2.71$) as that derived by \citet{Rafelski2015} using \bpz, indicating that the information contained in the SED is insufficient to provide an accurate photometric redshift estimate. Another galaxy (\#10496, $\zspec=1.10$) exhibits a similarly featureless continuum, with no ultraviolet observation. 

\begin{figure}
	\centering
	\resizebox{\hsize}{!}{\includegraphics{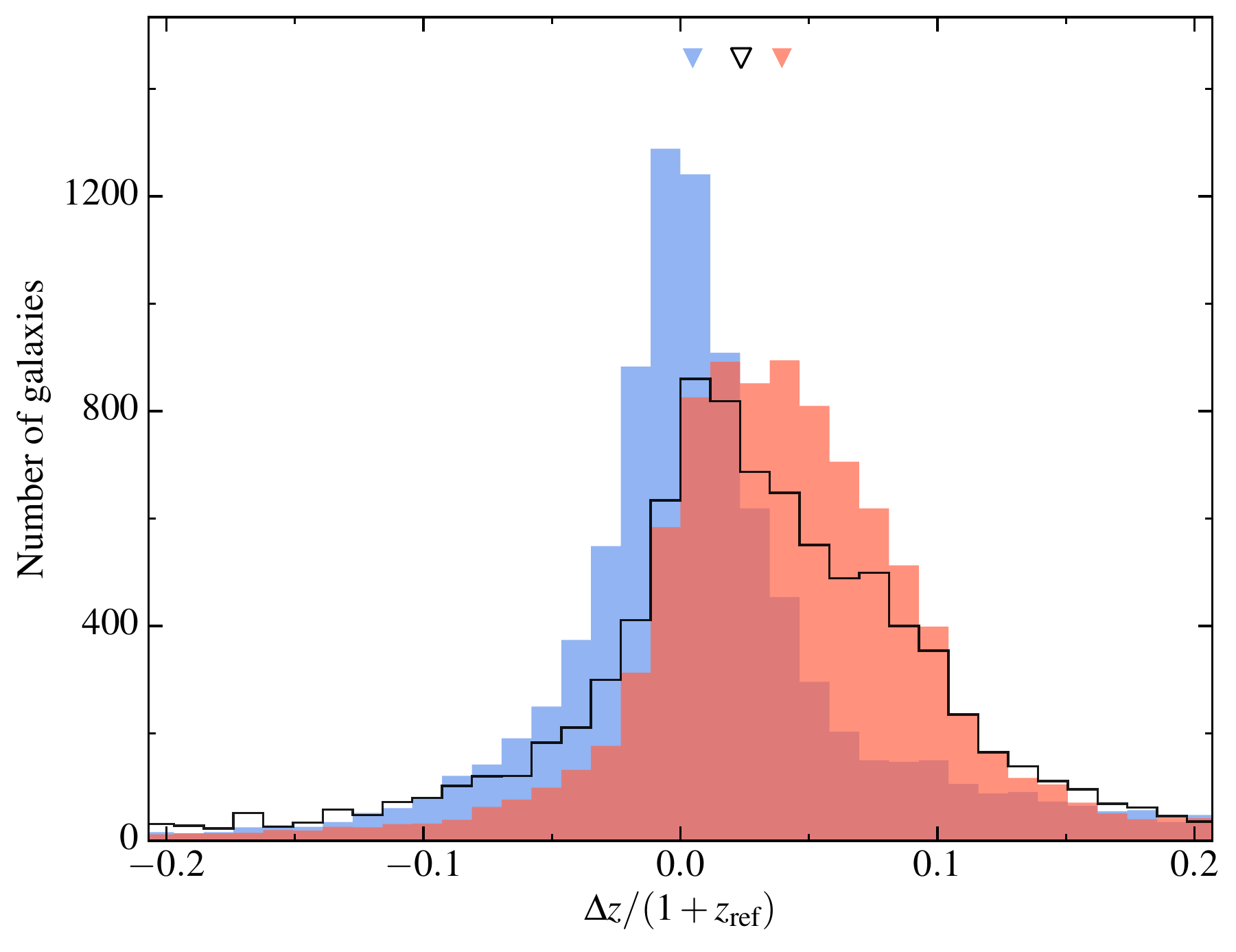}}
	\caption{Comparison of photometric redshifts estimated using different approaches (\zbeagle: this work; \zbpz\ and \zeazy: \citealt{Rafelski2015}), for the 9927 galaxies in the UVUDF catalogue. The distributions are those of $(\zbeagle-\zbpz)/(1+\zbpz)$ (blue filled histogram),  $(\zbeagle-\zeazy)/(1+\zeazy)$ (red filled histogram) and $(\zbpz-\zeazy)/(1+\zeazy)$ (black empty histogram). In each case, a triangle indicates the median of the distribution.}
	\label{fig:zbeagle_vs_zphot_hist}
\end{figure} 

In Fig.~\ref{fig:zbeagle_vs_zphot_hist}, we compare the photometric redshifts computed using the \beagle\ tool with those obtained by \citet{Rafelski2015} using the \bpz\ and \eazy\ codes. The corresponding normalised median absolute deviations and outlier fractions are reported in Table~\ref{tab:model1_photo_z}. Together, Fig.~\ref{fig:zbeagle_vs_zphot_hist} and Table~\ref{tab:model1_photo_z} show that the photometric redshifts computed with our approach and those derived by \citet{Rafelski2015} are globally consistent with each other, as $\sigNMAD$ and \txn{OLF} are typical of comparisons between different photometric redshift codes \citep[e.g., see section~4.2 of][]{Dahlen2013}. Fig.~\ref{fig:zbeagle_vs_zphot_hist} reveals a difference between the distribution of $(\zbeagle-\zbpz)/(1+\zbpz)$ (blue filled histogram), which is roughly centered around zero, and that of $(\zbeagle-\zeazy)/(1+\zeazy)$ (red filled histogram), which is highly skewed towards positive values. This is mainly because of the presence of a second peak including $\unsim$30 per cent of the objects around $\zbeagle-\zeazy\approx0.08(1+\zeazy)$ in the latter distribution. The similarity between \beagle- and \bpz-derived photometric redshifts implies that the distribution of $(\zbpz-\zeazy)/(1+\zeazy)$ is also skewed towards positive values, although in a less severe way (black empty histogram in Fig.~\ref{fig:zbeagle_vs_zphot_hist}).

\begin{figure*}
	\centering
	\begin{subfigure}{.70\hsize}
		\resizebox{\hsize}{!}{\includegraphics{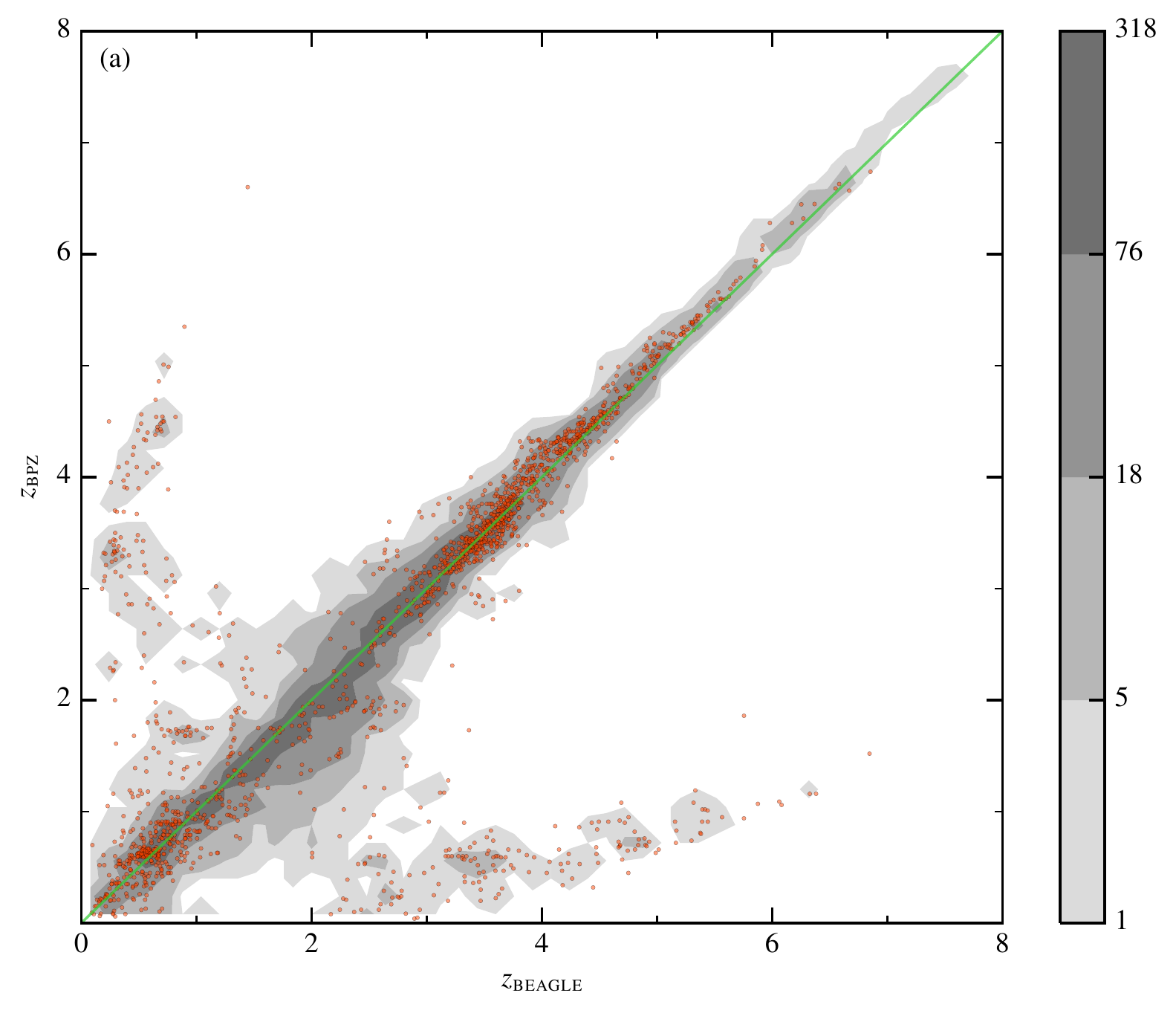}}
	\end{subfigure}
	\begin{subfigure}{.70\hsize}
		\resizebox{\hsize}{!}{\includegraphics{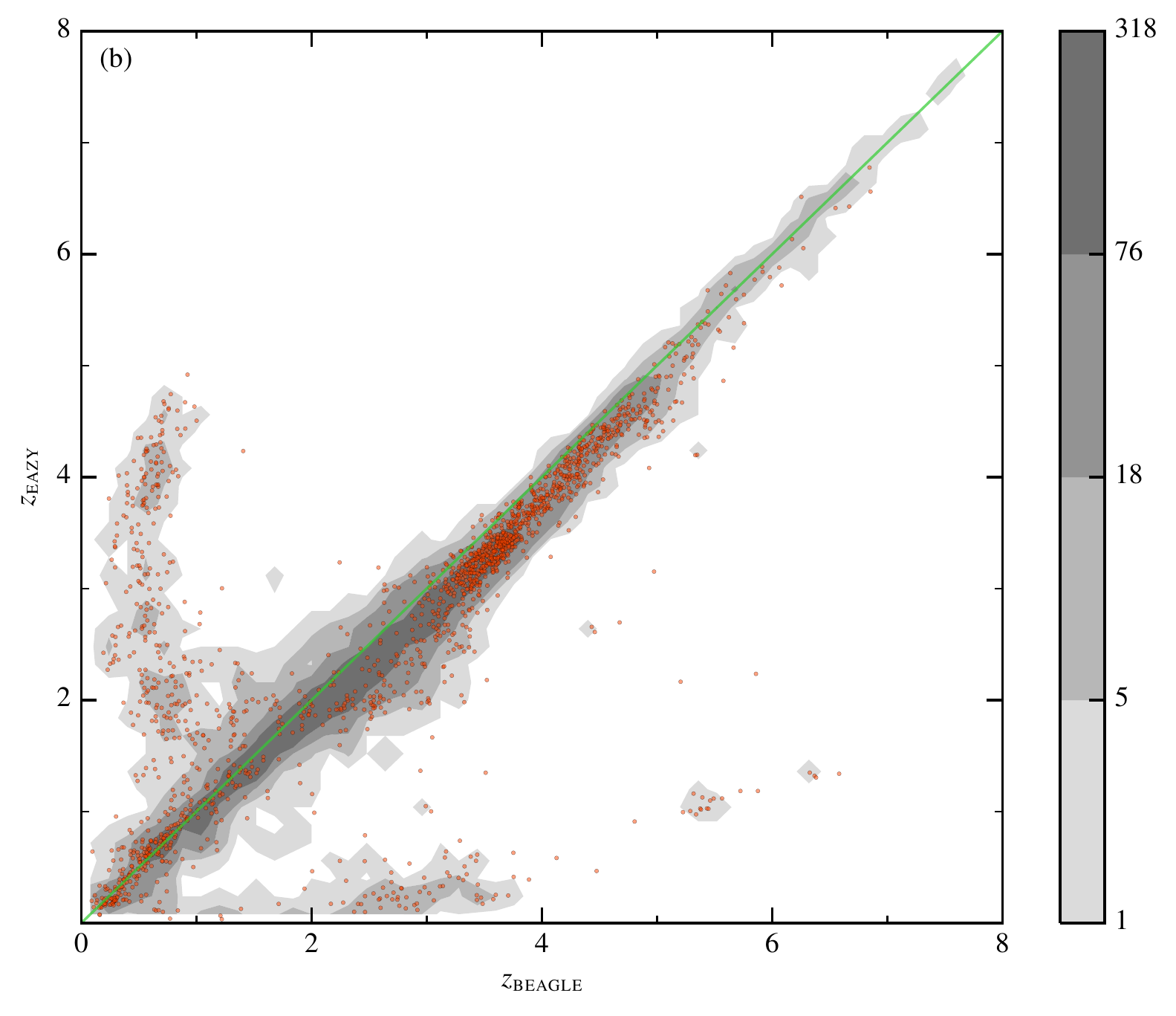}}
	\end{subfigure}
	\caption{(a) Photometric redshift \zbpz\ derived by \citet{Rafelski2015} using the \bpz\ code plotted against that derived in this work using the \beagle\ tool, \zbeagle, for the 9927 galaxies in the UVUDF catalogue. (b) Same as (a), but comparing the photometric redshift \zeazy\ derived by \citet{Rafelski2015} using the \eazy\ code to \zbeagle. In both panels, different gray levels correspond to different logarithmically-spaced galaxy densities (indicated on the right), red circles mark galaxies with multiple redshift solutions of comparable probability, and the green line shows the identity relation. See Section~\ref{sec:UVUDF_model} for detail.}
	\label{fig:zbeagle_vs_zphot}
\end{figure*} 

Fig.~\ref{fig:zbeagle_vs_zphot} shows a more detailed comparison of the photometric redshifts derived using the \beagle\ tool with those derived by \citet{Rafelski2015} using the \eazy\ (top panel) and \bpz\ (bottom panel) codes. In each panel, different gray levels correspond to different logarithmically-spaced galaxy densities, while red circles mark galaxies with multiple redshift solutions of comparable probability (see Section~\ref{sec:multiple_solutions} for an extended discussion). In Fig.~\ref{fig:zbeagle_vs_zphot}a, the vast majority of galaxies lie around the identity relation (green line) at all redshifts. A few galaxies with small \zbeagle\ ($\lesssim 1$) have large associated \zbpz, while a few with high \zbeagle\ have low \zbpz\ ($\lesssim 1$). The presence of red circles in these outlying regions suggests that multiple redshift solutions may be related to large discrepancies between photometric redshift estimates. In Fig.~\ref{fig:zbeagle_vs_zphot}b, most galaxies also lie close to the identity relation, but, as expected from Fig.~\ref{fig:zbeagle_vs_zphot_hist}, the estimates of \zeazy\ for a subtantial fraction of galaxies (at redshifts $2\lesssim \zeazy \lesssim 5$) are systematically smaller than those of \zbeagle. As in the case of Fig.~\ref{fig:zbeagle_vs_zphot}a, the outlying regions occupied by galaxies with discrepant \zbeagle\ and \zeazy\ are also populated by galaxies with multiple redshift solutions (see Section~\ref{sec:multiple_solutions}). We conclude from Figs~\ref{fig:zbeagle_vs_zphot_hist} and \ref{fig:zbeagle_vs_zphot} that the photometric redshifts estimated using the \beagle\ tool are in good general agreement with those estimated by \citet{Rafelski2015} using the \bpz\ and \eazy\ codes. The agreement between \zbpz\ and \zbeagle\ is good at all redshifts, while at $2\lesssim \zeazy \lesssim 5$, the redshifts estimated with the \eazy\ code are systematically lower, by $\unsim 0.08 (1+\zeazy)$, than those estimated using the \beagle\ and \bpz\ tools.

\subsection{Posterior probability distribution of model parameters}\label{sec:PDF}

\begin{figure*}
	\centering
	\resizebox{\hsize}{!}{\includegraphics{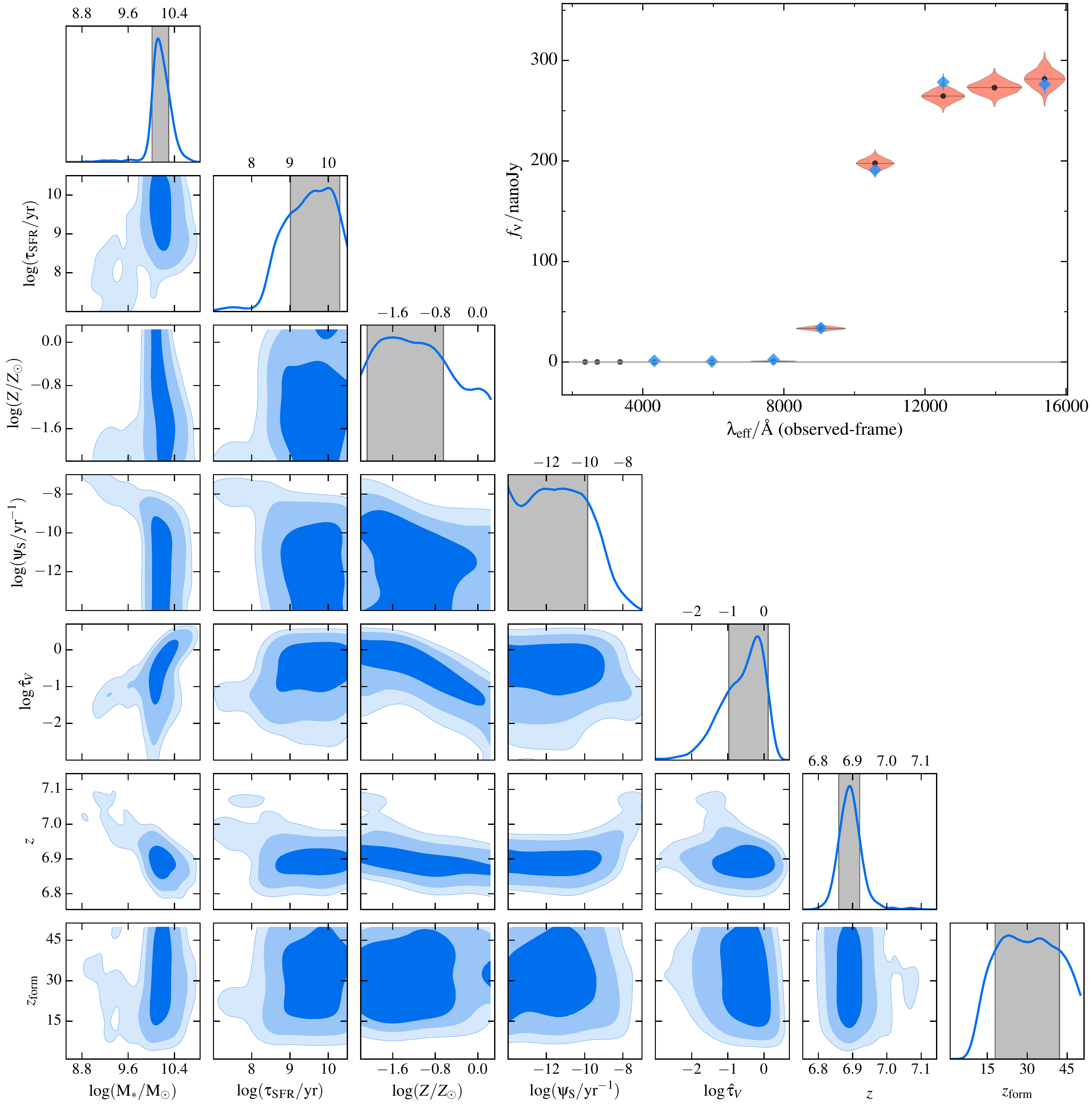}}
	\caption{Posterior probability distribution of the model parameters in Table~\ref{tab:model1_priors} for the UVUDF galaxy \#1021 (F850LP = 27.6 AB mag, $\pvalue = 0.13$). The diagonal panels show the marginal distributions of \Mstar, \tausfr, $Z$, $\psi_\txn{S}$, \tauV\, $z$ and \zform, and the off-diagonal panels the joint distribution of every pair of these parameters. The inset panel on the top right shows the observed SED of the galaxy (blue diamonds), along with the distribution of predicted fluxes (orange `violins') resulting from the posterior probability distribution of the model parameters (see Section~\ref{sec:PDF} for details). This object does not have any measurement in the F225W, F275W and F336W ultraviolet bands nor in the F140W near-infrared band.}
	\label{fig:posterior_A}
\end{figure*}

\begin{figure*}
	\centering
 	  \resizebox{\hsize}{!}{\includegraphics{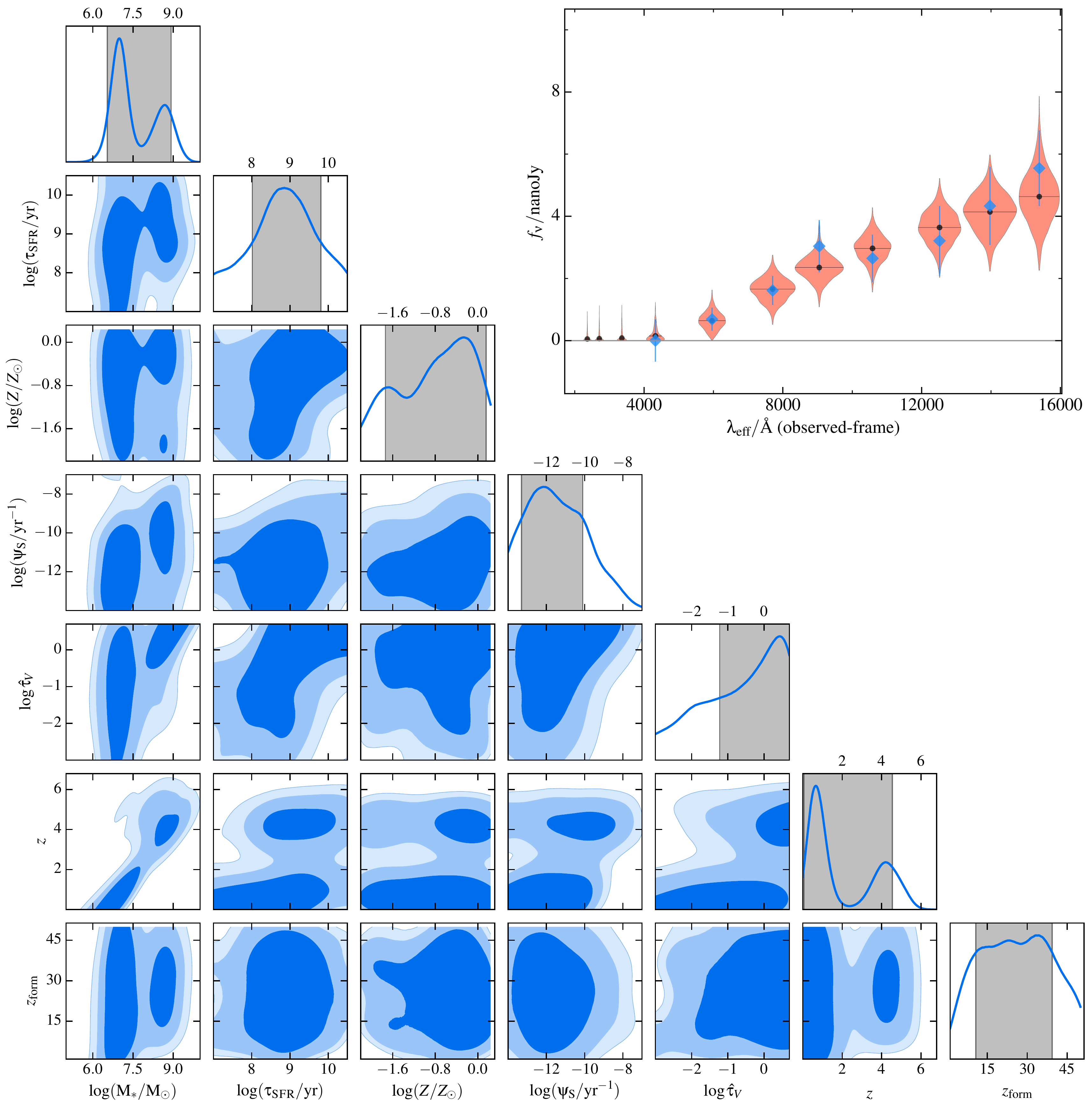}}
	\caption{Same as Fig.~\ref{fig:posterior_A}, but for the galaxy \#5866 (F850LP = 30.2 AB mag, $\pvalue = 0.78$). This object does not have any measurement in the F225W, F275W and F336W ultraviolet bands.}
	\label{fig:posterior_B}
\end{figure*} 

The advantage of the \beagle\ tool over dedicated photometric redshift codes, such as \bpz\ and \eazy, is that it also allows the derivation of rigorous statistical constraints on galaxy physical parameters. We compute the 1-dimensional (i.e. marginal) and 2-dimensional (joint) probability distributions of the model parameters in Table~\ref{tab:model1_priors} (and of other derived quantities) using the \getdist\ Python package, which we integrate into \pybeagle, our package for  post-processing \beagle\ results obtained with the \multinest\ algorithm (Appendix~\ref{app:multinest}). The \getdist\ package\footnote{Available at \url{https://github.com/cmbant/getdist}.} has been developed within \textsc{cosmomc},\footnote{Downloadable from \url{http://cosmologist.info/cosmomc/}.} a powerful Bayesian framework for the analysis of cosmological data originally presented in \citet{Lewis2002} and extensively exploited to interpret \Planck\ data. Both \multinest\ and \getdist\ are well documented and can be used as stand-alone packages. In particular, \textsc{getdist} allows one to compute continuous posterior probability distributions from the samples obtained through \multinest\ by means of 1-dimensional and 2-dimensional kernel density estimates. This presents several advantages over density estimation through standard histograms (e.g. continuity, well defined derivatives, no  requirement of a bin width). The major difficulty associated to the use of kernel density estimates lies in the handling of boundary effects, which are, however, rigorously treated in \textsc{getdist}, following the approach outlined in \citet{Lewis2015}.

We compute in this way the posterior probability distribution of model parameters and derived quantities for all galaxies in the UVUDF sample. For the sake of illustration, we focus here on two particularly instructive cases: a faint, high-redshift galaxy (\#1021); and a galaxy exhibiting multiple probability modes (\#5866), i.e., different regions of high posterior probability in the multi-dimensional parameter space of the model. The main panel of Fig.~\ref{fig:posterior_A} shows, on the diagonal, the marginal posterior probability distributions of the derived quantity \Mstar\ (the current stellar mass, which accounts for the fraction of mass returned to the ISM during stellar evolution; see footnote~\ref{foot:mass}) and the model parameters \tausfr, $Z$, $\psi_\txn{S}$, \tauV, $z$ and \zform\ for the galaxy \#1021 (F850LP = 27.6 AB mag), and off diagonal, the joint posterior probability distribution of every combination of these parameters. The small inset panel shows the observed SED of the galaxy (blue diamonds), along with the distribution of predicted fluxes (orange `violins') resulting from the posterior probability distribution of the model parameters. This was computed by considering the posterior probability and predicted broadband fluxes corresponding to each set of parameters in the posterior probability distribution sampled by \multinest. A kernel density estimate was then performed separately for each band to obtain a smooth flux distribution visualized by the violin (see Section~\ref{sec:PPC} for details about violin plots). The SED of object \#1021 reveals strong IGM absorption at $\lambda \lesssim 8000$ \AA, suggesting a high redshift for this galaxy.

The diagonal panels of Fig.~\ref{fig:posterior_A}, in which we show for each parameter the 68 per cent central credible interval as a grey shaded area, indicate that object \#1021 is a moderately massive galaxy [$\log(\Mstar/\Msun) \sim 10.2$] at redshift $z \sim 6.9$, with a moderate dust content ($\tauV \sim 0.4$). The model favours a long star formation timescale [$\log (\tausfr/\txn{yr}) \gtrsim 9$], high formation redshift [$\zform \gtrsim 20$], low metallicity [$\log (Z/\Zsun) \lesssim -0.7$] and low specific star formation rate [$\log (\psi_\txn{S} / \txn{yr}^{-1})\lesssim -10$], although the widths of the marginal posterior probability distributions of these parameters indicate that they are barely constrained by the observations. The off-diagonal panels illustrate the potential of a Bayesian approach to study degeneracies between model parameters: the three contour levels, showing the 68, 95 and 99 per cent central credible regions, reveal a correlation between stellar mass and dust attenuation optical depth, in the sense that larger \Mstar\ implies larger \tauV. This is because the enhanced flux produced by a more massive galaxy must be attenuated by more dust to produce the same observed flux, when all other parameters are fixed. The figure also shows a mild anti-correlation between dust attenuation optical depth and metallicity, a well known degeneracy resulting from the similar effects of an increase in $Z$ and \tauV\ on galaxy colours.

Fig.~\ref{fig:posterior_B} shows the analog of Fig.~\ref{fig:posterior_A} for the much fainter galaxy \#5866 (F850LP = 30.2 AB mag). The inset panel illustrates how the larger observational errors in this case allow much more extended distributions of the predicted fluxes. The marginal posterior probability distributions of $\tausfr$, $Z$, $\psi_\txn{S}$, \tauV\ and \zform\ in the diagonal panels of Fig.~\ref{fig:posterior_B} show that these parameters are only weakly constrained by the observations. Also, in contrast to Fig.~\ref{fig:posterior_A}, the marginal posterior probability distributions of \Mstar\ and $z$ show two peaks, indicating the presence of two solutions of comparable probability. The joint posterior probability distribution of \Mstar\ and $z$ further shows that these multiple solutions are correlated (i.e., the parameters are degenerate), since the low-redshift solution at $z \sim 0.8 $ favours a lower mass [$\log(\Mstar/\Msun) \sim 7 $] than the high-redshift one at $z \sim 4 $ [$\log(\Mstar/\Msun) \sim 8.7 $]. Such a correlation is expected because, at comparable mass-to-light ratio, a larger mass is required to produce the same apparent luminosity at high relative to low redshift. We note that the existence of this multi-modal solution arises primarily from the faintness of the galaxy \#5866: the low observational \SN\ ratio of this galaxy causes both the Balmer break at $\lambda \sim 3600$ \AA\ and the Lyman break at $\lambda \sim 1216$ \AA\ to be compatible with the observed drop in flux between the F435W and F606W bands. Fig.~\ref{fig:posterior_B} also shows correlations between dust attenuation optical depth and stellar mass, for the same reason as outlined above, and dust attenuation optical depth and redshift, for which solutions implying high values of \tauV\ correspond to higher $z$. Finally, the tighter constraint on \zform\ in the case of the high-redshift solution ($z\sim4$) results from the smaller age spread (and hence luminosity range) of stars in the galaxy in this case, implying less uncertainty on the age of the oldest stellar generation. 

It is also important to note that, while dedicated photometric redshift codes, such as those adopted in \citet{Rafelski2015}, can warn against the presence of multiple redshift solutions, the Bayesian approach implemented in the \beagle\ tool allows one to accurately characterise these solutions, for instance by providing their respective integrated probability. We return to this point in Section~\ref{sec:multiple_solutions} below, where we illustrate the power of the \beagle\ tool in such situations by performing a Bayesian model comparison of the different modes of a posterior probability distribution.

\section{Advanced statistical analysis with BEAGLE}\label{sec:discussion}

In the previous section, we have seen that the \beagle\ tool enables one to fit broadband galaxy SEDs to derive constraints not only on redshift, but also on other model parameters (e.g. stellar mass, star formation history, dust content) and derived quantities. We now describe another main feature of the \beagle\ tool, which is to allow advanced statistical analyses, such as the rigorous quantification of goodness of fit and the detailed study of potential correlations (i.e. degeneracies) between model parameters, to identify and characterise multi-modal solutions. 

\subsection{Posterior predictive checks}\label{sec:PPC}

The outcome of any Bayesian analysis is the posterior probability distribution of a set of parameters \emph{conditional} to a set of observations. Still, the posterior probability distribution alone does not allow one to determine whether the assumed model is a good description of the data, or whether it needs to be changed and improved. In a `frequentist' statistical framework, the goodness of a fit can be evaluated by comparing the value of a test statistics, such as $\chi^2$, to a reference distribution. This is often achieved via the computation of the \pvalue, i.e. the integrated tail probability. A major challenge in this case is to define the appropriate reference distribution for the adopted test statistics. For example, the reference distribution for a $\chi^2$ statistics is tied to the number of degrees of freedom of the assumed model. While this number can be known a priori in a few special cases, such as linear models, it is difficult to estimate in many practical situations. In contrast, in a Bayesian approach, one can design goodness-of-fit tests in which the reference distribution of the adopted test statistics is estimated straightforwardly using the model itself. Such tests, called posterior predictive checks \citep[e.g.][]{Guttman1967, Rubin1984, Gelman1996}, enable the probabilistic assessment of whether a model is a reasonable description of a dataset. 

The general idea motivating posterior predictive checks is the following: if the assumed model is a good description of a set of observations, the model should be able to produce `replicated observations' statistically indistinguishable from the true ones. These replicated data can be thought of as observations which \emph{could} have been measured, assuming that the variability in the data is  entirely captured by the adopted statistical model. The `posterior predictive probability distribution' of replicated data can be written as \citep[e.g.][]{Gelman1996}
\begin{equation}\label{eq:PPC}
P(\Db^\txn{rep} \mid \Db, H ) = \int P(\Db^\txn{rep} \mid \thetab, H ) \, P(\thetab \mid \Db, H) \, d\thetab \, ,
\end{equation} 
where $H$, $\thetab$ and \Db\ have the same meaning as in equation~\eqref{eq:bayes}, and $\Db^\txn{rep}$ is the replicated dataset. The first factor in the integral on the right-hand side of equation~\eqref{eq:PPC} is the probability distribution of replicated data conditional to the model parameters, while the second is the posterior probability distribution obtained by applying Bayes' theorem (equation~\ref{eq:bayes}). An advantage of a Bayesian goodness-of-fit test based on equation~\eqref{eq:PPC} is that, once a set of samples drawn from the posterior probability distribution is available, no further heavy computation is required.  

\begin{figure}
	\resizebox{\hsize}{!}{\includegraphics{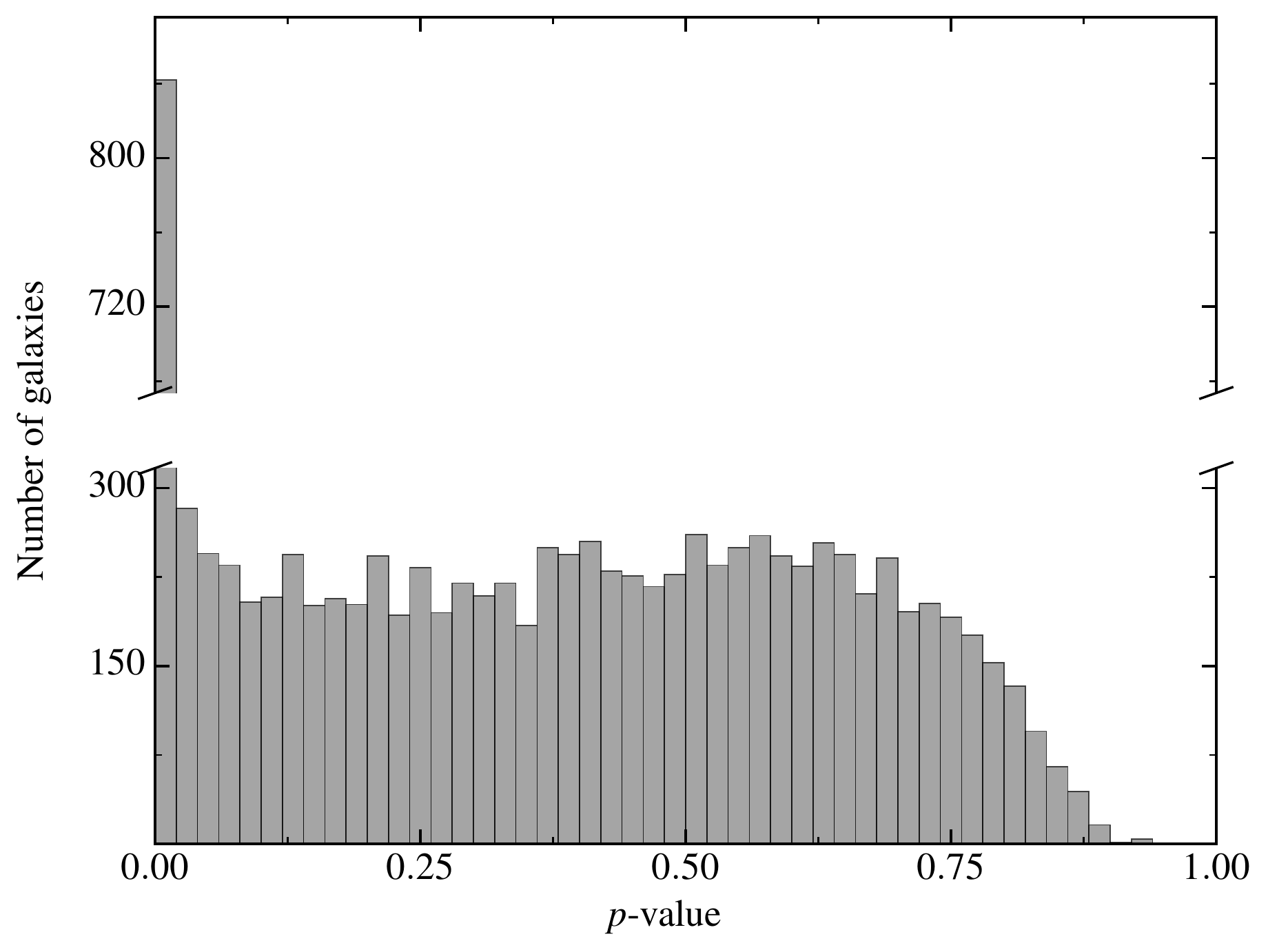}}
	\caption{Distribution of \pvalue s, computed according to equation~\eqref{eq:p-value}, for the 9927 galaxies in the UVUDF catalogue fitted with the model presented in Section~\ref{sec:UVUDF_model}.}
	\label{fig:p_value}
\end{figure}

We consider here two types of posterior predictive checks: a numerical one, based on the $\chi^2$ test statistics, and a graphical one. In both cases, for each galaxy in the UVUDF catalogue, we need a set of replicated data drawn from the probability distribution defined in equation~\eqref{eq:PPC}. For this, we start from the output of \multinest, which consists in an ensemble of $\Nout$ sets of parameters weighted by the posterior probability distribution (see Appendix~\ref{app:multinest}). Then, we draw $\Nrep =2000$ replicated datasets as follows (we have checked that adopting a larger $\Nrep$ has a negligible influence on the results):
\begin{description}
	\item[1)] we draw \Nrep\ sets of model parameters $\thetab^k$, with $1\leq k\leq\Nrep$, from the posterior probability distribution obtained with \multinest, using a `weighted sampling with replacement' scheme (see Appendix~\ref{app:weighted} for detail); 
	
	\item[2)] for each set of parameters $\thetab^k$, we draw the replicated data \yrep\ from a Gaussian distribution $\mathcal{N}[\bmath{\hat{y}}(\thetab^k), \mathbfss{\Sigma}]$, where $\bmath{\hat{y}}(\thetab^k)$ indicates the fluxes predicted by the model given the set of parameters $\thetab^k$, and $\mathbfss{\Sigma}$ is the covariance matrix of the data, i.e. in our case a diagonal matrix with elements $\mathbfss{\Sigma}_{i,i} = \sigma^2_i$ (as in Section~\ref{sec:UVUDF_model}).
\end{description}

To perform the numerical posterior predictive check, we adopt the $\chi^2$ test statistics as a measure of deviance between model predictions and data. For each galaxy in the UVUDF catalogue, we compute the $\chi^2$ deviance as 
\begin{equation}\label{eq:chi2}
\chi^2(\thetab^k) = \sum_i \left [ \frac{ y_i-\hat{y}_i(\thetab^k) }{\sigma_i} \right]^2 \, ,
\end{equation} 
where, as in equation~\ref{eq:likelihood}, $y_i$ indicates the observed flux in the $i$-th band, $\hat{y}_i(\thetab^k)$ the flux predicted by the model in the same band for a set of parameters $\thetab^k$, and $\sigma_i$ the error (which, as in equation~\ref{eq:likelihood}, includes both the observational error and the 2-per-cent relative error added in quadrature). We compute $\chi^2(\thetab^k)$ for all \Nrep\ sets of model parameters $\thetab^k$. Then, for each set of parameters, we substitute $y^\txn{rep}_i$ for $y_i$ in equation~\eqref{eq:chi2} and compute the corresponding $\chi^2_\txn{rep}(\thetab^k)$. In this way, we obtain two distributions of the $\chi^2$ statistics, one pertaining to the true data, and one to replicated data. To compare the two distributions, we adopt as \pvalue\ the fraction of replicated data with $\chi^2$ larger than the corresponding one obtained with the true data \citep{Gelman1996}, i.e.,
\begin{equation}\label{eq:p-value}
\pvalue = \frac{\txn{N}\left [ \chi^2_\txn{rep}(\thetab) > \chi^2(\thetab) \right]}{\Nrep} \, .
\end{equation} 

Fig.~\ref{fig:p_value} shows the distribution of \pvalue s computed in this way for all galaxies in the UVUDF catalogue. This reveals that most galaxies have \pvalue s in the range 0.05--0.95, indicating a satisfactory model fit, while 7 (13) per cent have \pvalue s less than 0.01 (0.05). Such very low \pvalue s suggests that there may be a problem with either the data or the model, or both, in the fit of these galaxies. We note that a way to discriminate between a data or model origin of a bad fit is to study the fit residuals in the observer and rest frames: while data-driven mismatches arise at the instrument (or reduction) level, hence acting in the observer frame, model-driven ones are caused by an inaccurate physical representation of galaxy SEDs, hence acting in the galaxy rest frame. 
 
\begin{figure*}
	\centering
	\resizebox{\hsize}{!}{\includegraphics{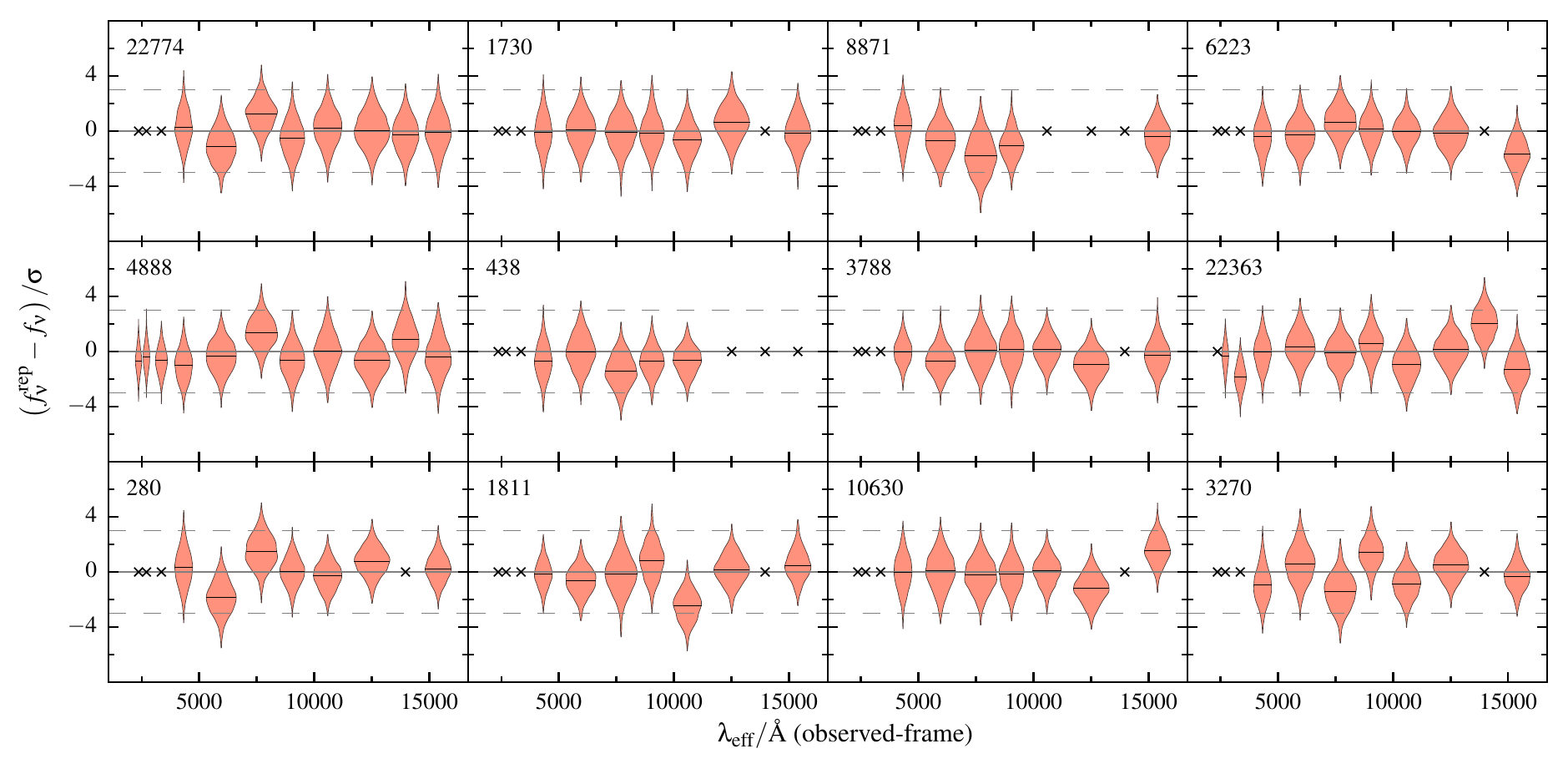}}
	\caption{Graphical posterior predictive check for 12 randomly selected galaxies with $\pvalue > 0.1$ (as computed using equation~\ref{eq:p-value} when fitting with the model presented in Section~\ref{sec:UVUDF_model}) in the UVUDF catalogue (the galaxy ID is indicated in the top right corner of each panel). In each panel, each `violin' shows the probability distribution of the residual between replicated and true data in a given photometric band, with the median marked by a black horizontal line. Crosses indicate bands with no measurement. The solid and dashed grey lines indicate zero and $\pm3\,\sigma$ residuals, respectively (see Section~\ref{sec:PPC} for details).}
	\label{fig:PPC_graph_high}
\end{figure*} 

\begin{table*}
\begin{threeparttable}
	\centering
	\begin{tabular}{C{0.09\linewidth-2\tabcolsep} C{0.075\linewidth-2\tabcolsep} C{0.075\linewidth-2\tabcolsep} C{0.075\linewidth-2\tabcolsep} C{0.075\linewidth-2\tabcolsep} C{0.075\linewidth-2\tabcolsep} C{0.075\linewidth-2\tabcolsep} C{0.075\linewidth-2\tabcolsep} C{0.075\linewidth-2\tabcolsep} C{0.075\linewidth-2\tabcolsep} C{0.075\linewidth-2\tabcolsep} C{0.075\linewidth-2\tabcolsep} C{0.075\linewidth-2\tabcolsep}}
\toprule

\multirow{2}{*}{Galaxy ID}	    &  \multicolumn{11}{c}{Photometric band} \\     

 \cmidrule{2-12} 

  					 &	 F225W  & F275W &	F336W & F435W  &	 F606W & F775W  &	F850LP &  F105W &	F125W & F140W &	F160W  & All\\     

\midrule

$22774$  &   &   &   &$0.53$  &$0.38$  &$0.30$  &$0.47$  &$0.54$  &$0.57$  &$0.53$  &$0.50$  & $0.41$  \\

$1730$  &   &   &   &$0.48$  &$0.53$  &$0.56$  &$0.55$  &$0.45$  &$0.53$  &   &$0.60$  & $0.59$  \\

$8871$  &   &   &   &$0.47$  &$0.44$  &$0.21$  &$0.32$  &   &   &   &$0.63$  & $0.24$  \\

$6223$  &   &   &   &$0.56$  &$0.58$  &$0.51$  &$0.64$  &$0.86$  &$0.69$  &   &$0.14$  & $0.58$  \\

$4888$  &$0.49$  &$0.75$  &$0.51$  &$0.38$  &$0.59$  &$0.22$  &$0.51$  &$0.55$  &$0.54$  &$0.38$  &$0.47$  & $0.53$  \\

$438$  &   &   &   &$0.43$  &$0.58$  &$0.23$  &$0.50$  &$0.52$  &   &   &   & $0.44$  \\

$3788$  &   &   &   &$0.97$  &$0.56$  &$0.55$  &$0.52$  &$0.76$  &$0.38$  &   &$0.60$  & $0.76$  \\

$22363$  &   &$0.72$  &$0.063$  &$0.65$  &$0.58$  &$0.71$  &$0.54$  &$0.39$  &$0.67$  &$0.072$  &$0.29$  & $0.21$  \\

$280$  &   &   &   &$0.46$  &$0.16$  &$0.21$  &$0.76$  &$0.76$  &$0.47$  &   &$0.78$  & $0.33$  \\

$1811$  &   &   &   &$0.86$  &$0.55$  &$0.51$  &$0.42$  &$0.031$  &$0.73$  &   &$0.58$  & $0.22$  \\

$10630$  &   &   &   &$0.51$  &$0.52$  &$0.59$  &$0.58$  &$0.79$  &$0.28$  &   &$0.14$  & $0.47$  \\

$3270$  &   &   &   &$0.36$  &$0.46$  &$0.25$  &$0.21$  &$0.44$  &$0.61$  &   &$0.68$  & $0.28$  \\

\bottomrule
\end{tabular}
\caption{Posterior predictive \pvalue s in the different photometric bands, and global \pvalue\ computed using equation~\eqref{eq:p-value} (rightmost column), for the 12 galaxies in Fig~\ref{fig:PPC_graph_high} (see Section~\ref{sec:PPC} for details).}
\label{tab:pvalue_high}	
\end{threeparttable}
\end{table*}

While Fig.~\ref{fig:p_value} summarises the global quality of photometric fits of UVUDF galaxies obtained with the model of Section~\ref{sec:UVUDF_model}, it does not allow us to characterise this performance in detail, i.e. band by band. For this reason, we also perform a graphical posterior predictive check. We randomly select 12 well-fitted galaxies with \pvalue s greater than 0.1, and 12 badly-fitted galaxies with \pvalue s less than 0.01. For each galaxy and each set of replicated data of that galaxy, we compute the residual between replicated and true data, $(y^\txn{rep}_i-y_i)/\sigma_i$, in each photometric band (where the symbols have the same meaning as in equation~\ref{eq:chi2}). Fig.~\ref{fig:PPC_graph_high} shows a `violin' plot of the resulting distribution of residuals for the 12  galaxies with \pvalue s greater than 0.1. Each violin was computed  by performing, for each band separately, a kernel density estimate to obtain a continuous distribution of the \Nrep\ residuals and then plotting the 0.997 central credible region of this distribution. As is customary in violin plots, the distribution was mirrored with respect to an axis parallel to the ordinate axis and the maximum width of each violin adapted to avoid overlap. Each violin of Fig.~\ref{fig:PPC_graph_high} therefore reflects the probability of obtaining a given residual in a given photometric band: the more extended the violin in the ordinate direction, the broader the distribution of the residual.

\begin{figure*}
	\centering
	\resizebox{\hsize}{!}{\includegraphics{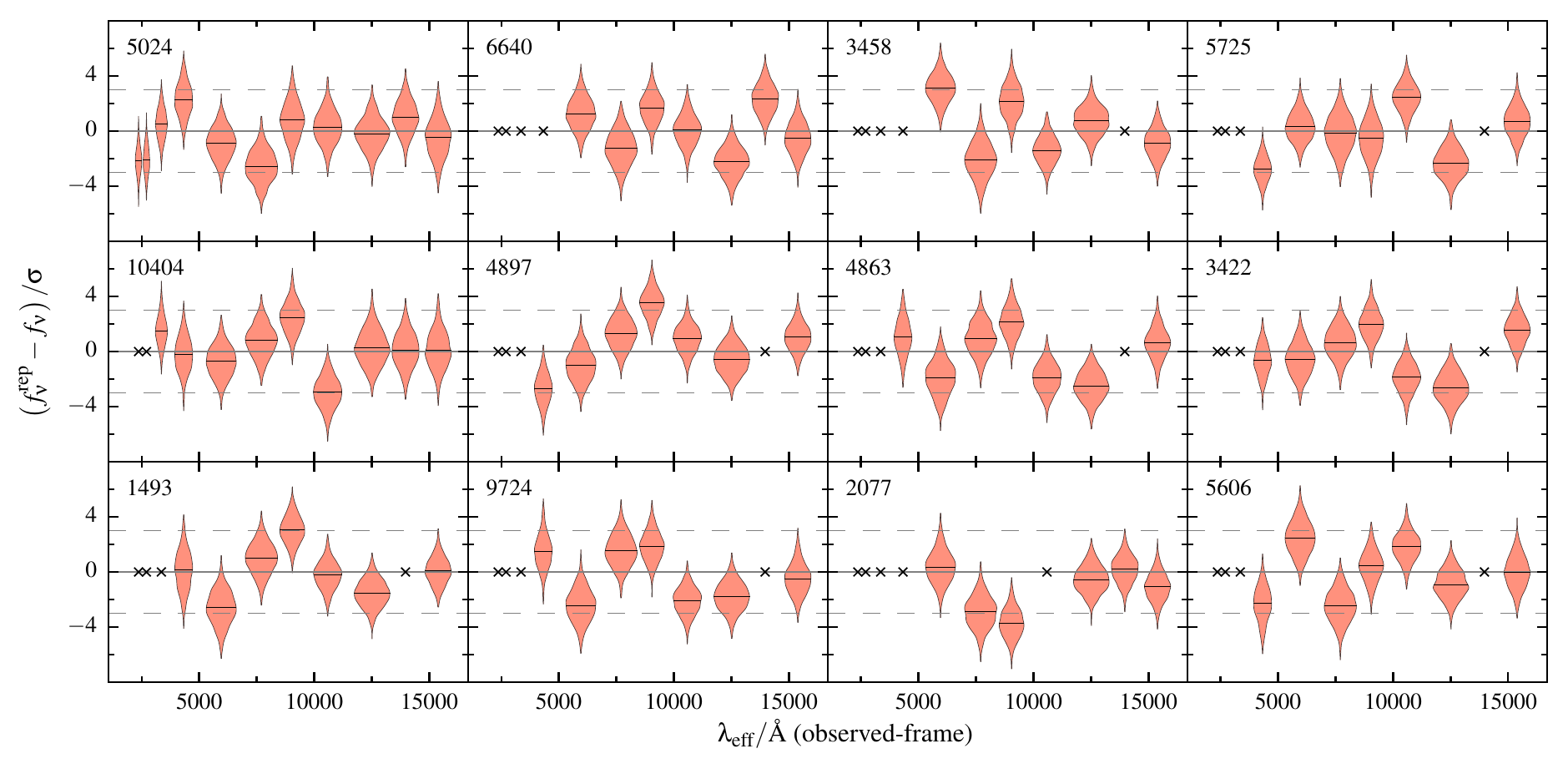}}
	\caption{Same as Fig.~\ref{fig:PPC_graph_high}, but for 12 randomly selected galaxies with $\pvalue \le 0.01$.}
	\label{fig:PPC_graph_low}
\end{figure*}

\begin{table*}
\begin{threeparttable}
	\centering
	\begin{tabular}{C{0.09\linewidth-2\tabcolsep} C{0.072\linewidth-2\tabcolsep} C{0.072\linewidth-2\tabcolsep} C{0.085\linewidth-2\tabcolsep} C{0.072\linewidth-2\tabcolsep} C{0.072\linewidth-2\tabcolsep} C{0.072\linewidth-2\tabcolsep} C{0.085\linewidth-2\tabcolsep} C{0.072\linewidth-2\tabcolsep} C{0.072\linewidth-2\tabcolsep} C{0.072\linewidth-2\tabcolsep} C{0.072\linewidth-2\tabcolsep} C{0.085\linewidth-2\tabcolsep}}
\toprule

\multirow{2}{*}{Galaxy ID}	    &  \multicolumn{11}{c}{Photometric band} \\     

 \cmidrule{2-12} 

  					 &	 F225W  & F275W &	F336W & F435W  &	 F606W & F775W  &	F850LP &  F105W &	F125W & F140W &	F160W  & All\\     

\midrule

$5024$  &$0.046$  &$0.065$  &$0.57$  &$0.062$  &$0.42$  &$0.042$  &$0.43$  &$0.56$  &$0.61$  &$0.35$  &$0.43$  & $0.0025$  \\

$6640$  &   &   &   &   &$0.28$  &$0.30$  &$0.13$  &$0.67$  &$0.046$  &$0.035$  &$0.55$  & $0.0095$  \\

$3458$  &   &   &   &   &$0.0035$  &$0.13$  &$0.076$  &$0.19$  &$0.48$  &   &$0.38$  & $0.0030$  \\

$5725$  &   &   &   &$0.0070$  &$0.66$  &$0.54$  &$0.47$  &$0.023$  &$0.038$  &   &$0.52$  & $0.0050$  \\

$10404$  &   &   &$0.19$  &$0.54$  &$0.49$  &$0.46$  &$0.031$  &$0.0090$  &$0.53$  &$0.51$  &$0.52$  & $0.0090$  \\

$4897$  &   &   &   &$0.018$  &$0.40$  &$0.24$  &$0.0005$  &$0.35$  &$0.58$  &   &$0.30$  & $0.0020$  \\

$4863$  &   &   &   &$0.35$  &$0.13$  &$0.39$  &$0.069$  &$0.082$  &$0.020$  &   &$0.48$  & $0.0030$  \\

$3422$  &   &   &   &$0.51$  &$0.55$  &$0.52$  &$0.069$  &$0.088$  &$0.020$  &   &$0.15$  & $0.0090$  \\

$1493$  &   &   &   &$0.43$  &$0.032$  &$0.37$  &$0.0035$  &$0.81$  &$0.13$  &   &$0.88$  & $0.0035$  \\

$9724$  &   &   &   &$0.23$  &$0.052$  &$0.19$  &$0.085$  &$0.037$  &$0.093$  &   &$0.49$  & $0.0040$  \\

$2077$  &   &   &   &   &$0.57$  &$0.016$  &$0.0020$  &   &$0.61$  &$0.85$  &$0.27$  & $0.0055$  \\

$5606$  &   &   &   &$0.072$  &$0.046$  &$0.042$  &$0.60$  &$0.082$  &$0.40$  &   &$0.60$  & $0.0025$  \\

\bottomrule
\end{tabular}
\caption{Same as Table~\ref{tab:pvalue_high}, but for the 12  galaxies in Fig~\ref{fig:PPC_graph_low}.}
\label{tab:pvalue_low}	
\end{threeparttable}
\end{table*}

As a complement to Fig.~\ref{fig:PPC_graph_high}, Table~\ref{tab:pvalue_high} lists for each galaxy the \pvalue\ computed using equation~\eqref{eq:p-value}. We also report the \pvalue\ computed in the same way for each photometric band separately, by considering the contribution of only that band to $\chi^2$ in equation~\eqref{eq:chi2}. In the case of (random) noise-driven residuals, and assuming that $\sigma$ (where we have dropped the band index) accounts for all possible sources of noise, we expect residual distributions centred around zero with a dispersion comparable to $\sigma$. This is what most violins reflect in Fig.~\ref{fig:PPC_graph_high}, although some bands for some galaxies display larger residuals ($\unsim2\,\sigma$;  e.g., \#22363: band F336W and F140W; \#1811: band F105W). Table~\ref{tab:pvalue_high} quantifies this information by providing the significance of residual deviations by means of the \pvalue : band F336W and F140W for object \#22363 and band F105W for object \#1811 all have \pvalue s in the range $\unsim[0.07,0.03]$. 

Fig.~\ref{fig:PPC_graph_low} and Table~\ref{tab:pvalue_low} show the analogs of Fig.~\ref{fig:PPC_graph_high} and Table~\ref{tab:pvalue_high} for the 12 randomly selected galaxies with \pvalue s less than 0.01. In contrast to Fig.~\ref{fig:PPC_graph_high}, Fig.~\ref{fig:PPC_graph_low} shows significant residual deviations between replicated and true data. These deviations differ from galaxy to galaxy, as Table~\ref{tab:pvalue_low} also highlights. Although a comprehensive analysis of fitting residuals from posterior predictive checks goes beyond the scope of the present paper, we stress that studying the distributions of these residuals in the observer and galaxy rest frames can help discriminate between a data or model origin (see above).

Hence, by means of posterior predictive checks, we have shown that the relatively simple model of Section~\ref{sec:UVUDF_model} provides a satisfactory fit to the photometry of a vast majority of UVUDF galaxies. We have illustrated how the combination of graphical and numerical posterior predictive checks can provide valuable insight into the origin of discrepancies between model and data. Distributions of residual deviance between model predictions and data, of the type shown in Figs~\ref{fig:PPC_graph_high} and \ref{fig:PPC_graph_low} for individual galaxies, can be straightforwardly extended to combine residuals from different objects, enabling the identification of data- and model-driven discrepancies too subtle to be detected in single galaxies. In this context, we believe that the \beagle\ tool will be valuable both to identify subtle systematics in observed datasets, and to characterise current limitations and drive future developments of spectral models.

\subsection{Multi-modal solutions}\label{sec:multiple_solutions}

\begin{table}
	\centering
	\begin{tabular}{C{0.16\columnwidth-2\tabcolsep} C{0.16\columnwidth-2\tabcolsep} C{0.24\columnwidth-2\tabcolsep} C{0.15\columnwidth-2\tabcolsep} C{0.155\columnwidth-2\tabcolsep} C{0.155\columnwidth-2\tabcolsep}}
\toprule

 Label & $2\, \ln \K$    & Kass \& Raftery 	& Total sample	&	\bpz\ outliers &	\eazy\ outliers	 \\     

 \midrule

A	& 	$[0,2]$	& `Barely worth mentioning'	&	$0.040$ & $0.19$ (160)	 &	 $0.21$ (155)	\\

B	&	$(2,6]$	& `Positive'				&	$0.093$ & $0.19$ (159)	  &	 $0.29$ (210)	\\

C	&	$(6,10]$ & `Strong'					&	$0.049$ & $0.043$ (35)	&	 $0.053$ (38)	\\

\\

\cmidrule{3-6}

A, B, C	& $\le 10$	& Not `very strong'			&	$0.18$ & $0.43$ (354)		&	 $0.56$ (403)	 \\

\\

D	&	$>10$	& `Very strong'				&	$0.15$ & $0.089$ (73)		&	 $0.13$ (92)	 \\

\bottomrule
\end{tabular}
\caption{Fractional distribution of the quantity $2\, \ln \K$, where \K\ is the Bayes factor (equation~\ref{eq:bayes_factor}), among the four `belief' categories defined by \citet[][regions A, B, C and D]{Kass1995}, of the total sample of 9927 UVUDF galaxies. The two rightmost columns indicate the fractional distribution among the same categories (with actual numbers in parentheses) of the outliers in the \zbeagle\ versus \zbpz\ and \zbeagle\ versus \zeazy\ comparisons (from Table~\ref{tab:model1_photo_z}).}
\label{tab:bayes_factor}	
\end{table}

So far, we have shown that: (i) multiband photometric fitting of UVUDF galaxies with the \beagle\ tool leads to photometric redshift estimates in good agreement with those derived from standard dedicated codes (Section~\ref{sec:photoz}); (ii) the accurate Bayesian characterization of the posterior probability distribution of model parameters with the \beagle\ tool allows a rigorous study of multiple solutions and degeneracies between model parameters (Section~\ref{sec:PDF}) ; and (iii) the relatively simple model of Section~\ref{sec:UVUDF_model} provides satisfactory fits to the photometry of most UVUDF galaxies (Section~\ref{sec:PPC}). In this section, we focus on the study of multi-modal solutions, which as noted above could be a major cause of discrepancy between photometric redshifts derived using different approaches (Fig~\ref{fig:zbeagle_vs_zphot}).

We start by noting that the \bpz\ and \eazy\ codes adopted by \citet{Rafelski2015} to estimate photometric redshifts provide a quantity, the `odds', which is sensitive to the occurrence of multiple redshift solutions. Both codes compute the marginal probability distribution of redshift and integrate this over some fixed range. In the \bpz\ code, this range is an interval of width $0.06\times(1+z)$ around the peak, while in the \eazy\ code, the interval is $0.2\times(1+z)$ wide. Such a fixed interval of integration can provide some indication about the concentration of the marginal probability distribution of redshift around the peak, but not a measure of the relative probabilities of different potential solutions. For reference, \citet{Rafelski2015} consider a redshift estimate to be reliable is the corresponding odds are greater than 0.9.

\begin{figure}
	\centering
	\resizebox{\hsize}{!}{\includegraphics{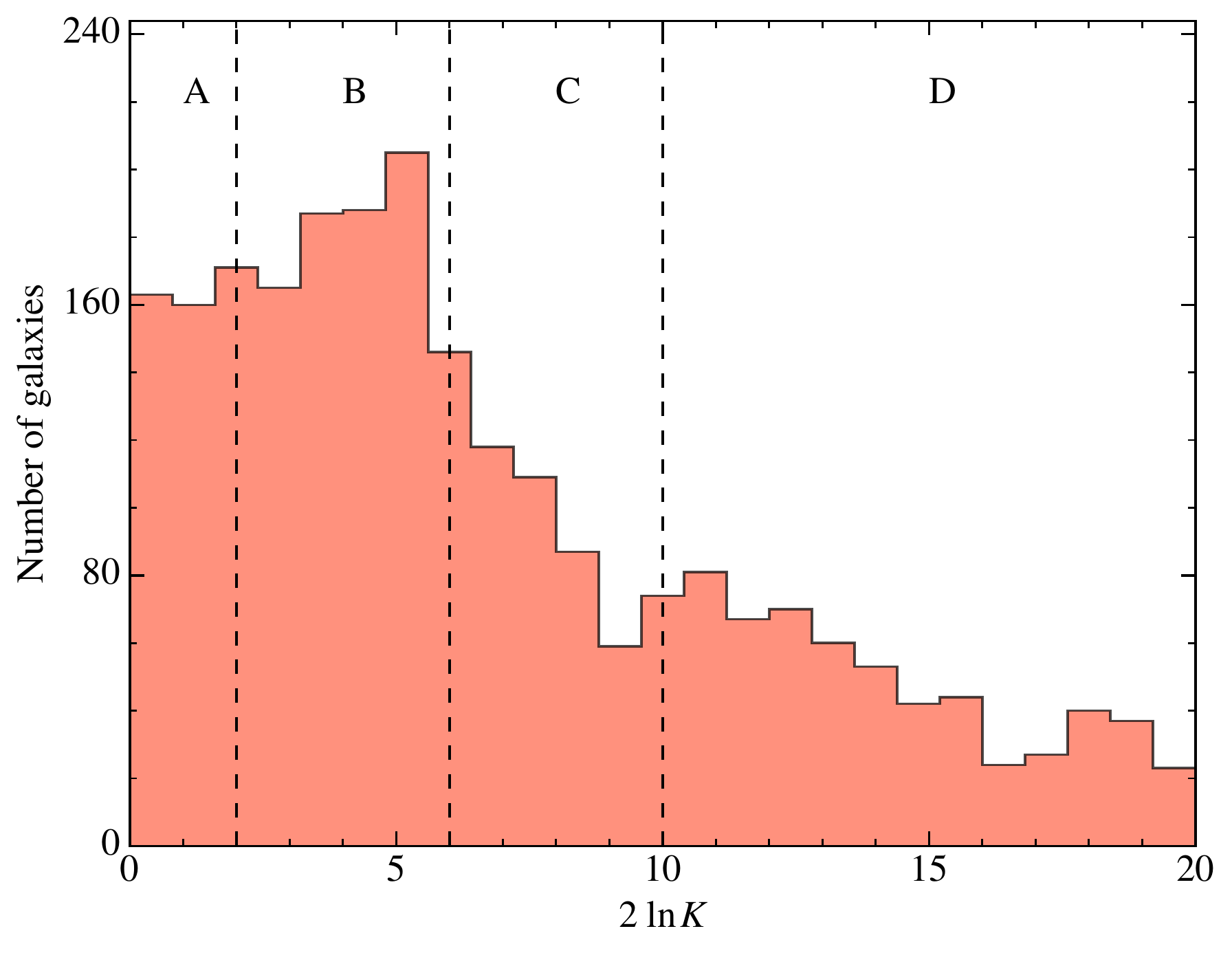}}
	\caption{Distribution of the quantity $2\, \ln \K$, where \K\ is the Bayes factor (equation~\ref{eq:bayes_factor}), for the 3243 UVUDF galaxies identified to have multiple redshift solutions separated by $\lvert \Delta z \rvert > 0.5$. The vertical lines mark the thresholds proposed by \citet{Kass1995} to separate different categories of `belief' (regions A, B, C and D), as specified in Table~\ref{tab:bayes_factor}. Note that  region D includes all galaxies with $2\, \ln \K > 10$, but for clarity we only show the plot up to  $2\, \ln \K=20$.}
	\label{fig:bayes_factor}
\end{figure}

As emphasized in Sections~\ref{sec:photoz} and \ref{sec:PDF}, our approach in the \beagle\ tool differs from that of most template-based photometric redshift codes, in that we consider redshift as just one of several model parameters influencing the predicted observables (Table~\ref{tab:adjust_par}). Thus, each value of photometric redshift is explored along with a set of other galaxy parameters. This implies that different redshift solutions also correspond to different solutions of other physical parameters, via the multi-dimensional posterior probability distribution of equation~\eqref{eq:bayes}. We note in passing that this may also potentially provide useful information to exclude some redshift solutions, for example, because of unlikely combinations of parameters (e.g., massive galaxy with very low metallicity). In the case of multi-modal solutions, the \beagle\ tool allows one to compute the probability associated to each mode and perform a Bayesian model comparison to interpret the results in a probabilistic way.
In practice, we identify different modes (higher probability contours in the multi-dimensional parameter space separated by lower probability valleys) in the posterior probability distribution of model parameters provided by \multinest\ (Appendix~\ref{app:multinest}) and compute the local evidence within each mode.  We then consider the two modes with largest local evidence, which we label $H_1$ and $H_2$ in order of decreasing local evidence. To assess the plausibility of these two solutions, we compute the `Bayes factor' \K\ given by the ratio of local evidences
\begin{equation}\label{eq:bayes_factor}
\K = \frac{P(\Db \mid H_1)}{P(\Db \mid H_2)} = \frac{\int P(\thetab_1 \mid H_1) \, P(\Db \mid \thetab_1, H_1)\,d\thetab_1}{\int P(\thetab_2 \mid H_2) \, P(\Db \mid \thetab_2, H_2)\,d\thetab_2} \, .
\end{equation} 
In this expression, the local evidence within each mode is the integral of the prior distribution times the likelihood (equation~\ref{eq:bayes}) in the subset of the entire parameter space occupied by that mode. We stress that this Bayesian framework enables the straightforward implementation of any type of prior distribution of model parameters (through equation~\ref{eq:bayes_factor}), such as those based on galaxy luminosity functions often adopted in photometric redshift codes \citep[e.g.][]{Benitez2000, Brammer2008}.

We analyse in this way the posterior probability distributions of all galaxies in the UVUDF catalogue and label as multi-modal those with redshift solutions separated by $\lvert \Delta z \rvert > 0.5$. This minimum $\Delta z$ threshold, which corresponds roughly to the typical posterior standard deviation of redshift for a faint source (with single solution) in the catalogue, allows us to remove multiple redshift solutions with statistically non-significant separation. To interpret the results of our analysis, we consider the quantity $2\ln \K$, which, according to \citet{Kass1995}, is better suited than the Bayes factor \K\ itself to drive the choice between different modes.\footnote{The choice of an optimal scale for the interpretation of the Bayes factor is a classical challenge in statistics, with no unique solution. The scale adopted here, firstly proposed by \citet{Kass1995}, is similar to the original scale of \citet{Jeffreys1961}, hence switching between the two scales will not significantly alter our conclusions.} Fig.~\ref{fig:bayes_factor} shows the distribution of $2\ln \K$ for the 3243 UVUDF galaxies identified to have multiple redshift solutions separated by $\lvert \Delta z \rvert > 0.5$. Dashed vertical lines divide the $2\ln \K$ axis into the four `belief' categories defined by \citet[][regions A, B, C and D]{Kass1995} and reported in Table~\ref{tab:bayes_factor}. As the table indicates, about 18 per cent (1800) of all galaxies in the UVUDF catalogue exhibit multiple redshift solutions separated by  $\lvert \Delta z \rvert > 0.5$, for which the Bayes factor does not allow a highly confident redshift selection. This is when adopting a high-confidence threshold, corresponding to $\K = \exp(5) \approx 150$ (i.e. $2\ln \K$=10), as recommended by \citet{Kass1995}. Adopting instead a lower threshold, corresponding to $\K = \exp(3) \approx 20$ (i.e. $2\ln \K$=6), would lower the fraction of galaxies with ambiguous redshift solutions to 13 per cent (Table~\ref{tab:bayes_factor}).

The distribution of Bayes factors in Fig.~\ref{fig:bayes_factor} and Table~\ref{tab:bayes_factor} may also be used to gain insight into the origin of the widely different redshift solutions found for some objects using different photometric redshift codes (illustrated by the outliers in Figs~\ref{fig:posterior_A} and \ref{fig:posterior_B}). In fact, as the rightmost columns of Table~\ref{tab:bayes_factor} show, about half of the outliers in the \zbeagle\ versus \zbpz\ and \zbeagle\ versus \zeazy\ comparisons, defined as galaxies with $\lvert \Delta z \rvert / (1+z) > 0.25$ (Section~\ref{sec:UVUDF_model} and Table~\ref{tab:model1_photo_z}), are objects with multiple, ambiguous redshift solutions ($2\ln \K \le 10$; see section~5  of \citealt{Ilbert2006} for a similar conclusion based on a comparison between spectroscopic and photometric redshifts, and the discussion of the `reliability parameter' in section~4.2 of \citealt{Brammer2008}). Moreover, for about 75 per cent of the outliers with ambiguous redshift solutions, the second redshift solution identified with the \beagle\ tool matches the redshift estimated by \citet{Rafelski2015} with the \bpz\ or \eazy\ codes. 

It is important to stress that insights of the kind provided by Fig.~\ref{fig:bayes_factor} and Table~\ref{tab:bayes_factor}, and the associated ability with the \beagle\ tool to reduce by over an order of magnitude (from 12 to 1) the number of outliers in the \zbeagle\ versus \zspec\ comparison through the exclusion of objects identified to have ambiguous redshifts (modulo a reduction of the sample size; see Section~\ref{sec:photoz}), represent a new, promising way to study the origin of photometric redshift outliers, from which future large photometric surveys, such as those planned with {\it Euclid} and the Large Synoptic Survey Telescope (LSST), can greatly benefit.

\section{Relation to existing SED fitting codes}\label{sec:comparison}

\renewcommand{\arraystretch}{1.5}
\begin{table*}
\begin{threeparttable}

	\centering
	\begin{tabular}{L{0.18\linewidth-2\tabcolsep} C{0.15\linewidth-2\tabcolsep} C{0.10\linewidth-2\tabcolsep} C{0.20\linewidth-2\tabcolsep} C{0.16\linewidth-2\tabcolsep} C{0.11\linewidth-2\tabcolsep} C{0.10\linewidth-2\tabcolsep} }
\toprule

\multirow{2}{*}{\textbf{Name and reference}}	    &  \multicolumn{3}{c}{\textbf{Non-stellar emission component}}	& \multicolumn{3}{c}{\textbf{Type of SED to be interpreted}} \\     

 \cmidrule{2-4}  \cmidrule{5-7} 

						    & Nebular & Dust & AGN & Photometry & Spectroscopy & Mix \\

\midrule

\textsc{cigale}\tnote{a} 		 	&    yes  \hskip1truecm (approximate)	&  yes 		   & partial  \hskip1.5truecm (accretion disc, dust torus) 	& yes  & no & no \\

\textsc{magphys}\tnote{b}  		&    H$\alpha$ and H$\beta$ only	&  yes 		   & no									& yes  & no & no \\

\textsc{galmc}\tnote{c}  		& yes  \hskip1truecm  (approximate)	  &  no 		   & no									& yes  (+ photo-$z$) & no & no \\

\textsc{bayesed}\tnote{d} 		&    no			&  no 		   & partial (dust torus)									& yes  (+ photo-$z$) & no & no \\

\beagle\ (this work)   		&    yes 			&  no 		   & no									& yes   (+ photo-$z$) & yes & yes \\

$\beagle\,2.0$  (in prep) 		&    yes 			&  yes 		   & yes									& yes   (+ photo-$z$) & yes & yes \\

\bottomrule
	\end{tabular}
\begin{tablenotes}
\item [a] \citep{Burgarella2005, Noll2009, Ciesla2015}
\item [b]  \citep{daCunha2008,daCunha2015}
\item [c]  \citep{Acquaviva2011, Acquaviva2015}
\item [d]  \citep{Han2012, Han2014} 
\end{tablenotes}
\caption{Main characteristics of different publicly available Bayesian SED fitting codes. In the case of photometric SED fitting, the ability to estimate photometric redshifts is indicated in parentheses.}
\label{tab:code_compar}	
\end{threeparttable}
\end{table*}

The interpretation of galaxy SEDs at ultraviolet, optical and infrared wavelengths in terms of physical parameters has been the subject of many studies, leading to the development of several public codes. In practice, most codes designed to fit galaxy spectra (e.g., \textsc{ppxf}, \citealt{Cappellari2004}; \textsc{starlight}, \citealt{Cid2005}; \textsc{steckmap}, \citealt{Ocvirk2006}; \textsc{ulyss}, \citealt{Koleva2009}) suffer from intrinsic limitations. For example, to select the best-fitting models, most codes rely on simple $\chi^2$ minimisation techniques, which do not allow the computation of realistic uncertainties in the derived physical parameters. Classical codes also tend to be programmed in `interpreted' (e.g., \textsc{idl}, Yorick, \textsc{matlab}) rather than `compiled' (e.g., \textsc{fortran}, \textsc{c}, \textsc{c}++) languages, at the expense of performance.\footnote{An exception is the public software \textsc{starlight} \citep{Cid2005} which is written in \textsc{fortran}, but the code is not open-source.} In addition, standard SED interpretation tools are usually tied to specific choices of built-in physical ingredients (e.g., stellar evolution and dust attenuation prescriptions), which cannot easily be changed, nor tested (e.g., by means of posterior predictive checks; see Section~\ref{sec:PPC}), nor extended to include new physical ingredients (e.g., emission from an AGN, neutral ISM absorption). An additional specificity of most current tools focused on the interpretation of photometric (rather than spectroscopic) galaxy SEDs is that these tend to be optimised for either redshift estimation (e.g., \textsc{hyperz}: \citealt{Bolzonella2000};  \textsc{lephare}: \citealt{Arnouts1999,Ilbert2006};  \textsc{kcorrect}: \citealt{Blanton2007};  \eazy: \citealt{Brammer2008}) or the determination of galaxy physical parameters (e.g., \textsc{cigale}: \citealt{Burgarella2005, Noll2009}; \textsc{magphys}: \citealt{daCunha2008,daCunha2015}; \textsc{fast}: \citealt{Kriek2009}), but not both simultaneously. 

In this context, and given the large number of existing SED fitting codes, we restrict our discussion in this section to the comparison of the \beagle\ tool with publicly available codes relying on an analogous Bayesian approach: \textsc{cigale}, \textsc{magphys}, \textsc{galmc} and \textsc{bayesed}, of which we report the main characteristics (and references) in Table~\ref{tab:code_compar}. As indicated, none of these codes is designed to interpret spectroscopic galaxy observations. Also, among them, only the \textsc{cigale} code includes at the same time nebular, dust and AGN emission, although in an approximate way. In this code, for example, nebular emission is incorporated using only two fixed emission-line templates to represent the emission from gas heated by young stars, ignoring the contributions by recombination continuum radiation and any AGN component. The \textsc{magphys} code has been designed to consistently interpret ultraviolet to far-infrared SEDs, by adopting an `energy budget' approach to account for dust emission \citep{daCunha2008}. This code does not include a full model of nebular emission, accounting for only H$\alpha$ and H$\beta$ emission, and it is based on a wide (albeit predefined) library of galaxy star formation histories. Finally, the  \textsc{bayesed} code does not include models for nebular and AGN emission, and it is based on a rigid model (exponentially declining function) to describe a galaxy star formation history.

The limitations of existing SED fitting codes mentioned above and in Table~\ref{tab:code_compar} are the main motivation for our development of the \beagle\ tool presented in this paper. This tool incorporates the most recent  prescriptions for stellar and nebular emission (and the dependence of these components on chemical composition), attenuation by dust, IGM absorption, etc., in a physically consistent and highly flexible way (Section~\ref{sec:model}): the modular design of the \beagle\ tool allows any of these prescriptions to be easily replaced by an alternative one. It also enables the straightforward implementation of additional physical ingredients, such as described in Section~\ref{sec:conclusions} for the \beagle\,2.0 version. The Bayesian approach adopted in the tool allows the user to rigorously quantify the uncertainties and degeneracies affecting model parameters. Finally, the use of MCMC techniques (implemented in the \multinest\ algorithm; see Appendix~\ref{app:multinest}) and a compiled language (\textsc{fortran} 2003) makes the exploration of complex, multi-dimensional parameter spaces much more efficient, and much less memory-demanding, than in conventional, grid-based approaches.

\section{Summary and future developments}\label{sec:conclusions}

We have introduced a novel tool, named \beagle, to model and interpret any combination of photometric and spectroscopic galaxy observation at ultraviolet to infrared wavelengths. Currently, the \beagle\ tool allows one to model the emission from stellar populations,  over wide ranges of age, metallicity and $\alpha$-element to iron abundance ratio, the emission from gas photo-ionized by young stars and the absorption of the light by interstellar dust and the intergalactic medium. The tool also includes a flexible parametrization of galaxy star formation and chemical enrichment histories, which can be drawn from analytic functions or from different flavours of galaxy formation models, such as phenomenological and semi-analytic models and hydro-dynamic simulations. A main strength of the \beagle\ tool with respect to other existing spectral analysis tools is the flexible, modular implementation of sophisticated prescriptions for the production of light and its transfer through the interstellar and the intergalactic media, in a physically consistent way. This enables one to adapt model complexity (i.e., the number of adjustable parameters let to vary freely; Section~\ref{sec:parametrization}) to the available observational constraints, without having to sacrifice the physical coherence of the model. Statistical inference on galaxy physical parameters from observations is achieved by means of a Bayesian approach. Unlike widely used statistical techniques focusing on simple point-wise estimates of best-fitting model parameters (e.g. minimum $\chi^2$) and confidence intervals (e.g. $\Delta\chi^2<1$), this approach allows the rigorous propagation of observational uncertainties into the output statistical constraints on model parameters. It can also reveal correlations (i.e. degeneracies) among model parameters, at the origin of multi-modal solutions, and provides a well-defined framework to account for parameter interdependency, to explore the properties of galaxy populations beyond those of single galaxies (through hierarchical modelling; Section~\ref{sec:statapp}) and to incorporate instrumental effects (Section~\ref{sec:instrum}).    

We presented a first application of the \beagle\ tool to interpret the photometric SEDs of 9927 galaxies in the redshift range $0.1\lesssim {z} \lesssim 8$ from the UVUDF sample of \citet{Rafelski2015}. Adopting a relatively simple model with 7 free parameters (stellar mass, star formation timescale, metallicity, specific star formation rate, attenuation by dust, redshift of observation and formation redshift of the oldest stars; see Table~\ref{tab:model1_priors}), we find that the photometric redshifts derived using the \beagle\ tool are globally consistent with the spectroscopic redshifts available for a small sub-sample of UVUDF galaxies and with the redshifts derived by \citet{Rafelski2015} for the full sample using two standard, dedicated photometric-redshift codes, \bpz\ and \eazy. The statistical sophistication of the \beagle\ tool allows us to gain unique quantitative insight into the origin of occasional discrepancies between photometric and spectroscopic redshifts, and between photometric redshifts estimated using different codes. Such outliers appear to arise mainly from the presence of multiple modes of comparable probability in the posterior probability distribution of model parameters, corresponding to different redshifts. In fact, the accurate Bayesian characterization of the posterior probability distribution of model parameters with the \beagle\ tool allows a rigorous study of multiple solutions and degeneracies between model parameters (Section~\ref{sec:PDF}). We have  illustrated the strength of posterior predictive checks in the framework of the \beagle\ tool to identify and interpret systematic offsets between models and data, pointing either to limitations in the data or necessary improvements of the models (Section~\ref{sec:PPC}). In the case of the UVUDF sample, a global (Bayesian) goodness-of-fit test indicates that the simplified model mentioned above reproduces well the photometric SEDs of 93 per cent of all galaxies in the catalogue. A complementary, graphical posterior predictive check further shows the potential of this approach to characterize systematic errors in the data, and limitations in the adopted physical model (Section~\ref{sec:PPC}).

The flexible \beagle\ tool is designed to evolved as more modules are incorporated to account for new physical ingredients (Fig.~\ref{fig:workflow}). The next version of the tool, currently in development ($\beagle\,2.0$), will include several novelties: (i) an enlarged grid of photo-ionization models describing the emission from gas in wider ranges of C/N/O abundance ratios, gas densities and IMF upper mass cutoffs than considered here \citep{Gutkin2016}. This is important to reproduce and interpret the emission from chemically pristine galaxies \citep[e.g.,][]{Erb2010,Stark2014}; (ii) a model to describe the emission from AGN narrow-line regions, fully consistent with the models of nebular emission from stellar populations \citep[][see in particular their section 2]{Feltre2016}. This will allow the reliable exploitation of emission-line diagnostic diagrams at rest-frame ultraviolet and optical wavelengths to interpret the emission from active and inactive galaxies \citep[e.g.,][]{Baldwin1981,Feltre2016}, hence opening a new window on studies of the co-evolution of black holes and galaxies; and (iii) a model to describe the ultraviolet and optical absorption features from stars and the neutral ISM in and around galaxies (Vidal-Garc\'ia et al., in preparation). Finally, to extend the capabilities of the \beagle\ tool at mid- and far-infrared wavelengths, we also plan to include models to describe the emission from dust heated by stars \citep{daCunha2008} and an AGN \citep{Fritz2006, Feltre2012}. 

With the addition of these and other future modules, the \beagle\ tool will incorporate a panchromatic, physically consistent description of galaxy SEDs. Together with the highly flexible implementation of star formation and chemical enrichment histories of galaxies, dust attenuation and IGM absorption, this will allow the coherent modelling and interpretation of any combination of photometric and spectroscopic galaxy observation, such as those gathered by modern ground-based (e.g., Atacama Large Millimeter Array, Extremely Large Telescopes) and space-based (e.g., {\it James Webb Space Telescope}) observatories, in terms of powerful constraints on galaxy formation models.

\section*{Acknowledgements}

We thank the anonymous referee for a thorough reading of the manuscript, which helped us improve this paper. We are also grateful to Emma Curtis-Lake and David Stenning for valuable comments on an early draft of the paper, which helped greatly improve our work. JC acknowledges support from FAPESP (project no. 2013/11837-9). JC and SC acknowledge support from the European Research Council via an Advanced Grant under grant agreement no. 321323 (NEOGAL).

\bibliographystyle{mnras}

\bibliography{Chevallard} 

\appendix

\section{Nested sampling with multinest}\label{app:multinest}

The \multinest\ package \citep{Feroz2008, Feroz2009} allows one to explore complex, multi-dimensional posterior probability distributions by appealing to the nested sampling algorithm of \citet{Skilling2006}. This algorithm was initially developed to calculate the Bayesian evidence (see equations~\ref{eq:bayes} and \ref{eq:evidence}), a quantity which requires long computational times when appealing to `standard' MCMC-based methods. The nested sampling algorithm transforms the problem of computing the multi-dimensional evidence integral  
\begin{equation}\label{eq:evidence}
\mathscr{Z} = \int \pi(\thetab) \, \mathcal{L}(\thetab) \, d\thetab   \, ,
\end{equation}
where $\pi(\thetab)$ and $\mathcal{L}(\thetab)$ refer to, respectively, the prior distribution and likelihood function for a model described by the parameters \thetab, into the computation of a 1-dimensional integral. 
This is based on the property that Bayesian evidence can be expressed as
\begin{equation}\label{eq:nested_Z}
\mathscr{Z}  = \int_0^1 \mathcal{L}(X) \, dX  \; ,
\end{equation}
where the function $\mathcal{L}(X) = F^{-1}\left [ X(l) \right]$ is the inverse of the prior hyper-volume $X$ over iso-likelihood hyper-surfaces $\mathcal{L}(\thetab) =l$, expressed by
\begin{equation}\label{eq:nested_X}
X(l)  = \int_{\mathcal{L}(\thetab) > l} \pi(\thetab) \, d\thetab \, .
\end{equation}
The domain of this integral is the volume enclosed by the iso-likelihood surface defined by the parameter $l$. Using the above transformation, the evidence can be computed in a straightforward way by numerically integrating equation~\eqref{eq:nested_Z} after calculating the function $X(l)$ for increasing likelihood thresholds $l$.

The difficulty of applying equation~\eqref{eq:nested_Z} to compute the evidence $\mathscr{Z}$ lies in the evaluation of the function $X(l)$ for increasing values of $l$, which correspond to iso-likelihood surfaces encompassing smaller and smaller regions of the prior volume. For this, \multinest\ employs the `simultaneous ellipsoidal nested sampling method': at iteration $i=1$, the algorithm starts by computing the likelihood function $ \mathcal{L}(\thetab^k)$ for $\N$ sets of parameters $\thetab^k$ (called `active points'), with $1\leq k\leq\N$, drawn randomly from the prior probability distribution. The point with lowest likelihood $l^i_\txn{min}$ is removed, and a new point $\thetab^\prime$ is drawn from the prior with the requirement that $\mathcal{L}(\thetab^\prime) > l_\txn{min}^i$. This is achieved by decomposing the prior volume into ellipsoids bounded by iso-likelihood surfaces corresponding to $\mathcal{L}(\thetab) = l_\txn{min}^i$, whose number and shapes are optimized via an `expectation-maximisation' method. At each iteration, the algorithm enables in this way the computation of $X(l^i_\txn{min})$ using equation~\eqref{eq:nested_X}, while the prior volume shrinks to regions of higher likelihood. The algorithm stops when the product of the shrunk prior volume $X(l_\txn{min}^i)$ and the likelihood $\mathcal{L} (\hat{\thetab})$, where $\hat{\thetab}$ is the active point with largest likelihood, falls below an adopted `evidence tolerance factor', i.e., $\mathcal{L}(\hat{\thetab}) \, X(l_\txn{min}^i) <  \mathscr{Z}_\txn{tol}$. 

As a side product of the evidence calculation, the \multinest\ algorithm computes the posterior probability $P(\thetab \mid \Db, H)$ of any set of parameters $\thetab$ ever drawn during an iteration, providing an ensemble of N$_\txn{out}^\txn{tot}$ sets of parameters weighted by the posterior probability distribution. This ensemble can be used to perform inference on model parameters, such as the computation of posterior means and marginal and joint distributions.

The interest of \multinest\ relative to other algorithms (such as MCMC) in the context of the current study is the possibility to identify multiple modes in the posterior distribution of model parameters, and to evaluate the `local' evidence in each mode (see Section~\ref{sec:multiple_solutions}). In practice, to analyse the SEDs of UVUDF galaxies with the model described in Section~\ref{sec:UVUDF_model} using \multinest, we must specify three parameters: the number of active points, the evidence tolerance factor and the `sampling efficiency'. The number of active points must be large enough to probe all potential modes of the posterior probability distribution, which we expect to be around 2 to 3 at most, mainly caused by degeneracies between redshift, mass and dust attenuation. We fix $\N=300$, similar to the value adopted by \citet{Feroz2009} to estimate cosmological parameters. We also fix $\mathscr{Z}_\txn{tol}=0.1$, lower than the value of 0.5 suggested by \citet{Feroz2009}. This is to accurately evaluate the evidence when computing the relative probabilities of different modes, for those galaxies exhibiting multi-modal solutions. The sampling efficiency is an additional factor introduced to account for the potentially inaccurate characterisation of the iso-likelihood surfaces by the ellipsoidal decomposition mentioned above. This parameter is the inverse of the factor by which the ellipsoids are `inflated' at each iteration, before a new candidate active point is drawn from these ellipsoids. A low sampling efficiency guarantees a more accurate evidence evaluation, but with a higher chance of drawing candidate active points that do not satisfy the criterion $\mathcal{L}(\thetab^\prime) > l^i_\txn{min}$, and hence must be rejected and redrawn. We adopt a sampling efficiency of 0.3, as recommended by \citet{Feroz2009} for an accurate estimate of the evidence. Finally, another adjustment of \multinest\ pertains to the `clustering algorithm' employed to search for different modes in the multi-dimensional posterior probability distribution of model parameters. Since this type of algorithm looses accuracy with increasing dimensionality, we restrict the clustering analysis here to those three parameters most correlated with one another, namely redshift, stellar mass and attenuation optical depth.

\section{Weighted sampling with replacement}\label{app:weighted}

We outline here the procedure used to produce `replicated observations' to perform posterior predictive checks (Section~\ref{sec:PPC}). For each observed galaxy analysed with the \beagle\ tool, \multinest\ outputs an ensemble of N$_\txn{out}^\txn{tot}$ sets of parameters weighted by the posterior probability distribution (Appendix~\ref{app:multinest}). We consider the subset $\Nout$ of these corresponding to posterior probabilities greater than $10^{-4}$ times that of the parameter set with largest posterior probability, noted $\tilde\thetab$, i.e., the $\Nout$ sets of parameters with $P(\thetab\mid \Db, H)>10^{-4} P(\tilde{\thetab} \mid \Db, H)$. From this ensemble of typically  $\Nout\sim10^3$ sets of model parameters, we draw \Nrep\ replicated parameter sets $\thetab^k$, with $1\leq k\leq\Nrep$, by appealing to a `weighted sampling with replacement’ algorithm. This means that the probability of drawing the parameter set $\thetab^i$ (with $1\leq i\leq\Nout$) is equal to the posterior probability $P(\thetab^i \mid \Db, H)$ (the `weight'). The drawn sets of parameters $\thetab^k$ are not removed from the ensemble of $\Nout$ points (hence the term `with replacement'), implying that any $\thetab^i$ can be drawn multiple times (in practice, this is the case only for those sets of parameters $\thetab^i$ with the largest weights).

\end{document}